\documentclass[numberedappendix]{emulateapj}
\usepackage{natbib}
\usepackage{amsmath}
\newcommand{\msol}{\,\textrm{M}_\sun}                
\newcommand{\kappagroup}{\overline{\kappa_{\rm group}}}

\shorttitle{Luminous and dark matter profiles from galaxies to cluster scales}
\shortauthors{Newman et al.}
\submitted{Accepted to ApJ: 2015 October 5}

\begin{document}
\title{Luminous and dark matter profiles from galaxies to clusters:\\
Bridging the Gap with Group-scale Lenses}
\author{Andrew B. Newman$^1$, Richard S. Ellis$^{2,}$\altaffilmark{4}, and Tommaso Treu$^3$}
\affil{$^1$ The Observatories of the Carnegie Institution for Science, 813 Santa Barbara St., Pasadena, CA 91101, USA; \href{mailto:anewman@obs.carnegiescience.edu}{anewman@obs.carnegiescience.edu}\\
$^2$ Department of Astrophysics, California Institute of Technology, MS 249-17, Pasadena, CA 91125, USA\\
$^3$ Department of Physics and Astronomy, University of California, Los Angeles, CA 90095, USA}
\altaffiltext{4}{Present address: European Southern Observatory (ESO), Karl-Schwarzschild-Strasse 2, 85748 Garching, Germany}

\begin{abstract}
Observations of strong gravitational lensing, stellar kinematics, and larger-scale tracers enable accurate measures of the distribution of dark matter (DM) and baryons in massive early-type galaxies (ETGs). While such techniques have been applied to galaxy-scale and cluster-scale lenses, the paucity of intermediate-mass systems with high-quality data has precluded a uniform analysis of mass-dependent trends. With the aim of bridging this gap, we present new observations and analyses of 10 group-scale lenses at $\langle z\rangle=0.36$ characterized by Einstein radii $\theta_{\rm Ein}=2\farcs5-5\farcs1$ and a mean halo mass of $M_{200}=10^{14.0} \msol$. We measure a mean concentration $c_{200}=5.0\pm0.8$ consistent with unmodified cold dark matter halos. By combining our data with other lens samples, we analyze the mass structure of ETGs in $10^{13}\msol-10^{15}\msol$ halos using homogeneous techniques. We show that the slope of the total density profile $\gamma_{\rm tot}$ within the effective radius depends on the stellar surface density, as demonstrated previously, but also on the halo mass. We analyze these trends using halo occupation models and resolved stellar kinematics with the goal of testing the universality of the DM profile. Whereas the central galaxies of clusters require a shallow inner DM density profile, group-scale lenses are consistent with a Navarro--Frenk--White profile or one that is slightly contracted. The largest uncertainties arise from the sample size and likely radial gradients in stellar populations. We conclude that the net effect of baryons on the DM distribution may not be universal, but more likely varies with halo mass due to underlying trends in star formation efficiency and assembly history.
\end{abstract}


\section{Introduction}

The distributions of dark and baryonic matter within galaxies of various masses is a key constraint on theories of galaxy formation and cosmology. In the standard cold dark matter (CDM) model, the distribution of dark matter (DM) across a wide range of scales is now well understood from $N$-body simulations \citep[e.g.,][]{NFW96,Diemand05,Gao12}. In realistic models of galaxy formation that include baryonic physics, however, the distributions of stars, gas, and DM depend on poorly understood processes such as gas cooling \citep[e.g.][]{Blumenthal86,Gnedin04}, thermal and mechanical feedback from supernovae \citep{Navarro96,Pontzen12} and active galactic nuclei \citep[AGN;][]{Martizzi13}, and dynamical heating in mergers \citep[e.g.,][]{ElZant01,Nipoti04,Tonini06,Laporte15}.

Detailed observations of the mass distribution therefore contain important information on the balance of these competing baryonic processes. Of particular interest is the radial density profile of DM on small scales, which is sensitive to this balance and may also constrain the microphysics of the DM particle \citep[e.g.,][]{Spergel00}. Since the relative importance of the various baryonic processes is expected to vary with a galaxy's mass and formation history, a valuable route to progress is to examine the distributions of dark and baryonic matter across galaxies, groups, and clusters, thereby spanning the full range of systems where the relevant observational techniques can be applied.

Strong gravitational lensing has emerged as a key technique for tracing the mass distribution for this wide range of systems, since it provides a geometric measure of the total mass within the Einstein radius (see, e.g., \citealt{Treu10ARAA} and \citealt{Treu14} for recent reviews). For the more massive systems, weak lensing allows the total mass to be traced to larger scales. Other observations, such as stellar kinematics on smaller scales where the stellar contribution is usually dominant, and satellite dynamics and X-ray emission on larger scales, enable the mass profile to be measured at several widely separated radii. This is essential to constrain multi-component models that separate the stellar and DM components.

The combination of strong lensing and stellar kinematics is now well established as a probe of the density profile of early-type galaxies (ETGs; \citealt{Treu02,Treu04,Jiang07,Auger10a,vandeVen10,Barnabe11,Barnabe13,Grillo13}). Based on more than 100 lenses discovered in the SLACS survey \citep{Bolton06,Bolton08,Shu15}, the logarithmic slope $\gamma_{\rm tot}$ (also denoted $\gamma'$) of the total density profile within the effective radius $R_e$, where $\rho_{\rm tot} \propto r^{-\gamma_{\rm tot}}$, has a mean value $\langle\gamma_{\rm tot}\rangle = 2.078 \pm 0.027$ and a fairly small scatter of $0.14 \pm 0.02$ \citep{Auger10a}. This has been interpreted as evidence for a ``conspiracy'' between DM and baryons that drives their combination to a nearly isothermal density profile \citep{Koopmans06,Koopmans09,Treu06,Gavazzi07}. The SLACS sample and more recent surveys (SL2S: \citealt{Ruff11,Sonnenfeld13a}; BELLS: \citealt{Bolton12}) have been used to constrain the stellar initial mass function (IMF) of massive ETGs \citep{Auger10b,Treu10} as well as the evolution of the density profile over cosmic time \citep{Sonnenfeld13b,Sonnenfeld14}. All of these studies pertain to \emph{galaxy-scale lenses} with Einstein radii in the range $1'' \lesssim \theta_{\rm Ein} \lesssim 2''$, velocity dispersions $\sigma \approx 250 \pm 40~{\rm km~s}^{-1}$, and halo masses $M_{200} \simeq 10^{13.2} \msol$ \citep{Gavazzi07}. 

Similar techniques have been extended to giant ellipticals in the centers of massive, relaxed clusters with $M_{200} \simeq 10^{15} \msol$. In a series of papers by \citet{Sand02,Sand04,Sand08} and \citet{Newman09,Newman11}, which culminated in a study of 7 systems \citep[][hereafter N13a, N13b, or N13 collectively]{Newman13a,Newman13b}, the average total density slope within $R_e$ was found to be $\langle \gamma_{\rm tot} \rangle = 1.16 \pm 0.05 {}^{+0.05}_{-0.07}$. This is much shallower than for galaxy-scale lenses and consistent with high-resolution \emph{dark matter only} simulations, despite the presence of significant stellar material on these scales. As a result, after decomposing the density profile into its stellar and DM components, N13 found the inner DM density profile to be \emph{shallower} than the canonical NFW slope, an intriguing result which several group of simulators have sought to explain \citep{Martizzi13,Laporte15,Schaller15b}.

Two natural questions are the origin of the wide range of total density profiles seen in ETGs, and whether the shallow DM profiles evident in brightest cluster galaxies (BCGs) are also present in lower-mass systems. The main goal of the present paper is to connect the trends observed in galaxy- and cluster-scale lenses by considering ETGs in \emph{intermediate mass halos} of $M_{200} \sim 10^{14} \msol$.

Several groups have recently performed systematic searches through wide-area imaging surveys to locate such lenses with intermediate Einstein radii $\theta_{\rm Ein} \simeq 2\farcs5-6''$. These include CASSOWARY \citep{Belokurov09,Stark13}, the Sloan Bright Arcs Survey \citep{Diehl09,Kubo09,Kubo10}, and the SL2S-ARCS sample \citep{Limousin09,More12}. We refer to these as \emph{group-scale} lenses in reference to their Einstein radii and halo masses (Section~\ref{sec:satellitekinematics}) that lie between galaxy- and cluster-scale lenses. This term does not imply that galaxy-scale lenses residing in lower-mass halos are not found in the enriched environments typical of massive galaxies, as many studies have shown \citep{Keeton00,Williams06,Auger07,Fassnacht08,Treu09}.

Follow-up studies of these group-scale lenses have focused on their halo masses and bulk mass-to-light ratio \citep{Thanjavur10,Munoz13}, scaling relations \citep{Foex13,Verdugo14}, and the mass--concentration relation (MCR) \citep{Verdugo11,Wiesner12,Auger13,Foex14}. However, very few of these group-scale lenses have been studied using the lensing and stellar dynamics approach to measure the mass distribution within $R_e$. \citet{McKean10} presented a detailed analysis of one radio-selected group-scale lens. Resolved stellar kinematic data are needed to separate the stellar and dark components, but to date only two group-scale lenses have been studied in this way \citep{Spiniello11,Deason13}.

In this paper we present new observations of a sample of 8 group-scale lenses using the DEIMOS spectrograph at the Keck~II telescope. The data allow us to the measure the radial stellar velocity dispersion profile of the brightest group galaxy (BGG) and to estimate the halo mass based on the kinematics of the satellite galaxies. In conjunction with strong lensing, these data provide mass measures at several widely separated radii. We combine the new sample with earlier data collected by \citet{Spiniello11} and \citet{Deason13} to create a sample of 10 well-studied lenses. The resulting sample fills in a long-standing gap in halo mass distribution of similarly analyzed lenses, enabling us to explore trends in the mass profiles of ETGs over halo masses of $10^{13} \msol - 10^{15} \msol$ using homogeneous techniques and data.

An outline of the paper follows. The reader interested in only the results and not the methodology may wish to begin in Section 7. We introduce the sample in Section~2. Sections 3--5 describe the observations and the associated constraints from kinematics (\S3), lensing (\S4), and stellar population synthesis (\S5). In Section 6, we describe our procedure for inferring mass models from these various data sets. In Section 7, we begin presenting our results with the MCR for our group-scale lenses. In Section 8, we move to smaller scales and examine the total density profile within $R_e$, combining our new group sample with earlier work on galaxy- and cluster-scale lenses to study trends over a factor of $\simeq 60$ in halo mass. In Section 9 we examine these trends using a set of CDM-motivated halo occupation models. We investigate trends in the DM density profile within $R_e$ using both these halo occupation models and direct mass modeling based on our resolved stellar kinematic data. Finally, in Section 10 we discuss the physical implications of our findings for models of ETG formation and the effect of baryons on their structure of their DM halos, before summarizing our key results in Section~11.

Throughout we use a $\Lambda$CDM cosmology with $\Omega_m = 0.3$, $\Omega_v = 0.7$, and $h = 0.7$. All magnitudes are expressed in the AB system. Stellar masses are based on a \citet{Salpeter55} IMF over $0.1-100 \msol$, and halo masses $M_{200}$ are defined relative to the critical density. We adopt a cosmological baryon fraction $\Omega_b / \Omega_m = 0.15$ where necessary \citep{Planck14}.

\begin{figure}
\centering
\includegraphics[width=\linewidth]{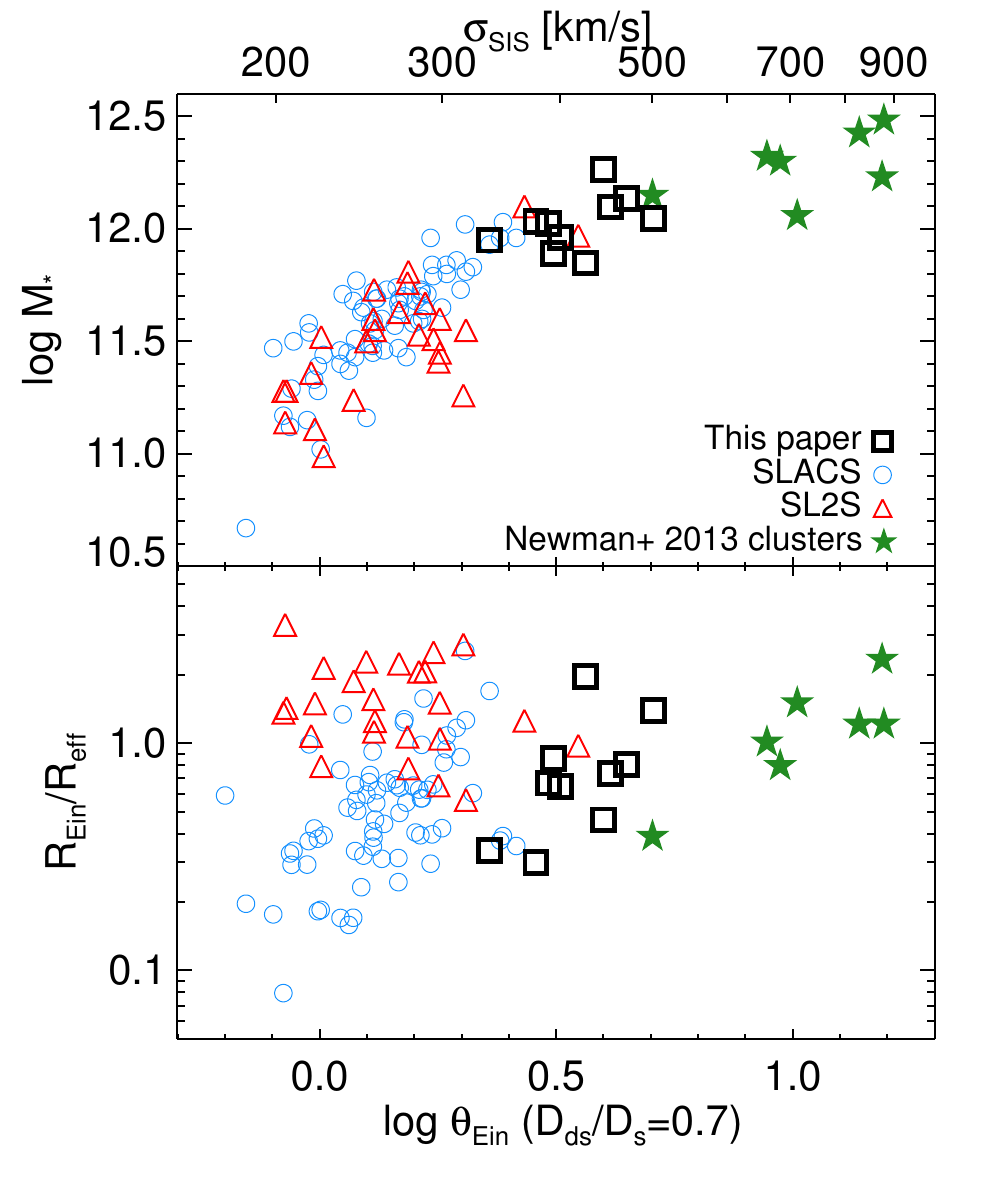}
\caption{Characteristics of ETGs for which strong lensing and stellar dynamics have been combined to measure the density profile within $\sim R_e$, as a function of angular Einstein radius $\theta_{\rm Ein}$. The present group-scale sample bridges the large samples of galaxy-scale lenses (SLACS, \citealt{Auger09}; SL2S, \citealt{Sonnenfeld13a,Sonnenfeld13b}) with the massive clusters analyzed by N13a, N13b. All $\theta_{\rm Ein}$ have been rescaled to a common $D_{\rm ds} / D_{\rm s} = 0.7$, and stellar masses $M_*$ have been homogenized to a Salpeter IMF. The top axis shows the velocity dispersion for a singular isothermal sphere with the indicated $\theta_{\rm Ein}$.\label{fig:rein}}
\end{figure}

\section{Group Lens Sample}
\label{sec:sample}

Our sample consists of 10 lenses selected to have Einstein radii $\theta_{\rm Ein}$ in the range $\approx 2\farcs5-5''$, whose basic characteristics and discovery references are listed in Table~\ref{tab:lenses}. For CSWA163 and CSWA1 we incorporated data published by \citet{Deason13} and \citet{Spiniello11}, while our analyses of the remaining 8 lenses are based on new observations. Figure~\ref{fig:rein} compares the present sample to the galaxy- and cluster-scale lenses that formed the basis of earlier joint analyses of strong lensing and stellar kinematics. The group-scale lenses have Einstein radii, stellar masses, and effective radii that bridge these earlier samples. In Section~\ref{sec:satellitekinematics} we show the average halo mass of our sample is $M_{200} = 10^{14.0} \msol$, indicative of a group or small cluster between the halo masses of the galaxy-scale ($M_{200} \simeq 10^{13.2} \msol$; \citealt{Gavazzi07}) and cluster-scale ($M_{200} = 10^{14.6-15.3} \msol$; N13a) lenses in Figure~\ref{fig:rein}.

\begin{figure*}
\centering
\includegraphics[width=0.48\linewidth]{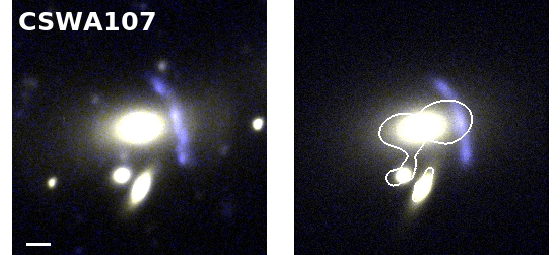} \hspace{1ex}
\includegraphics[width=0.48\linewidth]{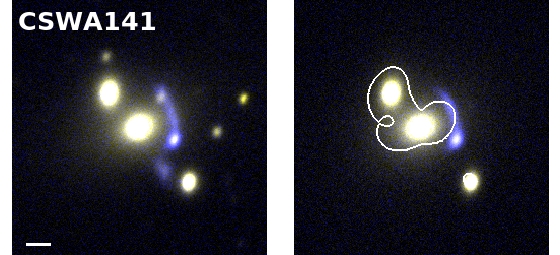} \\[1ex]
\includegraphics[width=0.48\linewidth]{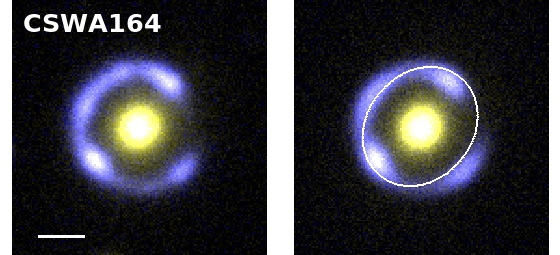} \hspace{1ex}
\includegraphics[width=0.48\linewidth]{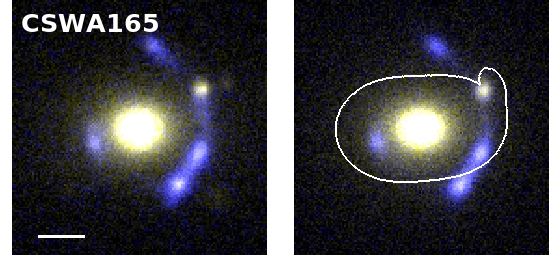} \\[1ex]
\includegraphics[width=0.48\linewidth]{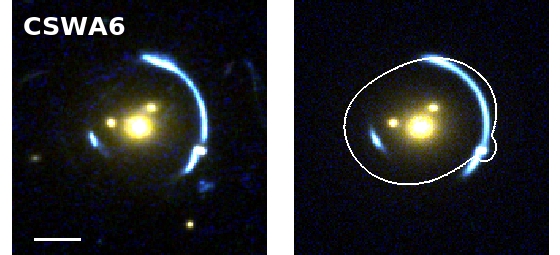} \hspace{1ex}
\includegraphics[width=0.48\linewidth]{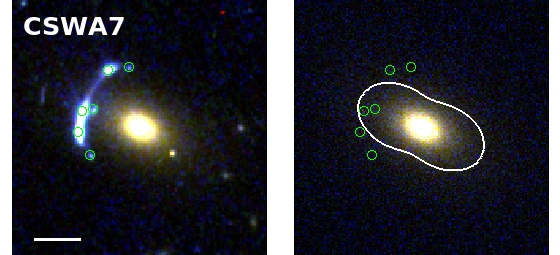} \\[1ex]
\includegraphics[width=0.48\linewidth]{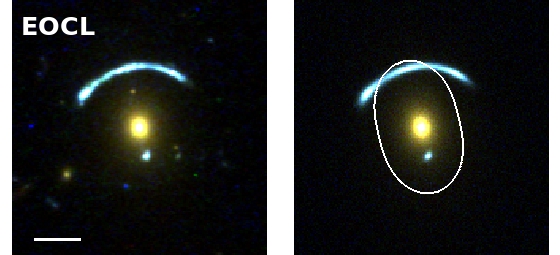} \hspace{1ex}
\includegraphics[width=0.48\linewidth]{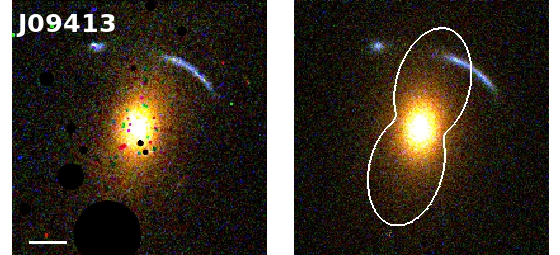} \\[1ex]
\includegraphics[width=0.48\linewidth]{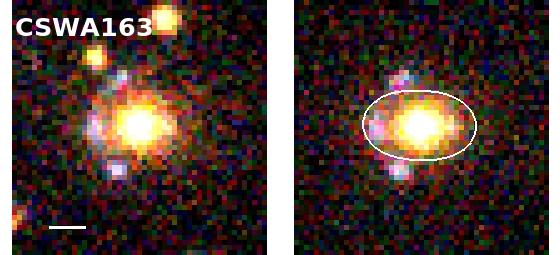} \hspace{1ex}
\includegraphics[width=0.48\linewidth]{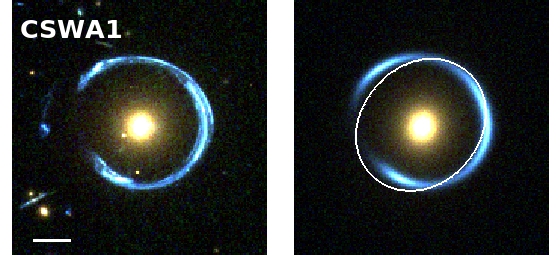}
\caption{\emph{Left panels:} Images of the inner regions of each lens from the sources described in Section~\ref{sec:imagingdata}. Rulers in the bottom-left corner have a length of 3''; note the scale varies among images. \emph{Right panels:} Images generated from the best-fitting analytic lens models described in Section~\ref{sec:lensing}. The solid line shows the outer critical curve. For the case of CSWA7, the positions of the compact multiple images (green circles) are used to constrain the lens model rather than the pixel-level data; this accounts for the lack of an arc image in its right panel.\label{fig:images}}
\end{figure*}

We required that our selected lenses be dominated by a BGG, since it is critical for our dynamical analysis that this galaxy be centrally located within the group-scale halo. Figure~\ref{fig:images} shows that the strong lensing region is dominated by a single deflector with the partial exception of CSWA141, which has a nearby satellite 1~mag fainter. On larger scales, we find that the central group galaxies are 0.9--1.7~mag brighter than the second-rank candidate group members.\footnote{We define these as the second-brightest galaxy within 500~kpc that has a photometric redshift in the Sloan Digital Sky Survey DR10 catalog \citep{Ahn14} within 0.1 of the lens galaxy. J09413 lies outside the SDSS footprint, so we instead use the images described in Section~\ref{sec:specobs} and compare to red sequence members.}

\begin{deluxetable*}{lllcccl}
\tablecaption{Lens Sample}
\tablehead{\colhead{Name} & \colhead{R.A.} & \colhead{Dec.} & \colhead{$z_L$} & \colhead{$z_S$} & \colhead{$\theta_{\rm Ein}$} & \colhead{References}}
\startdata
CSWA107 & 11:47:23.30 & +33:31:53.6 & 0.212 &1.205 & $2\farcs52$ & S13 \\   
CSWA141 & 08:46:47.46 & +04:46:05.1 & 0.241 & 1.425 & $3\farcs15$ & S13 \\  
CSWA164 & 02:32:49.87 & --03:23:26.0 & 0.450 & 2.518 & $3\farcs68$ & S13 \\  
CSWA165 & 01:05:19.65 & +01:44:56.4 & 0.361 & 2.127 & $4\farcs33$ & S13 \\   
CSWA6 (The Clone) & 12:06:02.09 & +51:42:29.5 & 0.433 & 2.00 & $4\farcs36$ & L09, S13 \\   
CSWA7 & 11:37:40.06 & +49:36:35.5 & 0.448 & 1.411 & $2\farcs73$ & K09, S13 \\  
8 O'Clock Arc (EOCL) & 00:22:40.91 & +14:31:10.4 & 0.380 & 2.73 & $3\farcs29$ & A07 \\  
J09413-1100 & 09:41:34.7 & --11:00:54.3 & 0.385 & \ldots & $4\farcs04$ & Li09 \\   
CSWA163 & 21:58:43.67 & +02:57:30.2 & 0.287 & 2.081 & $3\farcs49$ & S13, D13 \\
Cosmic Horseshoe (CSWA1) & 11:48:33.14 & +19:30:03.1 & 0.444 & 2.379 & $5\farcs08$ & B07, D08, S11
\enddata
\tablecomments{$\theta_{\rm Ein}$ is the Einstein radius as measured in Section~\ref{sec:lensing}. $z_L$ and $z_S$ are the lens and source redshifts, respectively. References: A07: \citet{Allam07}, D08: \citet{Dye08}, L09: \citet{Lin09}, Li09: \citet{Limousin09}, K09: \citet{Kubo09}, S11: \citet{Spiniello11}, D13: \citet{Deason13}, S13: \citet{Stark13}\label{tab:lenses}}
\end{deluxetable*}

\section{Spectroscopic Data}

Here we present spectroscopic observations of the BGGs and candidate group members. Our goals are (1) to measure the internal kinematics of the BGG through a radial velocity dispersion profile, and (2) to identify other group members whose velocities will constrain the halo mass. We also search for other line-of-sight structures that will inform our lensing analysis in Section~\ref{sec:lensing}.

\subsection{Observations and Reduction\label{sec:specobs}}

\begin{deluxetable}{llccccc}
\tablewidth{\linewidth}
\tablecaption{DEIMOS Spectroscopic Observations Log\label{tab:obslog}}
\tablehead{\colhead{} & \colhead{} & \colhead{$t_{\rm exp}$} & \colhead{} & \colhead{} & \colhead{P.A.} \\ \colhead{Name} & \colhead{Dates} & \colhead{(min.)} & \colhead{Masks} & \colhead{Redshifts} & (deg)}
\startdata
CSWA107 & 2013 Nov~27-28 & 180 & 2 & 102 & 83\\
CSWA141 & 2013 Nov~27-28 & 210 & 2 & 93 & 37\\
CSWA164 & 2013 Feb~10-11, & 242 & 0${}^{\rm a}$ & \ldots & $-74$\\
& 2013 Nov~28 & & & & \\
CSWA165 & 2013 Nov~27-28 &  196 & 2 & 78 & $-71$ \\
CSWA6 & 2013 Feb~11 & 180 & 1 & 60 & 25\\
CSWA7 & 2013 Feb~10 & 230 & 2 & 102 & 75 \\
EOCL &  2013 Nov~27 &  210 & 2 & 91 & 12 \\
J09413 &  2013 Feb~10-11 & 240 & 1 & 62 & $-30$
\enddata
\tablecomments{Exposure time $t_{\rm exp}$ refers to the total integration on the lens galaxy.\\${}^{\rm a}$ CSWA164 was observed using a long slit.}
\end{deluxetable}

We designed slit masks for the DEIMOS spectrograph \citep{Faber03} at the Keck~II telescope for the 8 groups listed in Table~\ref{tab:obslog}, targeting both the BGG and candidate satellites. Two masks were designed for 4 of the groups, with the satellite targets switched while the BGG slit was fixed on both masks. Three other groups were observed with a single slit mask, while the CSWA165 BGG was observed with a long slit. The 600 mm${}^{-1}$ grating was used in combination with the GG455 or GG495 blocking filters and a $1''$ slit. Candidate satellites were drawn from Sloan Digital Sky Survey (SDSS) catalogs \citep{Ahn14}, prioritizing red sequence members, followed by galaxies with photometric redshifts consistent with being members, and finally other bright galaxies in the field. Observations of the 8 groups were conducted over 4 nights in 2013 in clear conditions and seeing of $0\farcs7 - 0\farcs9$. Total exposure times on the BGGs were 3--4 hr.

The data were reduced using the {\tt spec2d} pipeline \citep{Cooper12,JNewman13}. Since the default sky subtraction routines are not appropriate for the extended BGGs, we adapted them to accommodate a more generous mask of the galaxy light. Redshifts of the satellite candidates were measured by cross-correlating with absorption- and emission-line templates using {\tt rvsao} \citep{rvsao} for 647 galaxies, whose velocities relative to the BGG were tabulated as $\Delta v = c (z - z_{\rm BGG}) / (1 + z_{\rm BGG})$.

\subsection{Lens Environments\label{sec:environments}}

\begin{figure*}
\includegraphics[width=\linewidth]{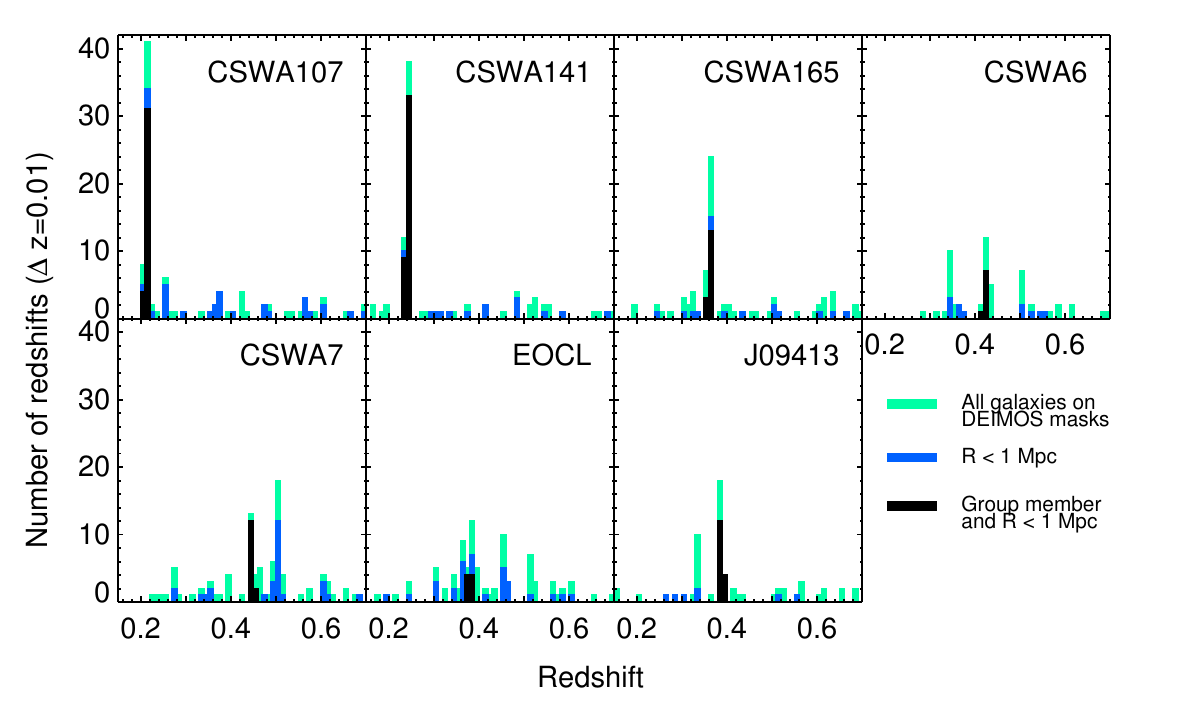}
\caption{Redshift distribution in the vicinity of each lens derived from our spectroscopic survey. All galaxies in the DEIMOS field of view are plotted. Dark colored histograms show the subset within $R < 1$~Mpc of the BGG, while the light blue histograms show the subset of these that are also identified as group members by the $R-v$ criterion shown in Figure~\ref{fig:caustics}.
 \label{fig:environment}}
\end{figure*}

We first use our redshift surveys to probe the large-scale environment of the groups and determine whether there are any line of sight structures relevant for our strong lensing analysis (Section~\ref{sec:lensing}), recognizing that our survey may be incomplete as it was intentionally biased toward group members. Nevertheless, structures are found in several cases. Figure~\ref{fig:environment} shows the redshift distribution around each of the 7 lenses that were observed in multi-slit mode.\footnote{Since the limiting magnitude of the SDSS catalogs reach fainter galaxies in lower redshift systems, this figure should not be used to gauge the relative richness of the groups.} CSWA165, CSWA107, and CSWA141 show no sign of additional structures in the field. J09413 is likewise dominated by the lensing group, with only a mild secondary peak that is located far from the lens (filled histograms show galaxies within 1~Mpc of the BGG). CSWA6 and CSWA7 each overlay comparably rich structures. EOCL is the most complex system, with three redshift peaks within 4000~km~s${}^{-1}$ of the lens. In Section~\ref{sec:lensing} we use these results to judge the contribution of external structures to our lens models. 

\subsection{Satellite Kinematics and Halo Masses}
\label{sec:satellitekinematics}

\begin{figure*}
\includegraphics[width=0.47\linewidth]{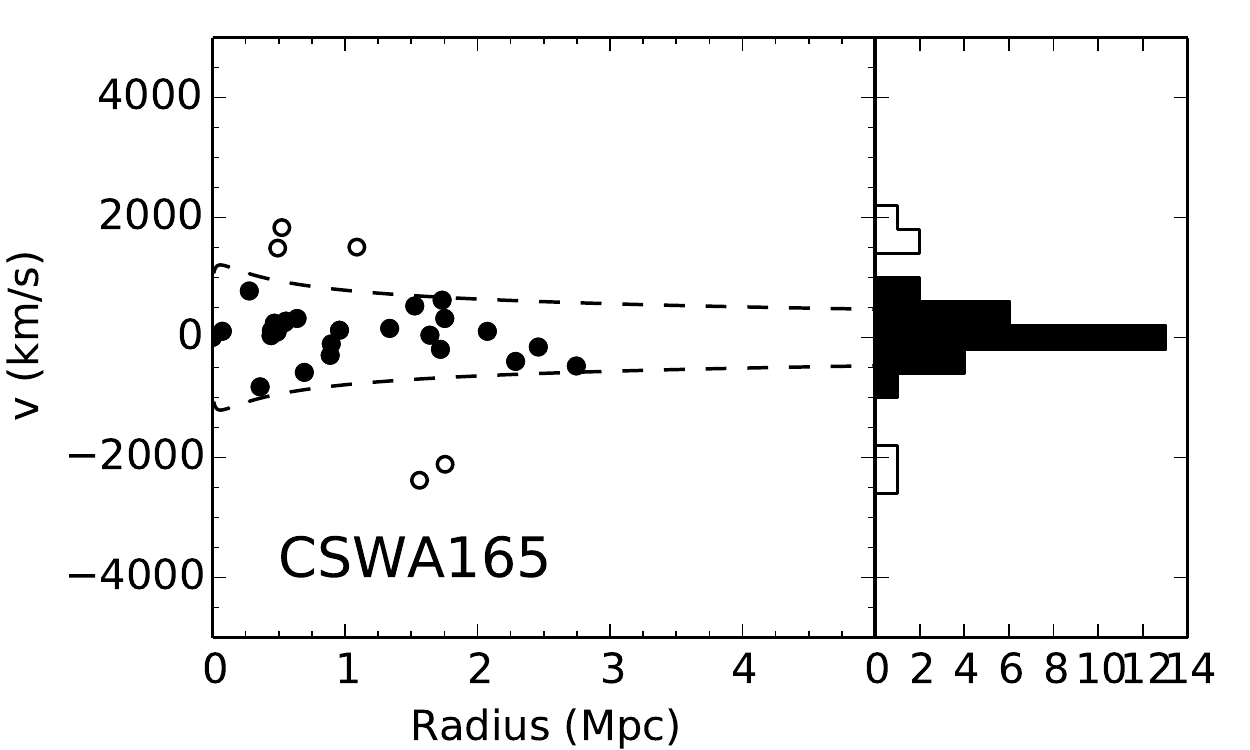}
\includegraphics[width=0.47\linewidth]{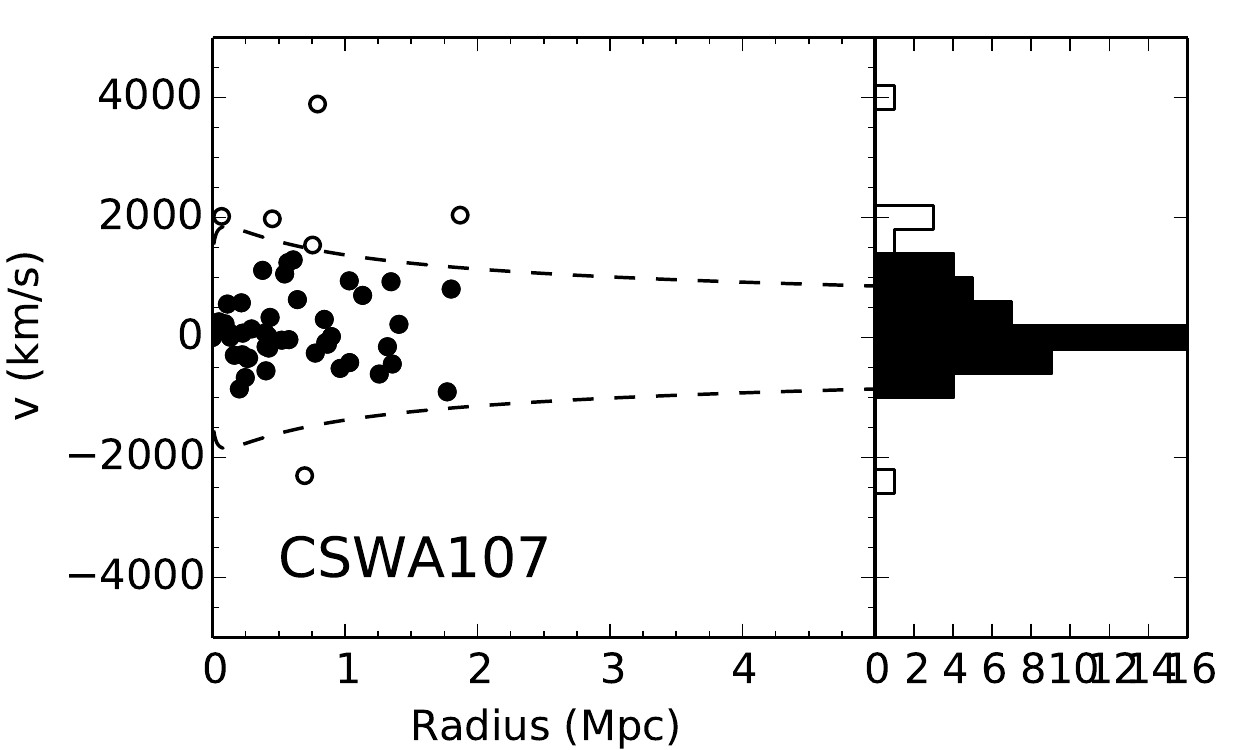}\\
\includegraphics[width=0.47\linewidth]{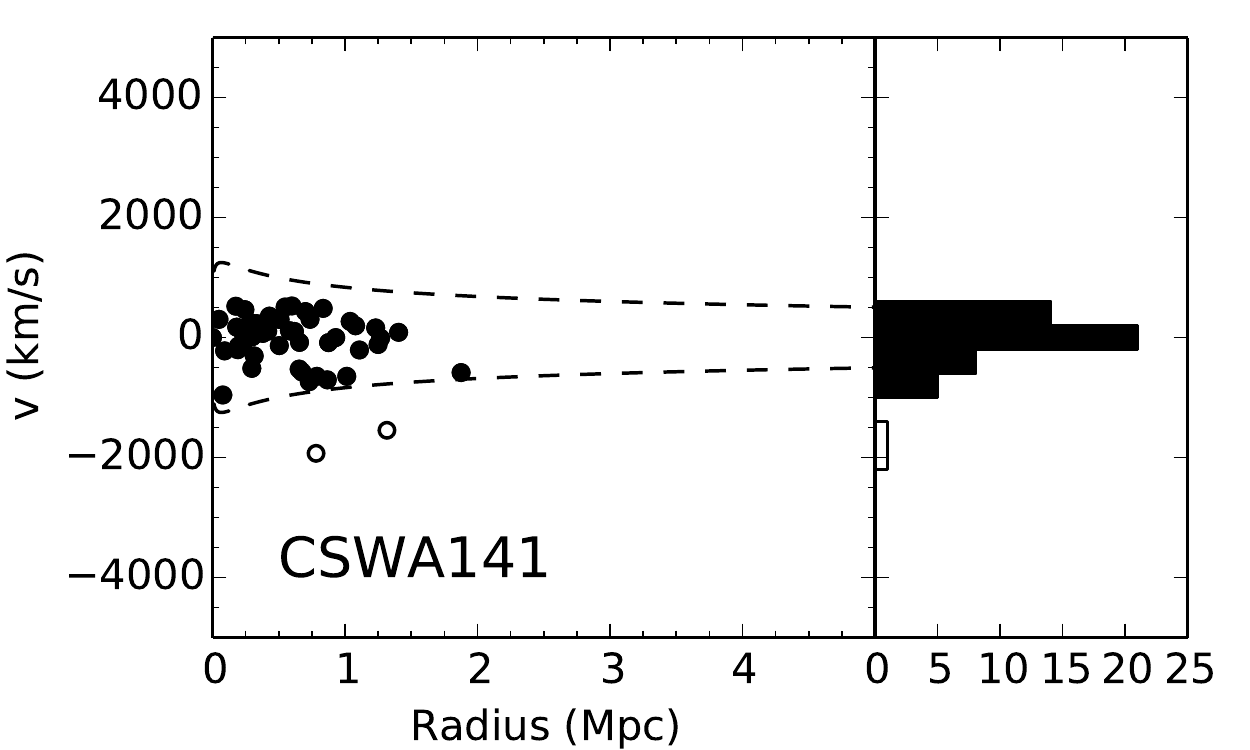}
\includegraphics[width=0.47\linewidth]{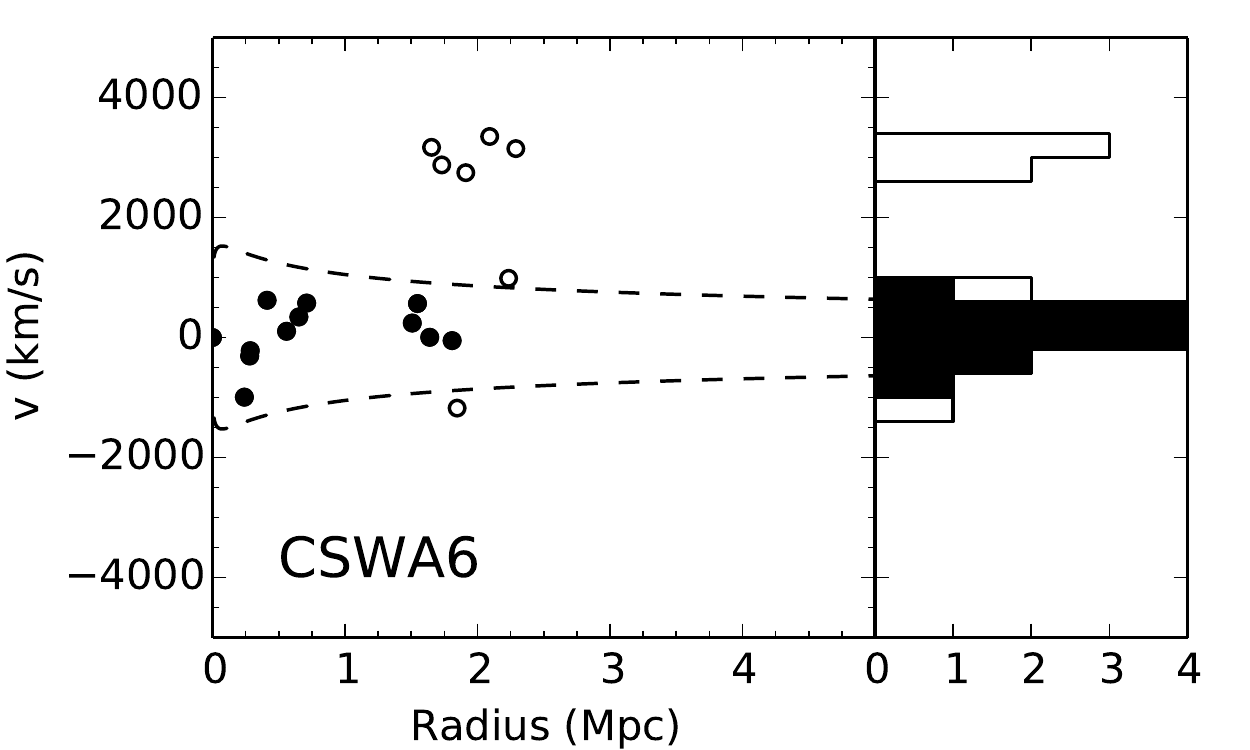}\\
\includegraphics[width=0.47\linewidth]{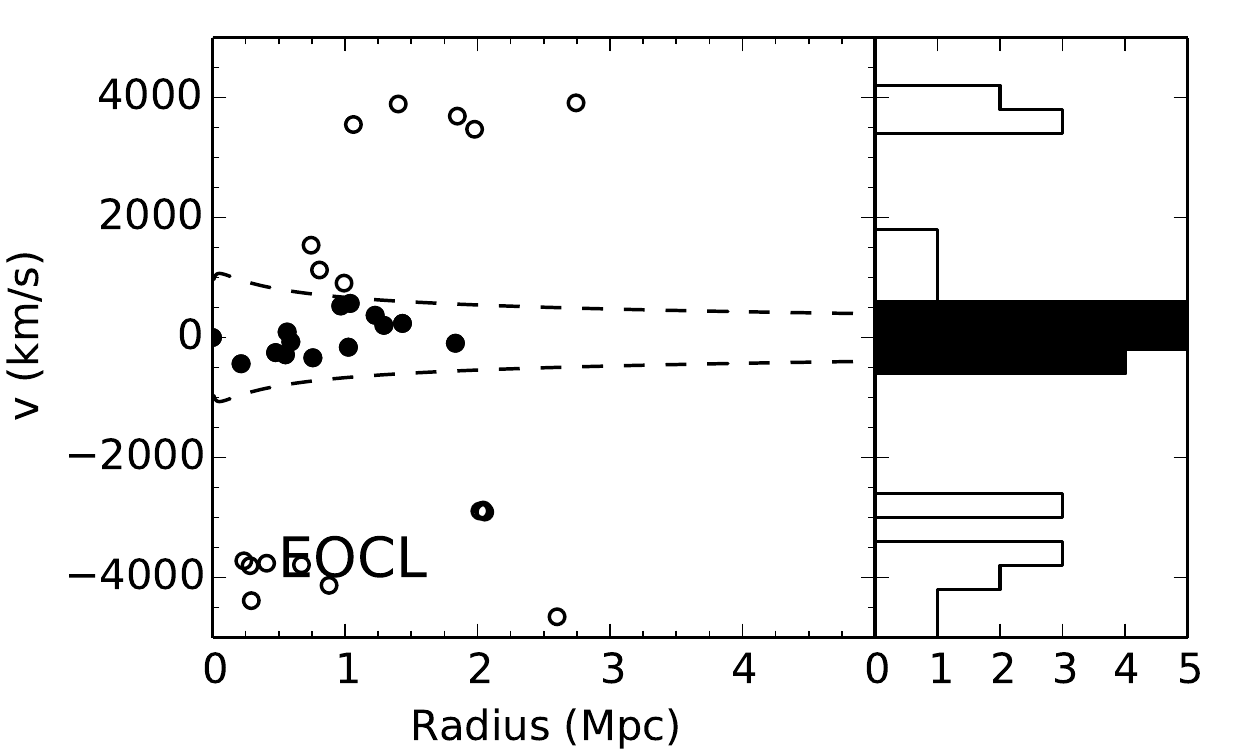}
\includegraphics[width=0.47\linewidth]{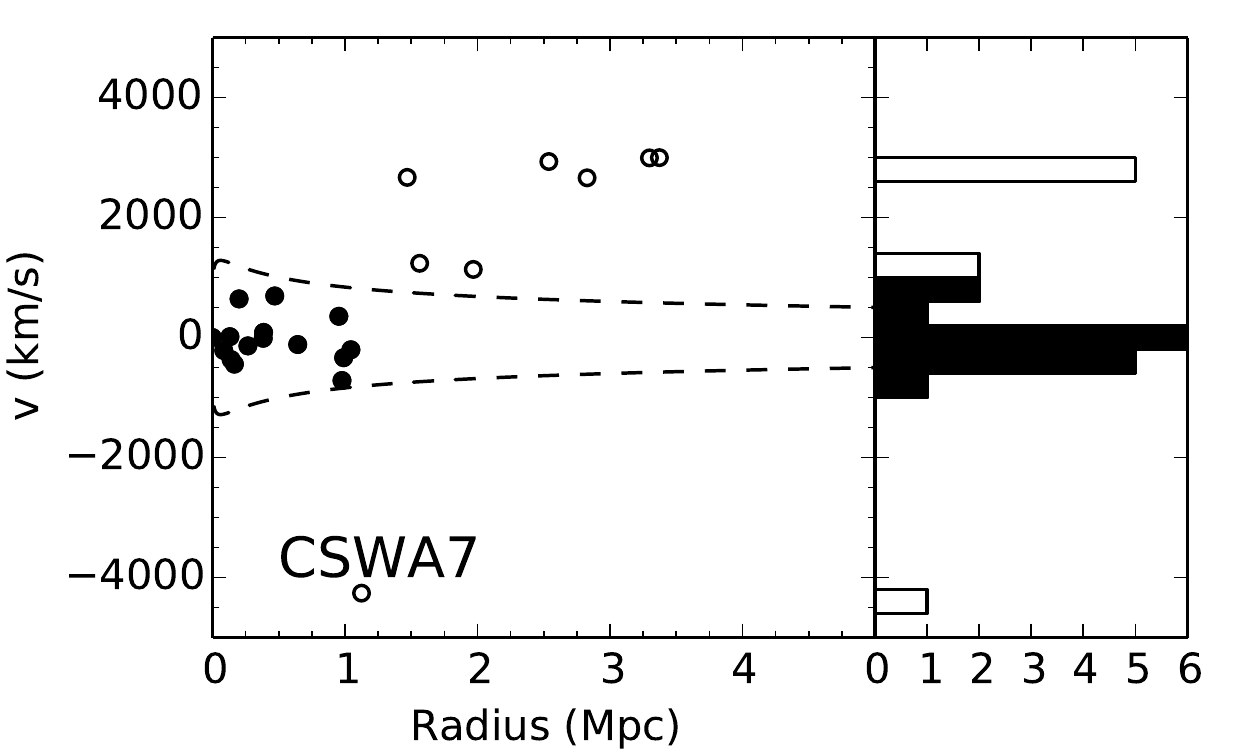}\\
\includegraphics[width=0.47\linewidth]{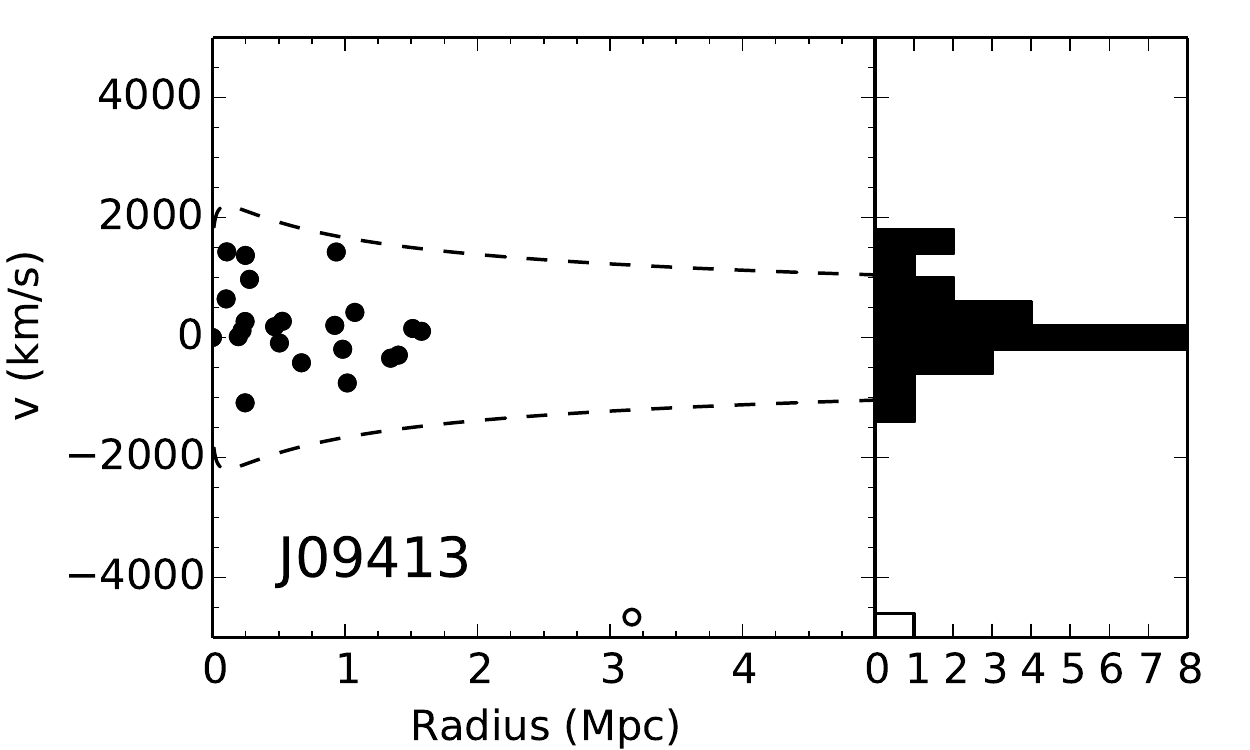}
\includegraphics[width=0.47\linewidth]{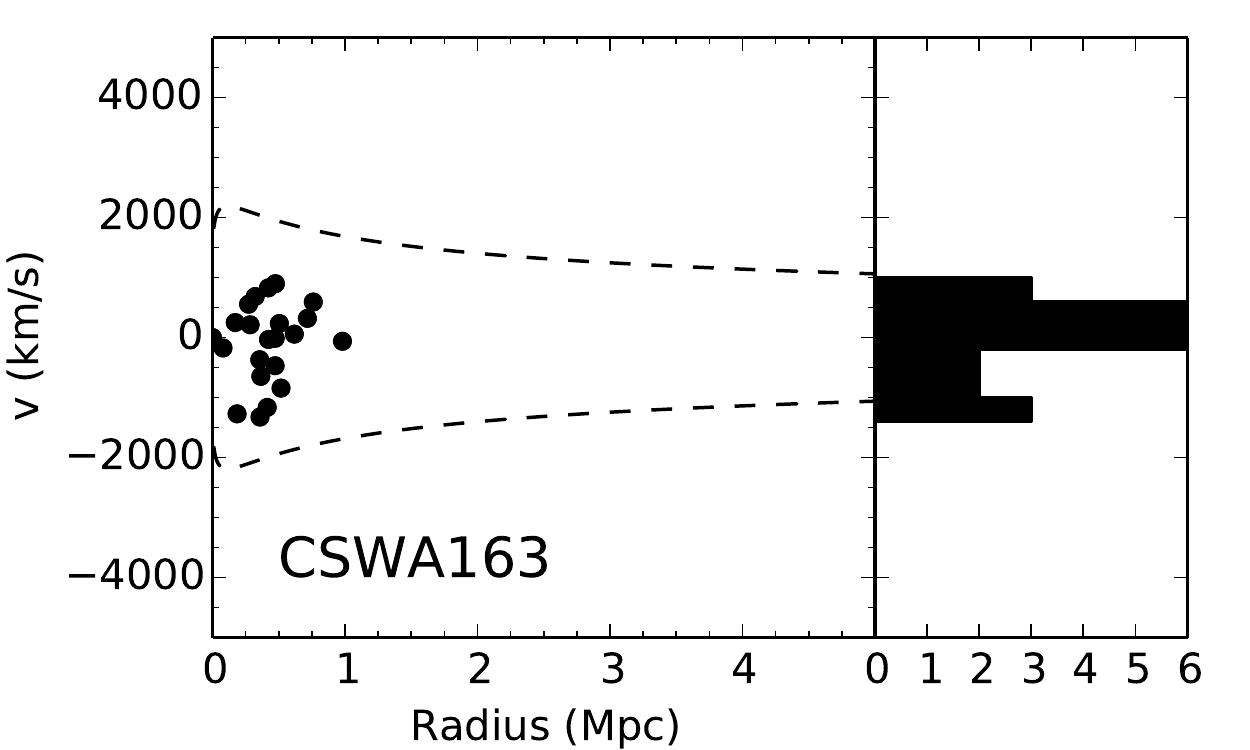}
\caption{Distribution of galaxy velocities relative to each BGG, $v = c(z - z_{\rm BGG})/(1+z_{\rm BGG})$, as a function of projected separation. Dashed curves show the region within which group members (filled circles) are selected using the iterative procedure described in Section~\ref{sec:satellitekinematics}. These members are used to calculate the satellite velocity dispersion $\sigma$, which in turn is used to estimate the halo mass $M_{200}$.\label{fig:caustics}}
\end{figure*}

\begin{deluxetable}{lcccc}
\tablecaption{Satellite Kinematics\label{tab:satellites}}
\tablehead{\colhead{Name} & \colhead{$N_{\rm gal}$} & \colhead{$\sigma$} & \colhead{$\langle v \rangle$} & \colhead{$\log M_{200} / \msol$} \\ \colhead{} & \colhead{} & \colhead{(km~s${}^{-1}$)} & \colhead{(km~s${}^{-1}$)} & \colhead{}}
\startdata
CSWA107 & 45 & $552 \pm 59$ & $38 \pm 91$ & $14.18 \pm 0.18$ \\
CSWA141 & 48 & $374 \pm 39$ & $45 \pm 64$ & $13.76 \pm 0.17$ \\
CSWA165 & 26 & $362 \pm 52$ & $82 \pm 61$ & $13.78 \pm 0.21$ \\
CSWA6   & 12 & $456 \pm 101$ & $124 \pm 175$ & $14.07 \pm 0.28$ \\
CSWA7   & 15 & $384 \pm 75$  & $-115 \pm 111$ & $13.89 \pm 0.25$ \\
EOCL    & 14 & $319 \pm 65$  & $3 \pm 124$ & $13.65 \pm 0.26$ \\
J09413  & 22 & $655 \pm 104$ & $144 \pm 118$ & $14.45 \pm 0.22$ \\
CSWA163 & 22 & $654 \pm 103$ & $-35 \pm 162$ & $14.40 \pm 0.22$
\enddata
\tablecomments{$N_{\rm gal}$ is the number of spectroscopically identified group members. $\langle v \rangle$ is the mean velocity of the satellites with respect to the BGG.}
\end{deluxetable}

The kinematics of the satellite galaxies of the group provide a measure of the mass on scales extending to virial radius. The first step in such an analysis is to identify the group members. We adopted an iterative cut in phase space that rejects galaxies with $|v - v_{\rm BGG}| > 3\sigma(R)$, where each velocity is compared to the local velocity dispersion $\sigma(R)$ appropriate to its projected group-centric radius $R$. For this purpose we set the shape of the $\sigma(R)$ profile using a fiducial NFW halo with concentration $c=4$ and isotropic orbits, while the overall velocity scale is then set by matching the aperture velocity dispersion within the virial radius to the measured value. \citet{Mamon13} advocate a similar cleaning procedure and provide useful analytic approximations (see also \citealt{Katgert04,Biviano06}).

Figure~\ref{fig:caustics} shows the $R-v$ plane for the eight groups with measured satellite kinematics (seven introduced in Section~\ref{sec:specobs} plus the \citealt{Deason13} data for CSWA163). The curves show the $3\sigma$ threshold for selecting group members identified with filled symbols. The effectiveness of this procedure is demonstrated by the rejection of some galaxies that might have been included by a simple velocity cut (in particular, several in CSWA7 and EOCL) which are found to be spatially coherent substructures. We then calculate the line-of-sight velocity dispersion $\sigma$ of the members using a simple standard deviation with an uncertainty estimated from Monte Carlo simulations. Table~\ref{tab:satellites} shows that $\sigma$ ranges from 319 to 655~km~s${}^{-1}$, with a median of 455~km~s${}^{-1}$. Furthermore, the BGGs are consistent with being at rest with respect to their satellites as expected if they are centrally located in their halos.

To estimate the halo mass, we use the scaling relation between $\sigma$ and $M_{200}$ determined by \citet{Munari13} in simulations:\footnote{We take the calibration in their Table~1 appropriate to galaxy tracers in simulations with AGN feedback. Varying the tracer and feedback physics changes the calibration by $\lesssim 0.08$~dex.} $\log h(z) M_{200} = 13.98 + 2.75 \log \sigma / (500~\textrm{km s}^{-1})$. We compared these masses with those obtained from two alternate approaches. \citet{Deason13} adapted the tracer mass estimator (TME) formulated by \citet{Evans03}, which they calibrated to $N$-body simulations. \citet{Zhang11} measured an empirical scaling relation between $\sigma$ and $r_{500}$ based on X-ray measurements.\footnote{Here we use their BCES bisector fit to the whole sample and convert $M_{200} = 1.38 M_{500}$; this conversion is exact for NFW halos with $c=5$ and depends weakly on concentration.} Using the TME or Zhang et al.~mass estimators shifts the halo masses systematically by $-0.05$~dex and $+0.08$~dex, respectively. We consider 0.08~dex as a reasonable estimate of the systematic uncertainty in the mass scale, which we add to the random errors in $\sigma$. The final halo masses and their uncertainties are listed in Table~\ref{tab:satellites} and span the range $\log M_{200}/\msol = 13.7-14.5$.

\begin{figure*}
\centering
\includegraphics[width=\linewidth]{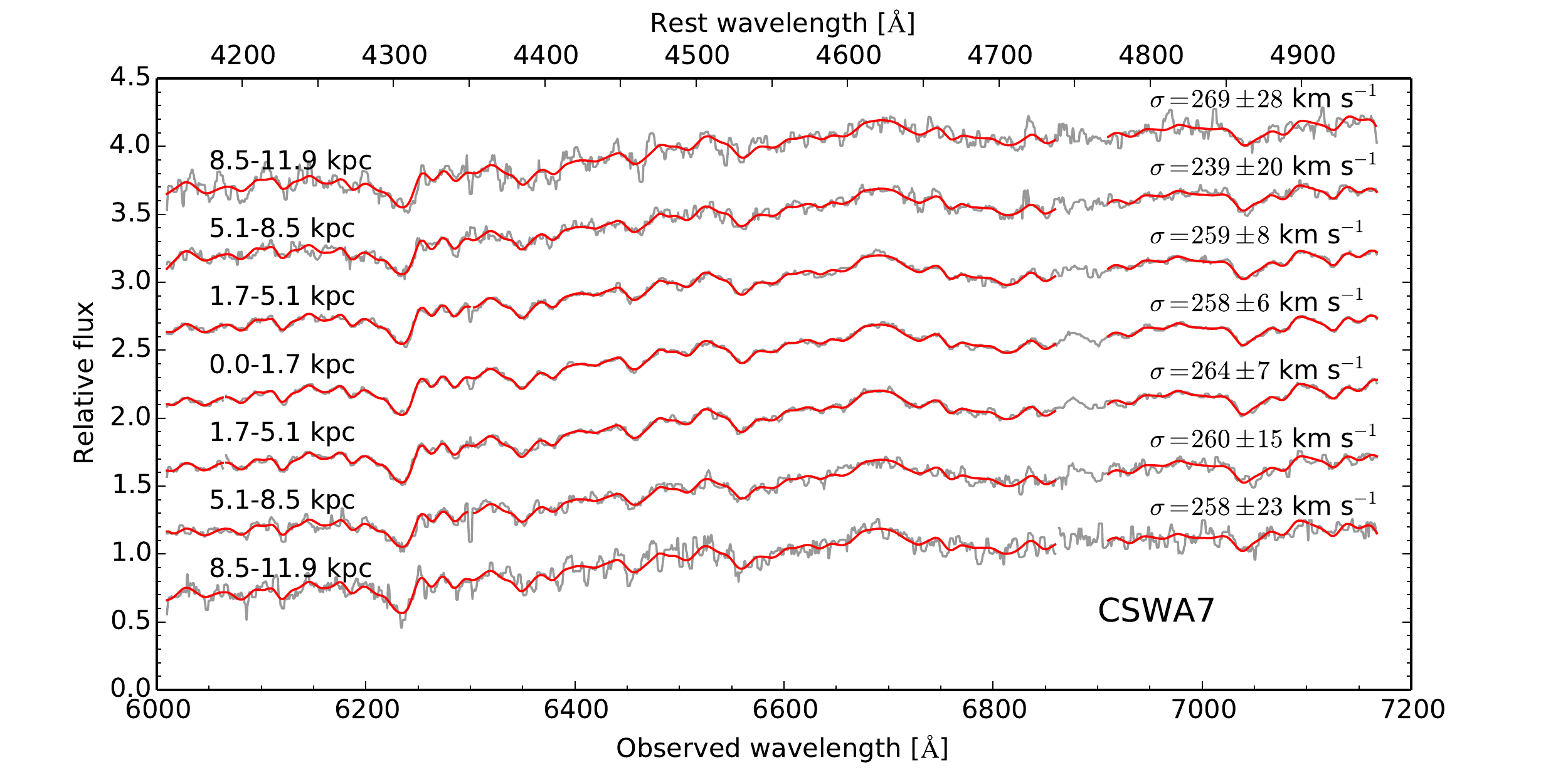}
\caption{Example of our resolved spectroscopy of BGGs in the case of CSWA7, showing the data (gray) and the convolved stellar template (red) used to extract the velocity dispersions in various spatial bins indicated in the figure. These bins are symmetric about the center of the galaxy. Spectra are smoothed with a 5~\AA~boxcar, normalized to a median flux of unity, and offset vertically for clarity. Spectral regions with uncertain calibrations were excluded from the fit and are not drawn in the model fits. Errors in $\sigma$ are statistical and do not include the systematic uncertainty of 5\% described in the text.\label{fig:specdemo}}
\end{figure*}

\subsection{Stellar Kinematics of the Central Galaxy\label{sec:vdprof}}

\begin{figure*}
\includegraphics[width=\linewidth]{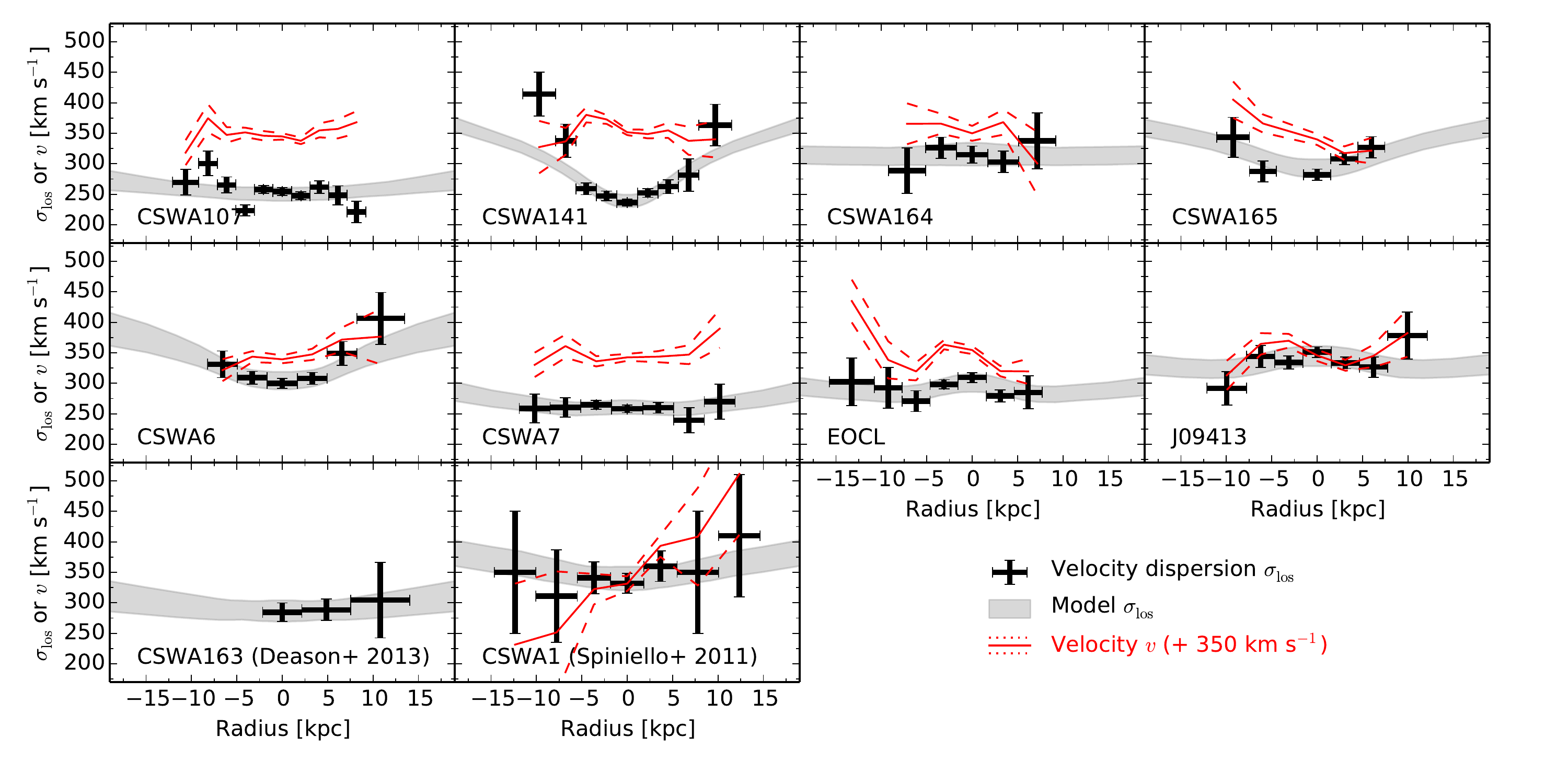}
\caption{Stellar kinematics of the BGGs extracted in spatial bins along the spectroscopic slit. Black crosses with error bars show the measured velocity dispersions $\sigma_{\rm los}$, while the red solid and dashed lines indicate the velocity profiles and their $\pm1\sigma$ uncertainties. Velocity profiles are shifted vertically so that the systemic velocity is at $+350$~km~s${}^{-1}$. The gray band shows the $\sigma_{\rm los}$ profiles of the fitted mass models, introduced in Section~\ref{sec:modeling}, and enclose 68\% of the posterior samples.\label{fig:dynamics}}
\end{figure*}

\begin{deluxetable*}{lccclccc}
\tablecaption{Velocity Dispersion Profiles of BGGs\label{tab:vdps}}
\tablewidth{0.75\linewidth}
\tablehead{\colhead{Name} & \multicolumn{2}{c}{Bin limits (arcsec)} & \colhead{$\sigma$ (km~s${}^{-1}$)} & \colhead{Name} & \multicolumn{2}{c}{Bin limits (arcsec)} & \colhead{$\sigma$ (km~s${}^{-1}$)}}
\startdata
CSWA107 &  $-3.50$ & $-2.67$ & $269 \pm 21$  & CSWA6 &  $-1.48$ & $-0.89$ & $331 \pm 21$ \\
\ldots &  $-2.67$ & $-2.07$ & $300 \pm 20$  & \ldots &  $-0.89$ & $-0.30$ & $309 \pm 10$ \\
\ldots &  $-2.07$ & $-1.48$ & $265 \pm 13$  & \ldots &  $-0.30$ & $0.30$ & $299 \pm 8$ \\
\ldots &  $-1.48$ & $-0.89$ & $223 \pm 9$  & \ldots &  $0.30$ & $0.89$ & $308 \pm 9$ \\
\ldots &  $-0.89$ & $-0.30$ & $257 \pm 6$  & \ldots &  $0.89$ & $1.48$ & $349 \pm 19$ \\
\ldots &  $-0.30$ & $0.30$ & $254 \pm 6$  & \ldots &  $1.48$ & $2.43$ & $406 \pm 42$ \\
\ldots &  $0.30$ & $0.89$ & $247 \pm 6$  & CSWA7 &  $-2.07$ & $-1.48$ & $258 \pm 23$ \\
\ldots &  $0.89$ & $1.48$ & $261 \pm 10$  & \ldots &  $-1.48$ & $-0.89$ & $260 \pm 15$ \\
\ldots &  $1.48$ & $2.07$ & $248 \pm 15$  & \ldots &  $-0.89$ & $-0.30$ & $264 \pm 7$ \\
\ldots &  $2.07$ & $2.67$ & $221 \pm 17$  & \ldots &  $-0.30$ & $0.30$ & $258 \pm 6$ \\
CSWA141 &  $-3.02$ & $-2.07$ & $414 \pm 36$  & \ldots &  $0.30$ & $0.89$ & $259 \pm 8$ \\
\ldots &  $-2.07$ & $-1.48$ & $337 \pm 27$  & \ldots &  $0.89$ & $1.48$ & $239 \pm 20$ \\
\ldots &  $-1.48$ & $-0.89$ & $258 \pm 9$  & \ldots &  $1.48$ & $2.07$ & $269 \pm 28$ \\
\ldots &  $-0.89$ & $-0.30$ & $247 \pm 8$  & EOCL &  $-3.02$ & $-2.07$ & $302 \pm 39$ \\
\ldots &  $-0.30$ & $0.30$ & $236 \pm 5$  & \ldots &  $-2.07$ & $-1.48$ & $292 \pm 33$ \\
\ldots &  $0.30$ & $0.89$ & $252 \pm 7$  & \ldots &  $-1.48$ & $-0.89$ & $270 \pm 17$ \\
\ldots &  $0.89$ & $1.48$ & $262 \pm 11$  & \ldots &  $-0.89$ & $-0.30$ & $298 \pm 7$ \\
\ldots &  $1.48$ & $2.07$ & $281 \pm 26$  & \ldots &  $-0.30$ & $0.30$ & $309 \pm 8$ \\
\ldots &  $2.07$ & $3.02$ & $363 \pm 34$  & \ldots &  $0.30$ & $0.89$ & $279 \pm 9$ \\
CSWA164 &  $-1.60$ & $-0.89$ & $288 \pm 37$  & \ldots &  $0.89$ & $1.48$ & $285 \pm 26$ \\
\ldots &  $-0.89$ & $-0.30$ & $326 \pm 17$  & J09413 &  $-2.31$ & $-1.48$ & $291 \pm 27$ \\
\ldots &  $-0.30$ & $0.30$ & $314 \pm 13$  & \ldots &  $-1.48$ & $-0.89$ & $344 \pm 17$ \\
\ldots &  $0.30$ & $0.89$ & $303 \pm 17$  & \ldots &  $-0.89$ & $-0.30$ & $334 \pm 10$ \\
\ldots &  $0.89$ & $1.60$ & $337 \pm 45$  & \ldots &  $-0.30$ & $0.30$ & $351 \pm 8$ \\
CSWA165 &  $-2.19$ & $-1.48$ & $343 \pm 32$  & \ldots &  $0.30$ & $0.89$ & $333 \pm 9$ \\
\ldots &  $-1.48$ & $-0.89$ & $287 \pm 17$  & \ldots &  $0.89$ & $1.48$ & $326 \pm 16$ \\
\ldots &  $-0.30$ & $0.30$ & $281 \pm 9$  & \ldots &  $1.48$ & $2.31$ & $378 \pm 38$ \\
\ldots &  $0.30$ & $0.89$ & $308 \pm 9$  &  & & & \\
\ldots &  $0.89$ & $1.48$ & $326 \pm 17$  &  & & & \\
\enddata
\tablecomments{Velocity dispersions are measured in rectangular apertures defined by the $1''$ slit width and the bin limits along the slit, which are tabulated relative to the galaxy center. Errors in $\sigma$ are statistical only and do not include the estimated 5\% systematic uncertainty. See \citet{Deason13} and \citet{Spiniello11}, respectively, for the CSWA163 and CSWA1 data.}
\end{deluxetable*}

Spatially resolved spectra of the BGGs were extracted and analyzed to derive a radial profile of the projected stellar velocity dispersion $\sigma_{\rm los}(R)$ following the procedures described in N13a. Briefly, extraction bins were constructed to ensure a minimum signal-to-noise ratio of 15~\AA${}^{-1}$ in the rest frame in the wavelength range $4150-4950$~\AA~around the $G$ band. Kinematics were measured using {\tt ppxf} \citep{Cappellari04}. Optimal stellar templates were constructed from a linear combination of spectra of G and K giants with metallicities near solar drawn from the MILES library \citep{MILES}. The templates were redshifted, convolved with a Gaussian (taking into account the instrumental resolution $\sigma = 78$~km~s${}^{-1}$), and added to a polynomial to filter the continuum, following well-established procedures. As a typical example, Figure~\ref{fig:specdemo} demonstrates the high quality of the resulting fits for CSWA7. We estimate a 5\% systematic uncertainty in $\sigma$ by varying the fitted wavelength region, the polynomial order, and the template library. Since the uncertainty is highly correlated amongst the spatial bins, we do not add this in quadrature to the random errors, but include a calibration factor with a Gaussian prior in our mass models (Section~\ref{sec:modeling}).  Figure~\ref{fig:dynamics} shows the derived velocity and velocity dispersion profiles. Rotational support is negligible or absent in every case, so we ignore it in our dynamical modeling. Our stellar velocity dispersion measurements are listed in Table~\ref{tab:vdps}.

\section{Lens Models and Galaxy Surface Photometry\label{sec:lensing}}

We now turn to our method for analyzing the strong lensing. After introducing the imaging data, we describe our technique for fitting the images at the pixel level using analytic models for the mass and light distributions of the lens and source. This allows a precise measurement of the Einstein radius. At the same time, we obtain multicolor surface brightness profiles of the BGGs, a key ingredient for our mass modeling in Section~\ref{sec:modeling}.

\subsection{Imaging Data\label{sec:imagingdata}}

CSWA6, CSWA7, and EOCL have been imaged by WFPC2 onboard the \emph{Hubble Space Telescope} (\emph{HST}) through the F450W, F606W, and F814W filters (program IDs 11167 and 11974, P.I.~Allam). J09413 was imaged using \emph{HST}/ACS through the F475W, F606W, and F814W filters (ID 10876, P.I.~Kneib). CSWA1 was observed with \emph{HST}/WFC3-UVIS through these same filters (ID 11602, P.I.~Allam). For CSWA163 we rely on $gri$ images from the SDSS. As part of the Keck/DEIMOS observations described in Section~\ref{sec:specobs}, we imaged CSWA107, CSWA141, CSWA164, and CSWA165 through the $B$ and $R$ filters. Exposure times ranged from 4 to 12~minutes with seeing of $0\farcs7-1\farcs0$. Astrometric and photometric solutions for the DEIMOS images were derived from stars in the SDSS catalog.

\subsection{Modeling the Lens Systems}\label{sec:lensing}

\begin{deluxetable*}{lcccccccc}
\tablecaption{Strong Lensing Constraints\label{tab:lensing}}
\tablehead{\colhead{Name} & \colhead{$\theta_{\rm Ein}$} & \colhead{$\overline{\kappa_{\rm group}}(\theta_{\rm Ein})$} & \colhead{$\gamma_{\rm SL}$} & \colhead{$q_{\rm mass}$} & \colhead{${\rm PA}_{\rm mass}$} & \colhead{$(\Delta {\rm RA}, \Delta {\rm Dec})$} & \colhead{$\Gamma_{\rm ext}$} & \colhead{$\theta_{\Gamma}$}} \startdata
CSWA107 & $2\farcs52$ & $0.96 \pm 0.10$ & 1.18 & 0.73 & $-89.6$ & \ldots & \ldots & \ldots \\
CSWA141 & $3\farcs15$ & $0.91 \pm 0.05$ & 1.23 & \multicolumn{2}{c}{(fixed to BGG)} & \ldots & \ldots & \ldots \\
CSWA164 & $3\farcs68$ & $1 \pm 0.05$ & 1.51 & 0.88 & $-18.5$ & $(0\farcs01, 0\farcs00)$ & 0.022 & 80.8 \\
CSWA165 & $4\farcs33$ & $0.97 \pm 0.05$ & 1.73 & 0.77 & $-72.3$ & $(0\farcs02, 0\farcs01)$ & 0.056 & $-66.6$ \\
CSWA6 & $4\farcs36$ & $0.79 \pm 0.10$ & 1.79 & 0.86 & $-60.2$ & $(-0\farcs27, 0\farcs30)$ & 0.012 & $-46.4$ \\
CSWA7 & $2\farcs73$ & $1 \pm 0.05$ & 1.85 & 0.67 & 61.6 & \ldots & 0.149 & 71.4 \\
EOCL & $3\farcs29$ & $1 \pm 0.05$ & 1.91 & 0.74 & 14.9 & $(0\farcs01, 0\farcs07)$ & 0.072 & 10.8 \\
J09413 & $4\farcs04$ & $1 \pm 0.07$ & 1.59 & 0.54 & $-14.2$ & \ldots & 0.087 & 57.0 \\
CSWA163 & $3\farcs49$ & $1 \pm 0.05$ & 1.74 & 0.71 & 87.9 & $(0\farcs02, 0\farcs02)$ & \ldots & \ldots \\
CSWA1 & $5\farcs08$ & $1 \pm 0.05$ & 1.66 & 0.90 & $-52.9$ & $(0\farcs03, 0\farcs06)$ & 0.022 & $-5.7$
\enddata
\tablecomments{$\overline{\kappa_{\rm group}}$ is the azimuthally averaged mean convergence of the main deflector measured within the Einstein radius $\theta_{\rm Ein}$. (For a single deflector, this is unity by definition, but differences arise when perturbing galaxies contribute convergence.) Offsets $(\Delta {\rm RA}, \Delta {\rm Dec})$ give the center of mass relative to that of light; where omitted, the center of mass is fixed. In models with external shear, $\Gamma_{\rm ext}$ and $\theta_{\Gamma}$ specify its amplitude and orientation (east of north).}
\end{deluxetable*}

We fit analytic models of the mass and light distribution to the multi-band data introduced above. By directly fitting the image pixels in several filters simultaneously, we naturally de-blend the lens and source galaxy light. Figure~\ref{fig:images} shows the regions around each lens used to fit the model, which roughly encompass the radii where uncertainties in the background level have a minimal effect. Light from the lens and source galaxies is modeled with a seven-parameter elliptical S\'{e}rsic profile. For the BGG, we let the magnitude and $R_e$ vary among filters to allow for color gradients, but we fit a common S\'{e}rsic index $n$, center $(x_0, y_0)$, position angle (P.A.), and axis ratio $q=b/a$. For the background source, only the magnitude is allowed to vary among filters.

The deflecting mass is modeled as a power law profile characterized by a slope $\gamma_{\rm SL}$, where $\rho \propto r^{-\gamma_{\rm SL}}$, and an Einstein radius $\theta_{\rm Ein}$. Ellipticity is introduced into the surface density following \citet[][see also \citealt{Keeton01}]{Schramm90}. The P.A., axis ratio, and center of the mass distribution are generally not tied to those of the light. The exceptions are the naked-cusp configurations (CSWA107, CSWA7, J09413) and CSWA141 (see Appendix~A) for which we found that the center and ellipticity cannot both be constrained. In these cases, we fixed the center of mass to that of the BGG. External shear with amplitude $\Gamma$ and orientation $\theta_{\Gamma}$ is also incorporated in the lens models where the data quality and image configuration can provide sufficiently useful constraints; this is the case except for CSWA107, CSWA141, and CSWA163.

In addition to the main deflector, we model the deflection from satellite galaxies in CSWA165, CSWA107, CSWA141, and CSWA6 where the perturbing galaxies are clearly visible in Figure~\ref{fig:images}. The satellite galaxy light was again modeled with a S\'{e}rsic profile, while the mass was treated as a singular isothermal ellipsoid (SIE) whose center, axis ratio, and P.A.~were fixed to those of the light. This leaves a single free parameter, $\sigma$. We place a Gaussian prior on $\sigma$ using the \citet{Faber76} relation, as measured by \citet{Bernardi03}. Similar procedures are commonly used in cluster lens modeling (see N13a and references therein).

We define the Einstein radius $\theta_{\rm Ein}$ such that $\overline{\kappa_{\rm tot}}(\theta_{\rm Ein}) = 1$, where $\overline{\kappa_{\rm tot}}$ is the azimuthally averaged mean convergence profile. We then define $\kappagroup$ as the convergence within $\theta_{\rm Ein}$ associated with the main deflector, i.e., the BGG and group-scale halo. This differs slightly from unity for the four lenses with perturbing satellites included in the model. $\kappagroup$ then serves as input to the mass models in Section~\ref{sec:modeling}.\footnote{Although the mass models in Section~\ref{sec:modeling} subdivide the main deflector into its stars and DM halo, the $\theta_{\rm Ein}$ derived from lensing using single power-law models is still valid, since $\theta_{\rm Ein}$ is known to be nearly independent of the mass profile (e.g., \citealt{Rusin03}, and tests in Section~4.2).} We describe our treatment of external convergence $\kappa_{\rm ext}$ along with our mass modeling procedure in Section~\ref{sec:kappaext}.

To fit the lens models, the parameter space is explored using {\tt MultiNest} \citep{Feroz09}, a Markov Chain Monte Carlo engine. For a given set of parameters, we generate galaxy images in the lens plane and ray-trace the source galaxy through the mass distribution. Images are generated for each observed filter and are convolved by the relevant point spread function (PSF). These model images are compared to the data to compute a likelihood $L \propto \exp(-\frac{1}{2}\chi^2)$. Figure~\ref{fig:images} shows that the best-fitting models provide acceptable fits. Although simple analytic models cannot be expected to trace the detailed source structure in all cases (e.g., CSWA1), they are adequate for measuring $\theta_{\rm Ein}$. Because its multiple images are virtually unresolved, CSWA7 was analyzed using a different technique that incorporates only the astrometric positions of these images (see Appendix~A).

With high-quality imaging data, the formal statistical errors on the mass model are very small; systematic differences are much more important. To assess these, we first modeled the lenses using a non-singular isothermal ellipsoid (NIE; \citealt{Kormann94}). We also tested models excluding external shear. Finally, we fit both NIE and power law models using the {\tt glafic} code \citep{Oguri10} and the positions of conjugate images as constraints, rather than the pixel-level data. By comparing these different methods, we find that $\kappagroup$ varies by less than about 0.05, which we take as a fiducial uncertainty. Higher uncertainties adopted in a few cases are discussed in Appendix A, where circumstances particularly to individual systems are reviewed and comparisons are made to models in previous publications.

The resulting lens model parameters are listed in Table~\ref{tab:lensing}. We generally find good alignment between mass and light: (1) spatial offsets are $< 0\farcs1$ for all systems except CSWA6, (2) the axis ratios $q$ agree on average, with a scatter of 0.1, (3) the position angles agree with a scatter of $16^{\circ}$. These comparisons support our assumption that the BGGs are centrally located within the group-scale halos.

\subsection{BGG Surface Photometry\label{sec:surfphot}}

\begin{deluxetable*}{lccccccccc}
\tablecaption{S\'{e}rsic Profile Fits to BGGs and Mass-to-Light Ratios from SPS Modeling\label{tab:bgg}}
\tablehead{\colhead{Name} & \colhead{$q$} & \colhead{P.A.} & \colhead{$n$} & \colhead{$R_{e, B}$} & \colhead{$L_{B}$} & \colhead{$R_{e, V}$} & \colhead{$L_{V}$} & \colhead{$\Upsilon^{\rm SPS}_{V,0}$} & \colhead{$\langle \Upsilon^{\rm SPS}_{V} \rangle$} \\
\colhead{} & \colhead{} & \colhead{(deg)} & \colhead{} & \colhead{(kpc)} & \colhead{($10^{11} L_{\odot}$)} & \colhead{(kpc)} & \colhead{($10^{11} L_{\odot}$)} & \colhead{$M_{\odot} / L_{\odot}$} & \colhead{$M_{\odot} / L_{\odot}$}}
\startdata
CSWA107 & 0.60 & 87.6 & 4.01 & 31.6 & 2.46 & 25.8 & 2.40 & 5.1 & 3.7 \\
CSWA141 & 0.79 & 54.3 & 5.84 & 56.0 & 3.31 & 40.3 & 2.90 & 5.1 & 3.7 \\
CSWA164 & 0.89 & -26.9 & 3.15 & 13.4 & 2.50 & 10.8 & 2.39 & 3.2 & 3.0 \\
CSWA165 & 0.78 & -74.6 & 6.10 & 42.1 & 3.32 & 29.6 & 3.29 & 4.9 & 3.8 \\
CSWA6 & 0.85 & -86.1 & 6.51 & 36.8 & 3.67 & 30.0 & 3.66 & 4.6 & 3.7 \\
CSWA7 & 0.65 & 63.8 & 6.07 & 29.9 & 2.49 & 24.2 & 2.47 & 4.5 & 3.7 \\
EOCL & 0.74 & 10.3 & 5.76 & 31.3 & 2.74 & 25.7 & 2.79 & 4.7 & 3.8 \\
J09413 & 0.68 & -6.5 & 6.89 & 53.4 & 4.00 & 46.1 & 4.60 & 5.2 & 4.0 \\
CSWA163${}^{\rm a}$ & 0.81 & -84.3 & $4^{\dagger}$ & 21.3 & 2.00 & 17.6 & 2.04 & 4.7 & 3.8\\
CSWA1 & 0.90 & -15.3 & 5.79 & 26.7 & 3.14 & 20.9 & 3.35 & 3.9 & 3.3
\enddata
\tablecomments{$B$ and $V$ refer to the rest frame. Radii are circularized. $\Upsilon^{\rm SPS}_{V,0}$ specifies $M_*/L_V$ measured within the central spectroscopic aperture ($1\farcs0 \times 0\farcs6$) based on SPS models and a \citet{Salpeter55} IMF. The light-weighted mean $M_*/L_V$ within the $V$-band effective radius is $\langle \Upsilon^{\rm SPS}_V \rangle$. See Sections~\ref{sec:surfphot} and \ref{sec:sps} for a discussion of the uncertainties.\\
${}^{\rm a}$ The S\'{e}rsic index was fixed for CSWA163 due to the poorer quality of the SDSS imaging.\label{tab:surfphot}}
\end{deluxetable*}

As described in Section~\ref{sec:imagingdata}, we fit S\'{e}rsic profiles to the BGG surface photometry in several filters as part of the lens modeling procedure.\footnote{For CSWA164, we found it necessary to perform a separate fit to the BGG light alone after masking the Einstein ring, as it is significantly blended with the BGG in the ground-based imaging. The S\'{e}rsic indices were also allowed to vary between filters.} These profiles were then interpolated to rest-frame $B$ and $V$ filters. The $B$ filter encompasses the $G$ band region where the BGG kinematics were measured, while the $V$ filter is the reddest generally available and should better trace the stellar mass; both are needed for our mass modeling. For each lens, we fit a linear relation to the $k$-correction derived from \citet[][BC03]{BC03} simple stellar population models as a function of the observed color, using the observed filter pair nearest to the redshifted $B$ or $V$ bands. After removing Galactic extinction following \citet{Schlafly11}, we apply this radially dependent $k$-correction and shift to the rest frame to obtain a surface brightness profile in units of $L_{\odot}$~kpc${}^{-2}$. This is then fit with a S\'{e}rsic model having a free $R_e$ and total luminosity, but with $n$ fixed to the value measured in the observed-frame fits. Table~\ref{tab:bgg} lists the rest-frame surface brightness profiles for each BGG. Errors in the S\'{e}rsic parameters are highly correlated. The most relevant measure of uncertainty for our analysis is the amplitude of systematic deviations between the observed and model surface brightness profiles, which is typically $< 0.15$~mag~arcsec${}^{-2}$ within $R < 8''$. 

\begin{figure}
\includegraphics[width=\linewidth]{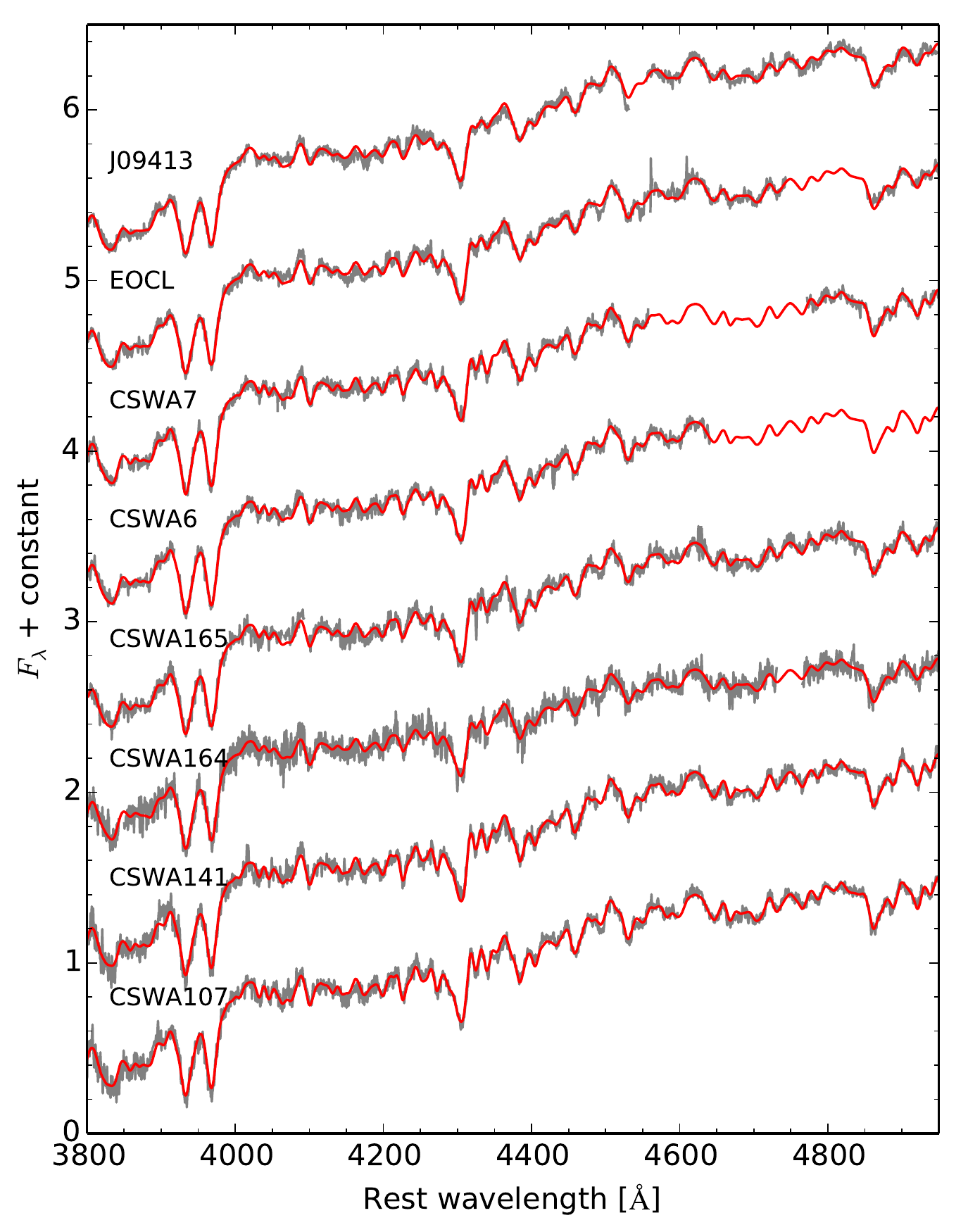}
\caption{Spectroscopy of the centers of the 8 BGGs observed with DEIMOS, extracted in a $1\farcs0 \times 0\farcs6$ aperture (gray) and fitted with SPS models (red) to estimate $\Upsilon^{\rm SPS}_{V,0}$ as described in Section~\ref{sec:sps}. The spectra are highly uniform. Regions contaminated by bright sky lines or uncertain relative flux calibration are not displayed.\label{fig:central_specfit}}
\end{figure}

\section{Stellar Population Synthesis and Radial Gradients}\label{sec:sps}

\begin{figure*}
\centering
\includegraphics[width=0.9\linewidth]{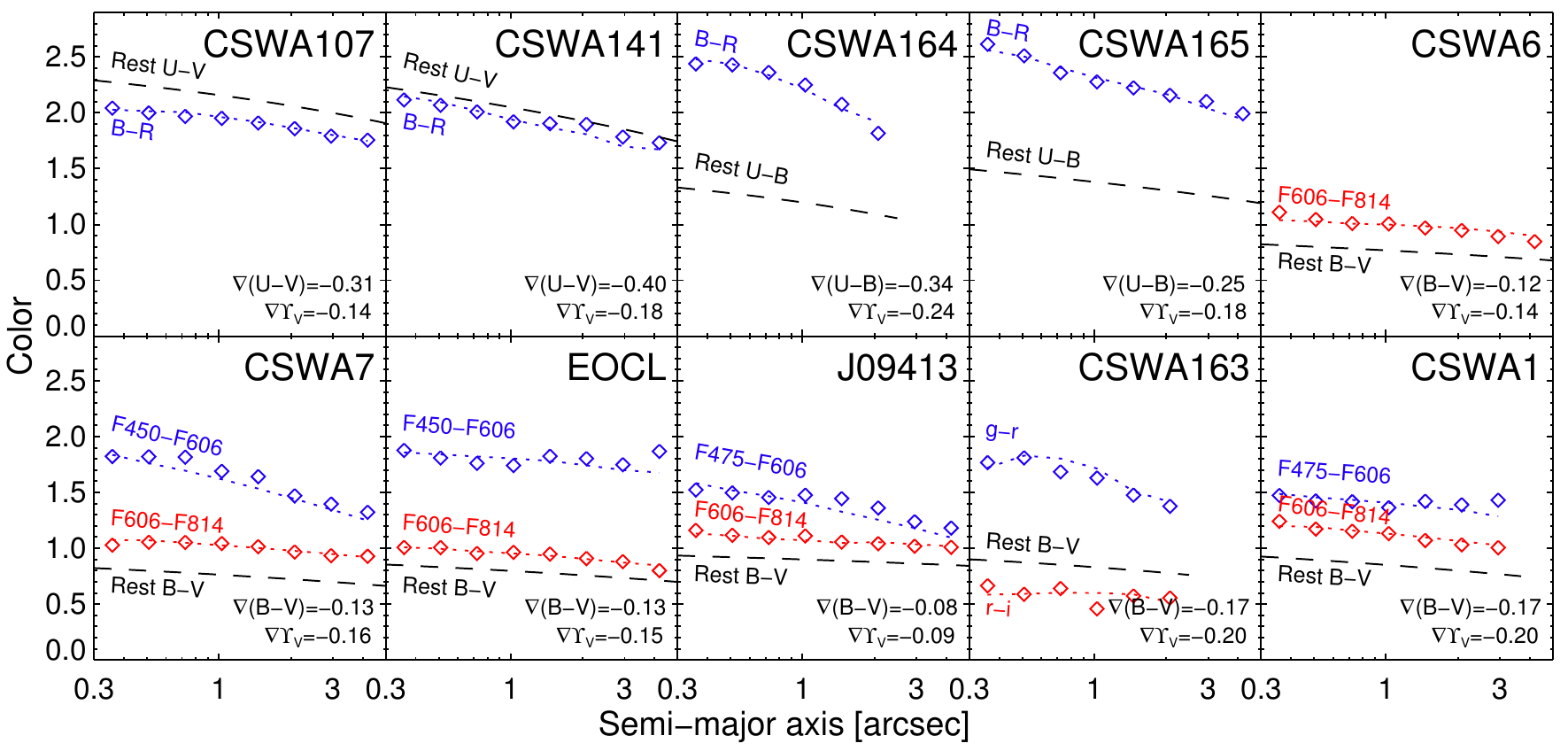}
\caption{Color gradients in our sample of BGGs. Each panel shows the radial color variation as measured in directly in the images (blue and red diamonds) and in the S\'{e}rsic model fits (dotted lines). Other galaxies, including the lensed object, are masked. Dashed lines show a rest-frame color obtained using the radially varying $k$-corrections described in the text. The slopes of the color ($\nabla {\rm color} = d{\rm color} / d \log R$) and $\Upsilon_V$ ($\nabla \Upsilon_V = d \log \Upsilon_V / d \log R$) gradients are given in lower-right corner of each panel.\label{fig:colorgrad}}
\end{figure*}

One goal of our analysis is to compare the stellar mass obtained from lensing and dynamics with that estimated with stellar population synthesis (SPS) models. In this section we analyze spectroscopic and photometric observations of the BGGs using SPS models to constrain the stellar mass-to-light ratio $\Upsilon^{\rm SPS}_{V} = M_*/L_V$. 

As a first step, we estimate a central value $\Upsilon^{\rm SPS}_{V,0}$. The DEIMOS spectrum extracted from the central $1\farcs0 \times 0\farcs6$ of each BGG is fit using the {\tt pyspecfit} code \citep{Newman14} and a suite of BC03 models based on a  fiducial Salpeter IMF. The models follow exponentially declining star formation histories $e^{-t/\tau}$, with uniform priors on $0 < \log {\rm age / Gyr} < 1$, $7 < \log \tau / {\rm yr} < 10$, and $0.01 < Z < 0.04$, where $Z$ is the metallicity. The redshift and velocity dispersion were also fitted simultaneously, and a 12th order multiplicative polynomial was used to filter the continuum. Figure~\ref{fig:central_specfit} shows the resulting fits and the uniformity of the BGG spectra. This leads to a narrow range of $\Upsilon^{\rm SPS}_{V,0}$ estimates listed in Table~\ref{tab:bgg}. The uncertainties are dominated by systematics discussed below.

A common approximation in lensing and dynamical studies of ellipticals is that stellar mass follows the optical light profile, with a radially invariant $M_*/L$. However, Figure~\ref{fig:colorgrad} demonstrates that all of the BGGs in our sample show negative color gradients, which implies that $M_*/L$ declines with increasing radius. To quantify this decline, we construct $U$, $B$, and $V$ band rest-frame S\'{e}rsic profile fits as described in Section~\ref{sec:surfphot}. For each BGG, we choose the rest-frame color most closely matching one of the observed pairs, which ensures that only a small color interpolation is needed for most systems (CSWA164 and CSWA165 are exceptions). The color gradient is then fitted by a linear slope $\nabla {\rm color} = d {\rm color} / d \log R$ over the interval $0\farcs3 < R < 5''$, which encompasses our lensing and dynamics constraints.

Metallicity gradients are generally found to be more significant than age gradients in massive, old ellipticals \citep[e.g.,][]{Tamura00,Mehlert03,Wu05,Rawle10,Tortora11,Greene13}. Therefore, to approximate the conversion from color to $M_*/L_V$ gradients, we use relations derived from BC03 simple stellar population models with a fixed age of 7 Gyr and varying metallicity: $d \log \Upsilon_V / d (B - V) = 1.2$, $d \log \Upsilon_V / d (U-B) = 0.72$, and $d \log \Upsilon_V / d (U-V) = 0.45$. (Note that with more extensive photometry extending into the the near-infrared, we could address this degeneracy directly in these objects.) For comparison, the \citet{Bell03} relation --- which is based on trends in age, dust, and metallicity found empirically from multi-band spectral energy distribution fitting of the overall galaxy population --- gives a very similar $d \log \Upsilon_V / d (B-V) = 1.305$.

The resulting gradients $\nabla \Upsilon_V = d \log (M_*/L_V) / d \log R$ are listed in Figure~\ref{fig:colorgrad}. The median is $\nabla \Upsilon_V = -0.15$ with a dispersion of only 0.03.\footnote{Here we have excluded CSWA164 and CSWA165, since their rest-frame colors required a significant extrapolation, as well as CSWA163 due to its lower data quality, but these exclusions turn out to have a minimal effect on the median.} In our fiducial mass models, we therefore place a  Gaussian prior on $\nabla \Upsilon_V$ with this mean and dispersion. The radial variation is then
\begin{equation}
\Upsilon^{\rm SPS}_{V}(R) = \Upsilon^{\rm SPS}_{V,0} \left(\frac{R}{0\farcs3}\right)^{\nabla \Upsilon_V},\label{eqn:grad}
\end{equation}
since the mean radius in the central DEIMOS aperture is $R=0\farcs3$. 
Using Equation~\ref{eqn:grad}, we can also compute the light-weighted mean $\langle \Upsilon^{\rm SPS}_V \rangle$ within the $V$-band effective radius, which we use in Section~\ref{sec:sys} to test the effect on our results of neglecting $M_*/L$ gradients.\footnote{For CSWA1 and CSWA163, since we do not have access to the resolved spectra to measure $\Upsilon^{\rm SPS}_{V,0}$, we instead used {\tt pyspecfit} to fit the SDSS $griz$ photometry measured in a 20~kpc aperture. In conjunction with the mean gradient $\langle \nabla \Upsilon_V \rangle$, we can then solve for the central $\Upsilon^{\rm SPS}_{V,0}$ that reproduces this aperture measurement.}

There are clearly significant systematic uncertainties in both the zeropoint of the $\Upsilon^{\rm SPS}_{V}$ estimates and their trend with radius. Concerning the former, we compared our measurements to those derived from the SDSS $griz$ colors using {\tt kcorrect} \citep{Blanton07}, shifted to a Salpeter IMF. Since the colors are measured in large apertures, we compare the results to our $\langle \Upsilon^{\rm SPS}_V \rangle$. This approach uses different data and models and so provides a useful route to estimate uncertainties. We find an rms difference of 0.05~dex between the methods, which we take as a fiducial random uncertainty; the systematic shift between the methods is only 0.03~dex.

Concerning the $M_*/L$ gradient, we have estimated a lower limit by supposing that the color gradient is dominated by metallicity variations: gradients in age or dust would give larger $M_*/L$ variations for a given color difference. However, the color gradients in our sample are larger than the typical values seen in other studies of elliptical galaxies. For example, \citet{Wu05} find a mean $\nabla (B-V) = -0.05 \pm 0.01$, compared to a mean $\nabla (B-V) = -0.13$ for the 6 galaxies with that color plotted in Figure~\ref{fig:colorgrad}. Similarly, \citet{Kuntschner10} measure a typical spectroscopic metallicity gradient of $d \log Z / d \log R = -0.25$ corresponding to $\nabla (B-V) = -0.07$ and $\nabla \Upsilon_V = -0.08$ in the BC03 models, which are about half our inferred values. It is not obvious why our sample would show much stronger stellar population gradients, although it may relate to differences in the assembly histories (e.g., properties of the cannibalized satellites) of central galaxies in $10^{14} \msol$ halos relative to typical ``field'' galaxies.

In our mass models, we use $\Upsilon^{\rm SPS}_{V,0}$ to inform broad priors on the stellar mass-to-light ratio, and we vary the radial gradient $\nabla \Upsilon_V$ within the range described above to test the sensitivity of our results (Section~\ref{sec:sys}).

\section{Modeling the Mass Distribution of Group-Scale Lenses}\label{sec:modeling}

With the observational constraints in hand, we now outline our method for inferring the mass distributions of the group-scale lenses. The key ingredients---the Einstein radius $\theta_{\rm Ein}$, the velocity dispersion profile $\sigma(R)$ of the BGG, and the halo mass $M_{200}$ of each lens---are used to constrain two-component mass models, consisting of a group-scale DM halo and the stellar mass of the BGG.

The BGG is modeled using the S\'{e}rsic fits presented in Section~\ref{sec:surfphot}. The stellar mass profile follows the $V$-band light multiplied by the radially dependent $\Upsilon^{\rm SPS}_V$ given by Equation~\ref{eqn:grad}. The luminosity profile at the wavelengths where kinematics are measured is also required to solve the Jeans equations, and for this we use the rest-frame $B$-band profile described in Section~\ref{sec:surfphot}. The DM halo is modeled as a generalized NFW (gNFW) profile
\begin{equation}
\rho_{\rm DM}(r) = \frac{\rho_0}{(r/r_s)^{\beta} (1 + r/r_s)^{3-\beta}},\label{eqn:gnfw}
\end{equation}
where the asymptotic inner slope $\beta = 1$ in the case of an NFW model. 

We compute the kinematic observables using the Jeans anisotropic modeling (JAM) routines \citep{Cappellari08}, which operate quickly on oblate, axisymmetric mass distributions. This allows us to move beyond spherically symmetric models often used in earlier work, including N13, and to test the effects of that assumption on our results. JAM requires that the surface density of mass and tracers be expressed as a multi-Gaussian expansion (MGE). For the S\'{e}rsic and gNFW profiles, we therefore determined the MGE coefficients as polynomial functions of $n$ and $\beta$, thus allowing the MGE to be quickly constructed for a given model.\footnote{For the gNFW profile, 18 Gaussian components were fit to $\rho(r/r_s)$ for many values of $\beta$. The amplitudes of these components were then fit by 12th order polynomials in $\beta$ while the widths $\sigma_i$ of the components were kept fixed. Over the range $10^{-2} < r/r_s < 10$ and $0 < \beta < 2$ relevant for this study (the smallest scales probed by the kinematic data are $R \simeq 1.5~{\rm kpc} \sim r_s / 100$), this parametrization accurately describes the enclosed mass $M(r)$ with an rms error of 0.1\% and a maximum error of 1.8\%, which is smaller than the uncertainties in $\sigma^2$. For the S\'{e}rsic profile, we fit 16 Gaussian components to the surface density $\Sigma(R/R_e)$ for many values of $n$, and then fit the amplitudes of the components with 4th degree polynomials in $n$. The result accurately describes $\Sigma(R/R_e)$ over the range $10^{-2} < R/R_e < 5$ and $3 < n < 7$ relevant for this paper with an rms error of 0.3\% and a maximum error of 0.9\%.}

The observed projected axis ratio $q$ and the inclination angle $i$ determine the intrinsic axis ratio $q' = \sqrt{q^2 - \cos^2 i} / \sin i$ of the three-dimensional mass distribution. We take a uniform prior $0 < \cos i < \cos i_{\rm min}$, where $i_{\min}$ is set such that $q' > 0.5$ as motivated by the absence of massive, non-rotating ellipticals with flatter shapes \citep[e.g.,][]{Weijmans14}. We further assume that the DM halo follows the ellipticity and orientation of the stellar ellipsoids. This is motived by the observed close correspondence between the projected axis ratio and position angle of mass and light in our lens models (Section~\ref{sec:lensing}).

Two free parameters other than inclination describe the stellar mass distribution. The first is an overall scaling
\begin{equation}
\alpha_{\rm SPS} = \Upsilon_V / \Upsilon^{\rm SPS}_V
\end{equation}
of the stellar mass relative to the SPS estimate, which is called the IMF mismatch parameter \citep[e.g.,][]{Treu10}. As in previous work, this parametrization allows us to constrain systematic offsets in the stellar mass scale of the SPS models, which are premised on a Salpeter IMF. The second parameter is the anisotropy $\beta_z$ of the velocity dispersion tensor in the meridional plane, for which we take a Gaussian prior $N(0.1, 0.2)$ based on the sample of ``slow rotator'' ETGs considered in \citet{Cappellari07}. 

In addition to the asymptotic inner density slope $\beta$, two scales are needed to describe the gNFW DM halo. Although the density profile is most easily written in terms of $\rho_0$ and $r_s$, for comparison with simulations it is more useful to adopt the parameters $M_{200}$ and $c_{-2}$. Although the DM halo is ellipsoidal in our models, we define $M_{200}$ in terms of a spherical overdensity, i.e., the mass within the sphere of radius $r_{200}$ that has a mean density equal to 200 times the critical density. The concentration $c_{-2} = r_{200} / r_{-2}$ is defined in terms of the radius $r_{-2} = (2-\beta)r_s$ at which the local logarithmic density slope is $-2$, which is identical to the scale radius $r_s$ for NFW models. In order to properly compare with $r_{200}$, we also use spherical measures of $r_s$ and $r_{-2}$.\footnote{Specifically, the sphericalized radii are defined as the geometric mean of the axes of the isodensity ellipsoid.}

For a given set of model parameters, we use JAM to construct the line-of-sight second velocity moment $\sigma^2_{\rm los}$ on a grid of radii and azimuthal angles. The surface brightness $I$ and $I\sigma^2_{\rm los}$ are then interpolated onto a rectangular grid, smoothed by the PSF of the spectroscopic observations, and binned in the same rectangular apertures used for spectral extraction (accounting for the slit width and orientation listed in Table~\ref{tab:obslog}). We thus generate a set of model velocity dispersions $\sigma_i^{\rm mod}$, which are compared with the data $\sigma_i^{\rm obs}$:
\begin{equation}
\chi^2_{\rm VD} = \sum_i \frac{(\sigma_i^{\rm obs} - g_{\rm VD} \sigma_i^{\rm mod})^2}{\Delta_i^2},
\end{equation}
where $\Delta_i$ is the uncertainty in measurement $i$. $g_{\rm VD}$ is a calibration factor that accounts for correlated systematic uncertainties in the velocity dispersion measures. Based on the tests in Section~\ref{sec:vdprof}, we place a Gaussian prior $N(1.0, 0.05)$ on $g_{\rm VD}$.

The mass within a cylinder of radius $\theta_{\rm Ein}$ is then computed and normalized by the critical surface density for lensing to obtain the model convergence $\overline{\kappa_{\rm group}^{\rm model}}$. This is compared to the measurements in Table~\ref{tab:lensing}:
\begin{equation}
\chi^2_{\kappa} = \left(\frac{\overline{\kappa_{\rm group}^{\rm obs}} - [\overline{\kappa_{\rm group}^{\rm model}} + \kappa_{\rm ext}]}{\sigma_{\kappa}}\right)^2.
\end{equation}
Here $\overline{\kappa_{\rm group}^{\rm obs}}$ is measured from the lens models with uncertainty $\sigma_{\kappa}$, and $\kappa_{\rm ext}$ is the external convergence arising form foreground or background structures, which is described separately below.  

The model likelihood is then $L \propto \exp[-(\chi^2_{\rm VD} + \chi^2_{\rm \kappa})/2]$. We incorporate the measurement of $M_{200}$ from satellite kinematics, where available (see Table~\ref{tab:satellites}), via a Gaussian prior. The parameter space is explored using the MCMC engine {\tt MultiNest}. Table~\ref{tab:priors} summarizes the parameters of our model and the priors.

\begin{deluxetable}{llc}
\tablecaption{Mass Model Parameters and Priors\label{tab:priors}}
\tablewidth{1\linewidth}
\tablehead{\colhead{Parameter} & \colhead{Description} & \colhead{Prior}}
\startdata
\multicolumn{2}{c}{\emph{BGG Stars}} \\
$\log \alpha_{\rm SPS}$ & IMF mismatch parameter & $U(-0.4, 0.4)$ \\
$\nabla \Upsilon_V$ & $d \log \Upsilon^{\rm SPS}_V / d \log R$ & $N(-0.15, 0.03)$ \\
$\beta_z$ & JAM anisotropy & $N(0.1, 0.2)$ \\
$\cos i$ & inclination & $U(0, i_{\rm min})$ (\S\ref{sec:modeling}) \\
\multicolumn{2}{c}{\emph{DM Halo}} \\
$\beta$ & inner slope & $U(0, 2)$ \\
$\log M_{200}$ & halo mass & Table~3 \\
$\log c_{-2}$ & concentration & $U(0, 1.3)$ \\
\multicolumn{2}{c}{\emph{Other}} \\
$\kappa_{\rm ext}$ & external convergence & See \S\ref{sec:kappaext} \\
$g_{\rm VD}$ & $\sigma$ calibration & $N(1.0, 0.05)$
\enddata
\tablecomments{$U(a, b)$ denotes a uniform prior over the interval $[a, b]$, and $N(\mu, \sigma)$ denotes a Gaussian prior with mean $\mu$ and dispersion $\sigma$. A Gaussian prior is placed on $\log M_{200}$ based on the measurement from satellite kinematics in Table~\ref{tab:satellites}; for the two groups that lack such a measurement, we take $U(13.5, 14.5)$.}
\end{deluxetable}

\subsection{External Convergence\label{sec:kappaext}}

Several lenses show signs of foreground or background structures, either through secondary peaks in the redshift distribution (Section~\ref{sec:environments}) or through external shear $\Gamma_{\rm ext}$ required by the strong lens models (Section~\ref{sec:lensing}). The three systems with the highest $\Gamma_{\rm ext}$ (CSWA7, J09413, EOCL) correspondingly show clear secondary peaks in the redshift distribution. CSWA6, which is fit with low $\Gamma_{\rm ext}$, does have secondary redshift peaks, but they are located far ($>1$~Mpc) from the lens. Reassuringly, of the three systems where the image configuration or data quality precluded a reliable estimate of $\Gamma_{\rm ext}$  (CSWA107, CSWA141, CSWA163), none shows a second redshift peak. Thus, the redshift survey and strong lensing analysis are broadly consistent indicators of the presence of significant external structures.

Strong lensing measures the mass within the Einstein radius, including any contribution from external convergence $\kappa_{\rm ext}$. To account for this, we introduce $\kappa_{\rm ext}$ as an additional parameter. For the 6 systems with a small external shear ($\Gamma_{\rm ext} < 0.04$) or no measurement, we place a prior on $\kappa_{\rm ext}$ of $N(0, 0.04)$ based on the small fluctuations expected from large-scale structure \citep[e.g.,][]{Takahashi11}. For the 4 systems with larger $\Gamma_{\rm ext}$ (CSWA7, J09413, EOCL, CSWA165), we use $\Gamma_{\rm ext}$ to inform a prior on $\kappa_{\rm ext}$. 
If the external contribution is dominated by a single halo, the relation between $\Gamma_{\rm ext}$ and $\kappa_{\rm ext}$ depends on the slope of the density profile, with $\Gamma_{\rm ext} = \kappa_{\rm ext}$ in the case of an isothermal slope $\rho \propto r^{-2}$. We therefore take a log-normal prior on $\kappa_{\rm ext}$ with a mean of $\Gamma_{\rm ext}$ but allow a broad dispersion of a factor of 2. An alternative approach is to estimate the distribution of $\kappa_{\rm ext}$ conditional on another lens observable, such as $\Gamma_{\rm ext}$ or the local galaxy overdensity, using cosmological simulations. For galaxies with $\Gamma_{\rm ext} \lesssim 0.1$, typical of our sample, this method yields yields a similar or slightly smaller dispersion in $\kappa_{\rm ext}$, giving confidence that our prior is reasonable \citep[e.g.,][]{Oguri05,Suyu10}. 

\begin{deluxetable}{lccc}
\tablecaption{Mass Model Fit Quality\label{tab:fitqual}}
\tablewidth{0.9\linewidth}
\tablehead{\colhead{Name} & \colhead{$\chi^2_{\rm VD}/N_{\rm VD}$} & \colhead{$\chi^2_{\kappa}$} & \colhead{$\chi^2_{M_{200}}$}}
\startdata
CSWA107 & 25.76/10 & 0.00 & 1.11 \\
CSWA141 & 17.07/9 & 0.85 & 1.06 \\
CSWA164 & 3.05/5 & 0.47 & \ldots \\
CSWA165 & 7.61/5 & 0.05 & 0.08 \\
CSWA6 & 5.48/6 & 0.04 & 0.05 \\
CSWA7 & 2.73/7 & 1.86 & 1.05 \\
EOCL & 6.85/7 & 0.18 & 0.64 \\
J09413 & 5.98/7 & 2.06 & 0.47 \\
CSWA163 & 0.92/3 & 1.46 & 0.01 \\
CSWA1 & 2.83/7 & 0.86 & \ldots
\enddata
\tablecomments{$\chi^2$ metrics are shown for the maximum \emph{a posteriori} probability model found using our fiducial modeling procedure. $N_{\rm VD}$ is the number of bins in the $\sigma(R)$ profile.}
\end{deluxetable}

\subsection{Fit Quality\label{sec:fitqual}}

Our mass models are generally flexible enough to fit all of the available data acceptably, as Table~\ref{tab:fitqual} demonstrates via $\chi^2$ metrics. The main deficiency is that the steep rise in the velocity dispersion profiles of CSWA141 and CSWA6 is not well fit (Figure~\ref{fig:dynamics}). This might signal that our mass models are not fully adequate in these cases. For the remainder of the paper, we bear this in mind when interpreting our results and pay attention to the influence of these two systems.

\subsection{Consistency Between Mass Models and Lensing\label{sec:consist}}

We have chosen to include only $\theta_{\rm Ein}$ as a lensing-based constraint in our mass models. However, strong lensing carries more information about the mass distribution. In particular, the radial magnification can constrain the logarithmic slope of the density profile $\gamma_{\rm SL}$ (see Table~\ref{tab:lensing}). This slope is affected by the well-known mass-sheet degeneracy and is therefore less robust than the mass within $\theta_{\rm Ein}$. Nevertheless, it is interesting to test whether our mass models are consistent with the available strong lensing data in a broader sense. With this aim we compared $\gamma_{\rm SL}$, which was derived solely from the strong lensing data (Table~\ref{tab:lensing}), to the total density slope $\gamma_{\rm LD, \theta_{\rm Ein}}$ of the mass models introduced in this section, which are fit to the stellar kinematics and $\theta_{\rm Ein}$. Since strong lensing is sensitive to the mass profile over the radial range where multiple images are formed, we measure $\gamma_{\rm LD, \theta_{\rm Ein}}$ locally around $R = \theta_{\rm Ein}$. CSWA7 was excluded from this comparison since the pixel-level data were not used in its strong lensing fit (see Section~6).

The mean density slopes recovered from the two methods agree very well: $\langle \gamma_{\rm SL} \rangle = 1.59 \pm 0.08$ and $\langle \gamma_{\rm LD, \theta_{\rm Ein}} \rangle = 1.57 \pm 0.05$. On an object-by-object basis, the scatter between the two methods is $\sigma_{\gamma} = 0.28$. Accounting for the statistical errors in $\gamma_{\rm LD, \theta_{\rm Ein}}$ estimated from the Markov chains, statistical agreement between the methods would require errors of $\simeq 0.2$ in $\gamma_{\rm SL}$. Although it is difficult to independently evaluate the uncertainties in $\gamma_{\rm SL}$, which are dominated by systematics, we note that the value of $\gamma_{\rm SL}$ derived for CSWA1 in a detailed study by \citet{Dye08} is 0.30 higher than ours, possibly due to differences in the source model and the subtraction of the lens galaxy and background. This comparison demonstrates that the mass models developed in this section have density slopes consistent with the strong lensing data for reasonable estimates of the uncertainties in both methods. 

The largest outlier in this comparison is CSWA107, with $\gamma_{\rm LD, \theta_{\rm Ein}} = 1.83 \pm 0.19$ but $\gamma_{\rm SL} = 1.18$. Since a shallow projected density profile is preferred to produce the naked cusp configuration, the disagreement in this system may reflect a degeneracy with the ellipticity, external shear, or the effect of the mass-sheet degeneracy. The good agreement between $\gamma_{\rm SL}$ and $\gamma_{\rm LD, \theta_{\rm Ein}}$ on average, however, shows that the mass-sheet degeneracy is generally not a limiting factor on these scales.

\section{The mass--concentration relation at the group scale}
\label{sec:mcr}

\begin{figure*}
\centering
\includegraphics[width=0.48\linewidth]{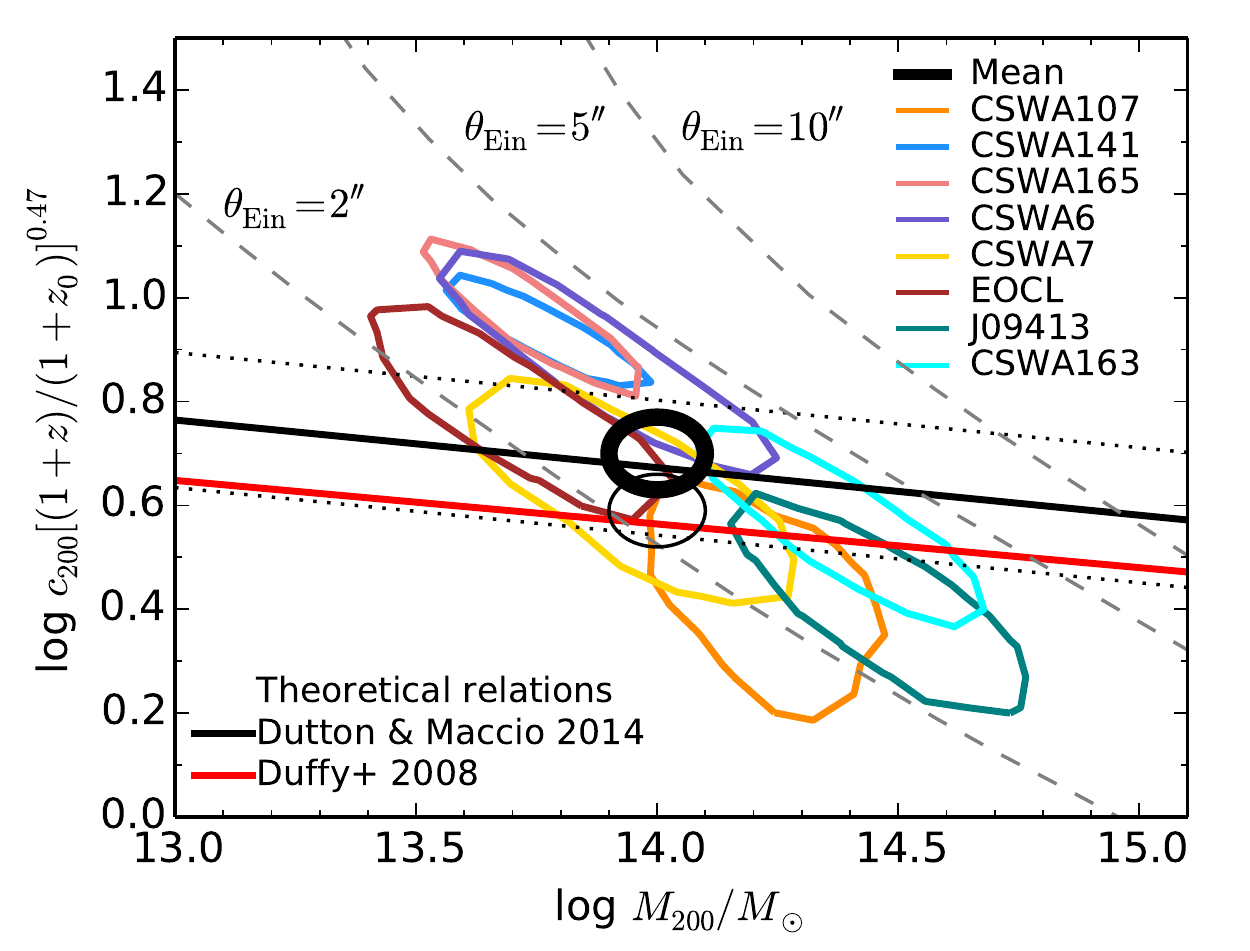}
\includegraphics[width=0.48\linewidth]{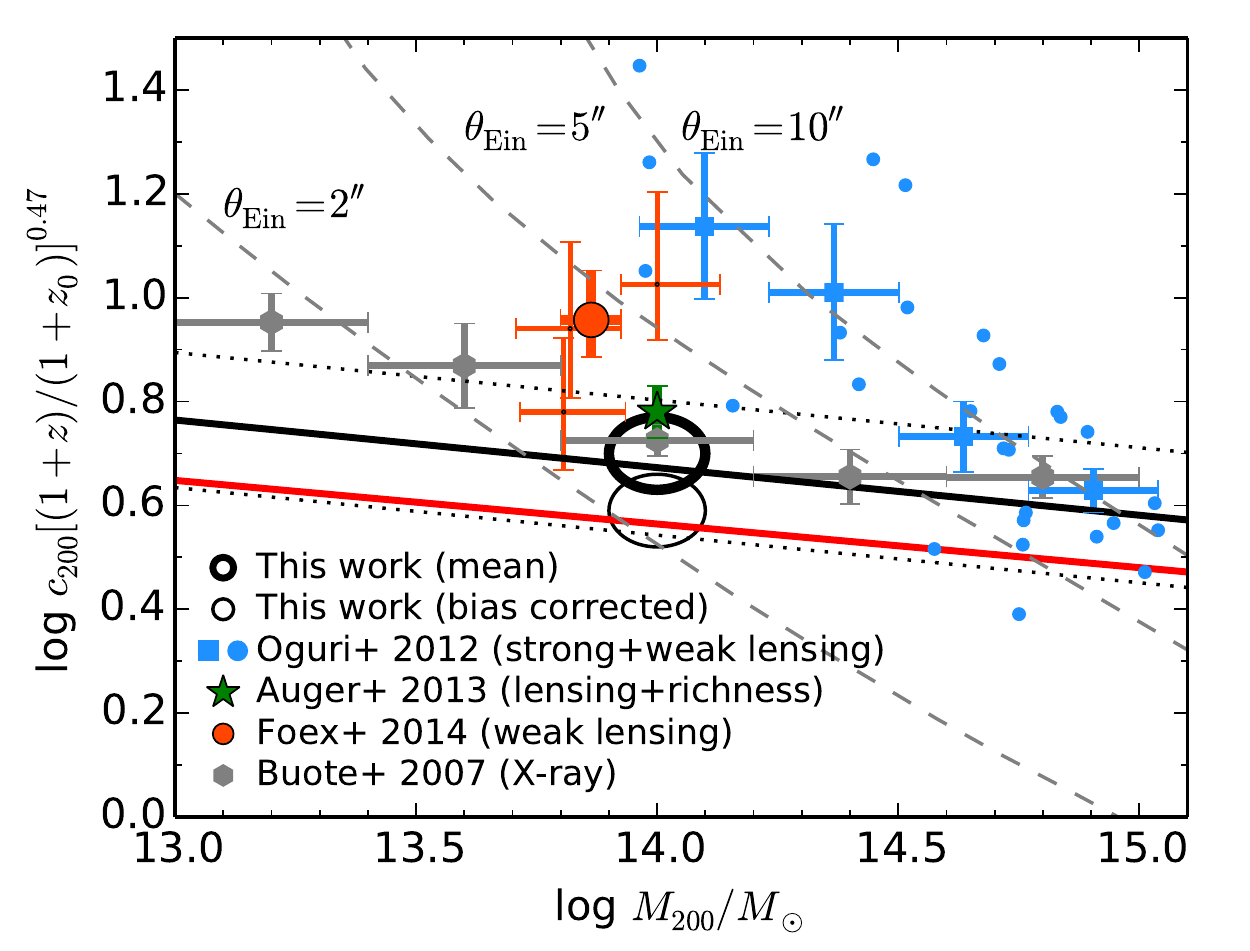}
\caption{\emph{Left:} Halo masses and concentrations normalized to $z_0 = 0.36$. Colored contours show the 68\% credible regions for individual lenses. The thick black ellipse shows our hierarchical inference for the mean halo mass and concentration of the group lenses and their $\pm1\sigma$ uncertainty. The thin ellipse represents our estimate of the underlying halo population after accounting for our selection in Einstein radius.  These are compared to the theoretical relations indicated in the lower caption. Dotted black lines indicate the $\pm1\sigma$ scatter in $\log c_{200}$ from \citet{Dutton14}. Contours of $\theta_{\rm Ein}$ are derived from our halo occupation model (Section~\ref{sec:cdmmodel}; these are insensitive to details of the stellar distribution for $\theta_{\rm Ein} \gtrsim 2''$.) \emph{Right:} Comparisons to published MCR constraints around $\simeq 10^{14} \msol$. The individual \citet{Oguri12} lensing measurements and the \citet{Buote07} X-ray data are binned to produce the blue and gray squares, respectively. Both the stack of the full \citet{Foex14} weak lensing sample (solid red circle) and their stacks in three bins of arc radius $R_A$ (red error bars) are shown: from low to high $c_{200}$, $2''<R_A<3\farcs5$, $3\farcs5<R_A<5\farcs5$, and $5\farcs5<R_A<20''$.\label{fig:mc}}
\end{figure*}

The various data sets assembled in this paper constrain the group mass distribution on multiple scales. In this section, we consider the large-scale mass distribution as quantified by the halo mass--concentration relation (MCR) and compare it to theoretical relations obtained from CDM simulations. Here we consider only the 8 lenses in our sample for which we have measured the halo mass $M_{200}$ from satellite dynamics (Section~\ref{sec:satellitekinematics}). Since theoretical studies of the MCR are usually premised on an NFW DM profile, our results in this section only are derived using mass models that assume NFW halos (i.e., $\beta = 1$ and $c_{-2} = c_{200}$). Furthermore, because the MCR evolves with redshift, we evolve all measured concentrations---including those drawn from the literature---to the mean redshift $z_0 = 0.36$ of our sample using the \citet{Duffy08} scaling $c_{200} \propto (1+z)^{-0.47}$.

The left panel of Figure~\ref{fig:mc} shows our constraints on the halo mass and concentration of each lens. Lines of constant $\theta_{\rm Ein}$ are overlaid. As expected, the covariance between $M_{200}$ and $c_{200}$ follows the slope of these lines, since $\theta_{\rm Ein}$ is more precisely measured than $M_{200}$. Constraints are broad for individual lenses, primarily due to the uncertainties in the satellites' velocity dispersion and hence $M_{200}$. Therefore, we combine results from the 8 groups using a hierarchical Bayesian method that allows us to infer the mean $\langle \log M_{200} \rangle$, the mean concentration $\langle \log c_{200}' \rangle$ at $M_{200} = 10^{14} \msol$, and the intrinsic scatter in both distributions. The mathematical details of this framework are given in Appendix~B.

The resulting constraints are $\langle \log M_{200} \rangle  = 14.0 \pm 0.1$ and $\langle \log c_{200}' \rangle = 0.70 \pm 0.07$ (thick black ellipse in the left panel of Figure~\ref{fig:mc}). This mean concentration is entirely consistent with the theoretical expectation for unmodified NFW halos derived by \citet{Dutton14}, who predict a mean $\log c_{200} = 0.67$ at $\log M_{200} = 14$ and $z = 0.36$. This concentration is slightly higher than some earlier theoretical MCRs, which \citet{Dutton14} attribute to their use of the \citet{Planck14} cosmological parameters. To illustrate the difference compared a theoretical MCR based on the WMAP5 cosmology, \citet{Duffy08} found $\log c_{200} = 0.56$ at the same mass and redshift when considering all halos and $\log c_{200} = 0.62$ when restricting to the relaxed ones. We also find that the intrinsic dispersion in concentrations, $\sigma_{\log c'} = 0.14 \pm 0.07$, is consistent with the range predicted in CDM simulations.

Lens selection can have an important effect on the concentrations of a sample. The selection function of surveys that search for strong lenses is complex, but the Einstein radius $\theta_{\rm Ein}$ is the single most important variable \citep[e.g.,][]{Gavazzi14}. As Figure~\ref{fig:mc} shows, contours of constant $\theta_{\rm Ein}$ are diagonal in the mass--concentration plane. This implies that selecting lenses within a particular range of $\theta_{\rm Ein}$ (or with a non-uniform weighting) will lead to biases in the slope and intercept of the MCR relative to that of the underlying halo population, unless the selection is taken into account. We estimate this effect using a halo occupation model. The details are introduced in Section~\ref{sec:cdmmodel}, but for the present purpose it is sufficient to consider ETGs residing in a cosmological distribution of halos following the \citet{Dutton14} MCR. From such a mock sample, we weight galaxies to match the distribution of $\theta_{\rm Ein}$ in our group lens sample and compute the mean halo mass and offset from the MCR. We find $\langle \log M_{200} \rangle = 14.0$, in agreement with our dynamical measurement, and $\langle \Delta \log c_{200} \rangle = 0.11$.\footnote{Here we populate NFW halos with galaxies having a Salpeter IMF, but variations to the IMF and inner DM profile, described in Section~\ref{sec:cdmmodel}, affect this correction only at the $\simeq 0.03$~dex level.} In other words, lenses with $2\farcs5 < \theta_{\rm Ein} < 5\farcs1$ follow a MCR that is somewhat offset from that of the parent ETG population. Applying this estimated correction to our inference for the group lenses yields $\langle \log c_{200}' \rangle = 0.59 \pm 0.07$ for the mean concentration of the underlying halo population at $M_{200} = 10^{14} \msol$, which is shown by the thin ellipse in Figure~\ref{fig:mc}.

This correction for the Einstein radius selection has the same magnitude as the differences among current theoretical MCRs ($\simeq 0.1$~dex). We conclude that, within the present uncertainties, our group-scale lenses have a mean concentration consistent with unmodified, cosmologically motivated halos.

\subsection{Comparison to Published Concentrations\label{sec:compareconc}}

Since the concentrations of groups and low-mass clusters are contentious, here we briefly compare our results to other studies. We include only those that extend as low as $M_{200} \simeq 10^{14} \msol$, convert published masses and concentrations to our overdensity definition, and evolve concentrations to $z_0 = 0.36$ using the scaling described above. Since selection effects will prove to be important, it is useful to have a model for the relative numbers of strong lenses having different Einstein radii. \citet{More12} showed that the image separation distribution (ISD) roughly follows $dP/d\theta \propto \theta^{-2.8}$ for image separations $\theta \simeq 2\theta_{\rm Ein} = 3''-30''$.

Some authors studying group-scale strong lenses have found ``normal'' concentrations consistent with our measurements. \citet{Auger13} analyzed 26 strong lenses (median $\theta_{\rm Ein} = 4\farcs0$) selected from the CASSOWARY catalog, the source of many of the lenses in our group sample. They combined $\theta_{\rm Ein}$ with an estimate of $M_{200}$ for each lens based on \emph{galaxy richness}, using a scaling relation calibrated to X-ray masses.\footnote{Comparing the richness-based masses with our dynamical masses for the 7 systems in common, we find agreement within the uncertainties in 5 cases. The richness-based masses are much higher (0.8 dex) for EOCL---likely explained by the presence of multiple structures at similar redshifts artificially boosting the richness---and CSWA141.} The right panel of Figure~\ref{fig:mc} shows that the Auger et al.~concentration at $M_{200} = 10^{14} \msol$ (green star) agrees with our value. \citet{Deason13} analyzed CSWA163 using similar methods to our own. As mentioned in Section~\ref{sec:sample}, we have incorporated their data into our sample, and we find consistent values of $M_{200}$ and $c_{200}$.

Other authors have claimed evidence for overconcentrated halos at the scale of groups and low-mass clusters. In an thorough analysis of 28 lenses discovered in the Sloan Giant Arcs Survey (SGAS), \citet{Oguri12} inferred a very steep slope for the MCR, $c_{\rm vir} \propto M_{\rm vir}^{-0.59\pm0.12}$ (c.f.~$c_{\rm vir} \propto M_{\rm vir}^{-0.1}$ in CDM simulations). Oguri et al.~modeled two selection effects: the probability for a cluster to produce an arc with a length-to-width ratio $l/w > 5$ (i.e., the arc cross-section), and the selection function of the SGAS among such lenses, which they approximated as proportional to $\sqrt{\theta_{\rm Ein}}$. Since these effects were insufficient to reconcile CDM-only simulations with the steep observed slope, Oguri et al.~suggested that halos are significantly modified on large scales by baryon cooling, even at fairly high masses $\gtrsim 10^{14} \msol$.
We suggest that the steep MCR found by Oguri et al.~arises from the SGAS sample selection. In particular, there are almost no lenses with $\theta_{\rm Ein} < 5''$ (see blue points in Figure~\ref{fig:mc}, right panel), even though the number of strong lenses in the universe increases rapidly toward smaller $\theta_{\rm Ein}$ \citep[e.g.,][]{More12}. This implies a rather hard cutoff in the selection function. As Figure~\ref{fig:mc} shows, this corresponds to a diagonal cut in the mass--concentration plane that will induce a steep slope, which matches that of the Oguri et al.~MCR almost exactly. 
Samples of lenses with $\theta_{\rm Ein} > 5''$ are adequate to measure the concentrations of  massive clusters with $M_{200} \gtrsim 10^{14.7} \msol$, but one must probe smaller $\theta_{\rm Ein}$ to reach halos with ``normal'' concentrations and smaller masses. Similar considerations apply to the \citet{Wiesner12} analysis of the Sloan Bright Arcs Survey sample. 

\citet{Foex14} used a stacked weak lensing analysis to measure the mean mass and concentration of 80 strong lensing groups in the SARCS sample having $2'' \lesssim \theta_{\rm Ein} \lesssim 20''$ (mean $\simeq 4''$). Their mean halo mass is comparable to our group sample, but their mean concentration is significantly higher (orange circle in Figure~\ref{fig:mc}, right panel). By stacking in bins of arc radius $R_A$, a proxy for $\theta_{\rm Ein}$, they note a strong increase in concentration, but not halo mass, with increasing $R_A$ (orange crosses in figure). The concentration in their lowest bin $2'' < R_A < 3\farcs5$ is consistent our measurement, suggesting that the cause of the discrepancy in the full stack can be traced to the lens sample, not a difference in the mass probes used. Although the origin of the difference in concentration relative to our sample is not fully clear, it is possible that higher-$\theta_{\rm Ein}$ lenses are somewhat over-represented in the Foex et al.~stacks.\footnote{Approximately equal numbers of lenses with $R_A = 2''-3\farcs5$ and $R_A = 3\farcs5-5\farcs5$ are present in their stacks, whereas the More et al.~ISD implies that the former should be 3 times more numerous.}

Finally, it is interesting to compare to results from X-ray studies. Figure~\ref{fig:mc} shows that for halo masses $M_{200} \gtrsim 10^{14} \msol$, the \citet{Buote07} compilation agrees with both our group sample and the \citet{Dutton14} theoretical MCR. The Buote et al.~concentrations exceed the theoretical relation only at lower masses.

In summary, we find ``normal'' concentrations in $10^{14} \msol$ halos indicating that baryons have little effect on the DM distribution around the scale radius. This echoes some earlier work at this mass scale, but not all. Differences in lens sample selections are likely to explain at least some of the differences. In Section~\ref{sec:discussion} we discuss our results in the context of hydrodynamical simulations.

\begin{deluxetable}{lcccc}
\tablecaption{Total Density Profile Slopes and Stellar Fractions of ETG Lenses in Massive Halos\label{tab:slopes}}
\tablewidth{\linewidth}
\tablehead{\colhead{Name} & \colhead{$R_e$ (kpc)} & \colhead{$\log M_*/M_{\odot}$} & \colhead{$\gamma_{\rm tot}$} & \colhead{$f_{*,\rm Salp}^{\rm 2D}$}}
\startdata
\cutinhead{Group-scale Lenses (BGGs)} \\
CSWA107 & 25.8 & 11.95 & $1.63 \pm 0.19$ & $0.16 \pm 0.07$ \\
CSWA141 & 40.3 & 12.03 & $1.50 \pm 0.11$ & $0.09 \pm 0.03$ \\
CSWA164 & 10.8 & 11.85 & $1.84 \pm 0.13$ & $0.30 \pm 0.10$ \\
CSWA165 & 29.6 & 12.10 & $1.62 \pm 0.10$ & $0.15 \pm 0.05$ \\
CSWA6 & 30.0 & 12.13 & $1.52 \pm 0.12$ & $0.16 \pm 0.05$ \\
CSWA7 & 24.2 & 11.96 & $1.62 \pm 0.15$ & $0.18 \pm 0.05$ \\
EOCL & 25.7 & 12.03 & $1.68 \pm 0.14$ & $0.19 \pm 0.05$ \\
J09413 & 46.1 & 12.26 & $1.57 \pm 0.15$ & $0.12 \pm 0.03$ \\
CSWA163 & 17.6 & 11.89 & $1.74 \pm 0.13$ & $0.19 \pm 0.03$ \\
CSWA1 & 20.9 & 12.05 & $1.66 \pm 0.10$ & $0.16 \pm 0.02$ \\
\cutinhead{Clusters-scale Lenses (BCGs)} \\
A611 & 52.0 & 12.43 & $1.26 \pm 0.04$ & $0.06 \pm 0.01$ \\
A383 & 44.0 & 12.30 & $1.14 \pm 0.08$ & $0.06 \pm 0.01$ \\
A2667 & 54.0 & 12.15 & $0.96 \pm 0.08$ & $0.04 \pm 0.01$ \\
MS2137 & 31.0 & 12.23 & $1.25 \pm 0.05$ & $0.07 \pm 0.01$ \\
A963 & 34.0 & 12.32 & $1.27 \pm 0.07$ & $0.10 \pm 0.02$ \\
A2390 & 28.0 & 12.06 & $1.34 \pm 0.05$ & $0.07 \pm 0.01$ \\
A2537 & 59.0 & 12.48 & $1.02 \pm 0.04$ & $0.05 \pm 0.01$
\enddata
\tablecomments{All stellar masses refer to a Salpeter IMF. For the BGGs, $R_e$ is measured in rest-$V$ band and $M_* = L_V \langle \Upsilon_V^{\rm SPS} \rangle$ (Table~\ref{tab:bgg}). The BCGs have been fit with de Vaucouleur's profiles and so differ from published $R_e$ and $M_*$ in \citet{Newman13a}. Random errors in $M_*/L_{\rm SPS}$ are estimated as 0.05~dex (Section~\ref{sec:sps}) for the BGGs and 0.07~dex (N13a) for the BCGs. The random error in $f_{*, \rm Salp}^{\rm 2D}$ includes both the projected mass and $M_*/L_{\rm SPS}$.}
\end{deluxetable}

\section{The total mass density profile within the effective radius}
\label{sec:totslope} 

Moving to smaller scales, we now consider the density structure within the effective radius. We begin with the average logarithmic slope $\gamma_{\rm tot}$ of the total density profile within $R_e$, where $\rho_{\rm tot} \propto r^{-\gamma_{\rm tot}}$. This is one of the simplest and most robust quantities that can be inferred from a lensing and dynamics analysis. Furthermore, since it requires only a measure of $\theta_{\rm Ein}$ and a single aperture velocity dispersion, $\gamma_{\rm tot}$ has been measured for large samples of galaxy-scale lenses and so is particularly useful for examining trends across a wide range of ETG properties. In this section we combine our group-scale lenses with earlier data on galaxy- and cluster-scale lenses. We examine empirical trends in $\gamma_{\rm tot}$ within and among these samples, which collectively span a factor of $\simeq 60$ in halo mass. In Section~\ref{sec:separation} we will interpret these trends using CDM-motivated models and focus on the more challenging goal of separating the luminous and DM density profiles.

\begin{figure*}
\centering
\includegraphics[width=0.9\linewidth]{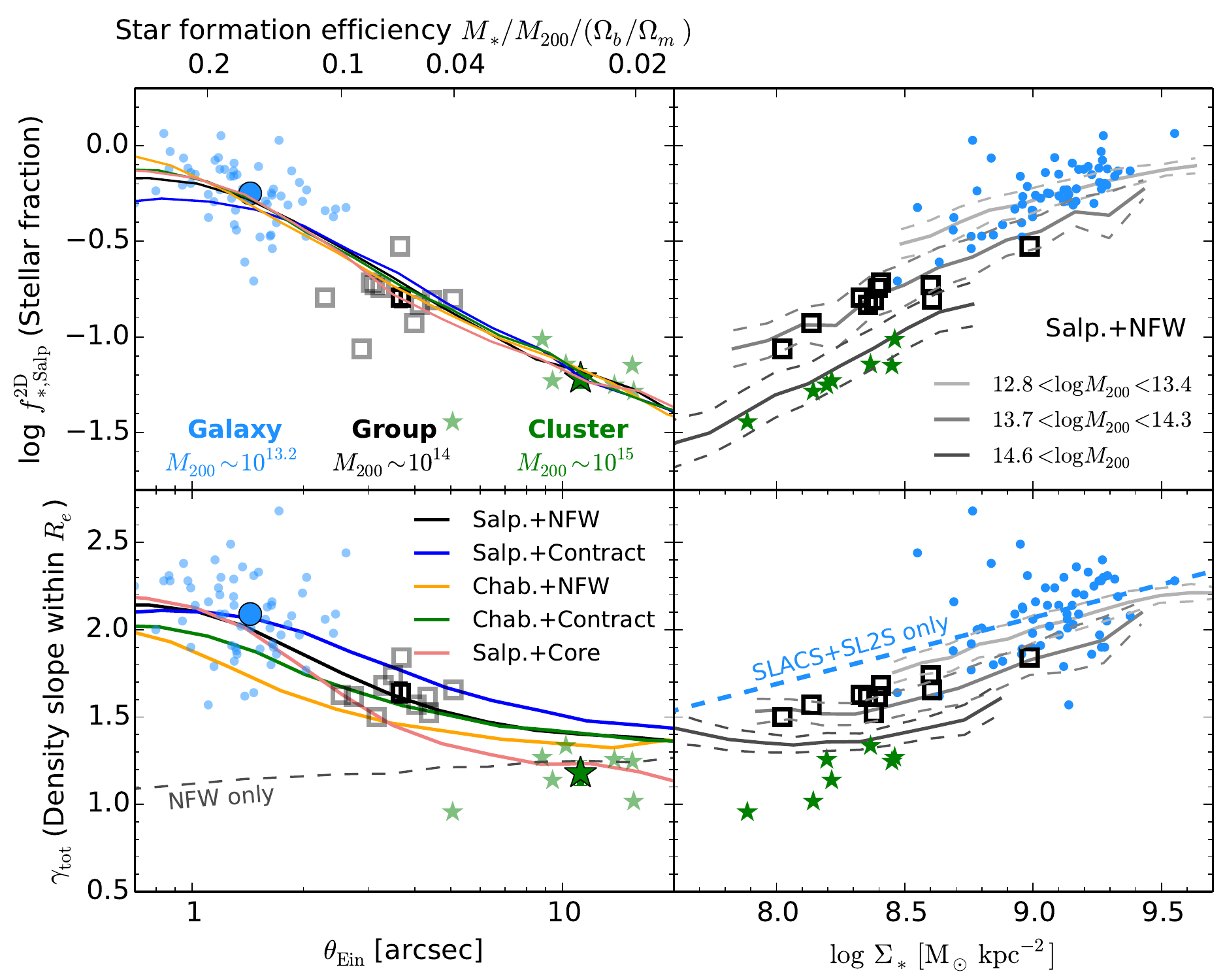}
\caption{Trends in the total density slope $\gamma_{\rm tot}$ and projected stellar mass fraction $f^{\rm 2D}_{*, \rm Salp}$ within $R_e$ are plotted as a function of Einstein radius $\theta_{\rm Ein}$ and mean stellar surface density $\Sigma_*=M_*/(2\pi R_e^2)$ for three ETG lens samples spanning $10^{13} \msol-10^{15} \msol$ in halo mass: galaxy-scale lenses drawn from the SLACS survey (blue circles), the present group-scale sample (black squares), and the BCGs of massive clusters studied by N13 (green stars). Darker symbols show the mean of each data set. \emph{Left panels:} Sold lines indicate the mean trends for the 5 halo occupation models introduced in Section~\ref{sec:cdmmodel}, which adopt different IMFs and inner DM density profiles. For comparison, the dashed ``NFW only'' line corresponds to an unmodified NFW halo with no stars. The top axis delineates the average star formation efficiency $M_*/M_{200}/(\Omega_b/\Omega_m)$ as a function of $\theta_{\rm Ein}$, which is taken from the Salpeter+NFW halo occupation models. \emph{Right panels:} In three bins of halo mass, lines indicate the trend in the Salpeter+NFW halo occupation model along with its $1\sigma$ scatter (dashed). \label{fig:totslopes}}
\end{figure*}

Since our mass models have separate DM and stellar components, the total density profile is not explicitly parameterized. Instead, following \citet{DuttonTreu14}, we define $\gamma_{\rm tot}$ as the mass-weighted mean density slope within $R_e$:
\begin{equation}
\gamma_{\rm tot} = -\frac{1}{M(r)} \int_0^{R_e} 4\pi r^2 \rho(r) \frac{d \log \rho}{d \log r} dr = 3 - \frac{4\pi R_e^3 \rho(R_e)}{M(R_e)},\label{eqn:gammatot}
\end{equation}
where $\rho(r)$ and $M(r)$ are the spherically averaged density and enclosed mass profiles. The posterior distribution of $\gamma_{\rm tot}$ for each lens can then be evaluated from its Markov chains. Heuristically, the data constrain the density slope most robustly between the Einstein radius and the velocity dispersion aperture. Since the Einstein radius is typically $\simeq 0.5-1 R_e$ (see Figure~\ref{fig:rein}), $R_e$ is a convenient location for defining $\gamma_{\rm tot}$.\footnote{While this represents a slight extrapolation for our group sample, where $\theta_{\rm Ein} \simeq 0.7 R_e$ on average, quantitative differences in the slope defined over slightly different radial ranges are small. For example, the mean $\langle\gamma_{\rm tot}\rangle$ would change by only 0.06 if we were instead to define it within $0.7 R_e$.}

To compare our group-scale lenses with other samples of ETG lenses, we have gathered data from the following sources:

\begin{itemize}
\item \emph{Galaxy-scale lenses:} We select 59 lenses from the SLACS survey with measured $R_e$, stellar masses, and density slopes \citep{Auger09,Auger10a}. Since Auger et al.~use single-component power law mass models, their slope $\gamma'$ is nearly equivalent to our $\gamma_{\rm tot}$ (see \citealt{Sonnenfeld15}). The average $\theta_{\rm Ein}$ is $1\farcs4$ (rescaled here, as throughout, to a lensing distance ratio $D_{\rm ds}/D_{\rm s}=0.7$). Although the halo mass cannot be measured for individual SLACS lenses, a stacked weak lensing analysis of a subset indicated a mean $\langle \log M_{200} / M_{\odot} \rangle = 13.2 \pm 0.2$ (\citealt{Gavazzi07}, converted to our adopted overdensity; see also \citealt{Auger10b}). Based on the distribution of $\theta_{\rm Ein}$ and our halo occupation model, introduced below, we estimate the scatter in $M_{200}$ within this sample to be $\sim0.3$~dex.

\item \emph{Group-scale lenses:} We use the 10 lenses analyzed in the present paper having $\theta_{\rm Ein} = 2\farcs5-5\farcs1$. As shown in Section~\ref{sec:mcr}, the average halo mass is $\langle \log M_{200} / M_{\odot} \rangle = 14.0 \pm 0.1$ with an intrinsic scatter of 0.2~dex. The relevant measurements for the group- and cluster-scale lenses are listed in Table~\ref{tab:slopes}.

\item \emph{Cluster-scale lenses:} For the central galaxies of massive clusters, we use the 7 BCGs analyzed by \citet{Newman13a,Newman13b}. In order to compare more consistently with the other lens samples, we remeasured the BCGs' $R_e$ and $M_*$ using a de Vaucouleurs' profile (rather than the dPIE profile used by N13) and also recomputed $\gamma_{\rm tot}$ following Equation~\ref{eqn:gammatot}. This yields a mean $\langle \gamma_{\rm tot} \rangle = 1.18 \pm 0.07 {}^{+0.05}_{-0.07}$; although this definition of $\gamma_{\rm tot}$ differs formally from that adopted in N13a, in practice the difference is quite small (c.f.~$\langle \gamma_{\rm tot} \rangle = 1.16$ in N13a). The BCGs have a typical $\theta_{\rm Ein} = 11''$ and occupy halos with a mean mass of $\langle \log M_{200} / M_{\odot}\rangle = 14.9$ and a scatter of 0.3~dex.
\end{itemize}

The group sample is situated at only slightly higher mean redshift ($\langle z \rangle = 0.36$) than the SLACS ($\langle z \rangle = 0.20$) and N13 ($\langle z \rangle = 0.25$) samples, so evolutionary differences among the samples are expected to be minimal.

In addition to $\gamma_{\rm tot}$, we measure the projected stellar fraction within $R_e$ for each lens:
\begin{equation}
f_{*, \rm Salp}^{\rm 2D} = M_*^{\rm 2D}(R_e)/M^{\rm 2D}_{\rm tot}(R_e),
\end{equation}
where $M^{\rm 2D}_{\rm tot}(R_e)$ is the projected mass within $R_e$ derived from the lensing and dynamics model, and $M_*^{\rm 2D}(R_e)$ is the stellar mass in the same aperture estimated using SPS models and a Salpeter IMF.

Figure~\ref{fig:totslopes} illustrates the trends in $\gamma_{\rm tot}$ and $f_{*, \rm Salp}^{\rm 2D}$ among the three observational samples as a function of $\theta_{\rm Ein}$ (left panels) and stellar surface density $\Sigma_* = M_* / (2 \pi R_e^2)$ (right panels). The first trend clearly visible in the data is the decline in $\gamma_{\rm tot}$ and $f_{*, \rm Salp}^{\rm 2D}$ with increasing $\theta_{\rm Ein}$ and $M_{200}$. The mean total density slope in the group-scale lenses is $\langle \gamma_{\rm tot} \rangle = 1.64 \pm 0.05 \pm 0.07$ (the second error is our estimate of the systematic uncertainty; see Section~\ref{sec:sys}). This is significantly shallower than that of the galaxy-scale lenses, $\langle \gamma_{\rm tot} \rangle = 2.09 \pm 0.03$, and steeper than that of the BCGs, $\langle \gamma_{\rm tot} \rangle = 1.18 \pm 0.07 {}^{+0.05}_{-0.07}$. The intrinsic scatter within all samples is much smaller than the systematic variation among them: $\sigma_{\gamma} = 0.07 \pm 0.05$ (groups), $0.19 \pm 0.03$ (galaxies), and $0.17 \pm 0.08$ (clusters).  As we will see, these trends arise from a dependence of $\gamma_{\rm tot}$ and $f_{*, \rm Salp}^{\rm 2D}$ on halo mass. The relatively small scatter \emph{within} each sample arises from the similarly narrow \emph{ranges} of halo mass that they span, while the systematic trend \emph{among} samples arises from the fact that they occupy nearly disjoint ranges of $\theta_{\rm Ein}$ and $M_{200}$.

The stellar fraction also declines sharply with increasing Einstein radius and halo mass: $\langle f_{*,\rm Salp}^{\rm 2D} \rangle = 0.60$ for the galaxy-scale lenses, 0.17 for groups, and only 0.06 for clusters. 
To first order, this trend drives the decline in $\gamma_{\rm tot}$: since the DM distribution is more extended than the stars, $\gamma_{\rm tot}$ will be shallower in systems with lower $f^{\rm 2D}_{*, \rm Salp}$. The main question is whether the trend in $\gamma_{\rm tot}$ can be ascribed entirely to the declining stellar fraction, or whether a non-universal IMF or dark matter profile is required. We return to this question in Section~\ref{sec:separation}.

In addition to the broad trends in $\gamma_{\rm tot}$ and $f^{\rm 2D}_{*, \rm Salp}$ evident from comparing the galaxy, group, and cluster lens samples with one another, there are also correlations \emph{within} each sample. For the galaxy-scale lenses it has been found that $\Sigma_* = M_* / (2\pi R_e^2)$, the mean stellar surface density within $R_e$, is correlated with $\gamma_{\rm tot}$. This is natural, since galaxies with more concentrated stellar distributions should have steeper density profiles. Furthermore, the dependence on $\Sigma_*$ appears to be fundamental, since  there is no residual correlation with $M_*$ or $R_e$ individually \citep{Sonnenfeld13b}.

Our second important conclusion that we draw from Figure~\ref{fig:totslopes} (right panels) is that although the central galaxies of groups and clusters have lower $\Sigma_*$ and shallower $\gamma_{\rm tot}$ than the galaxy-scale lenses, they do not lie on a simple extension of the scaling relation seen in the galaxy-scale samples. The dashed line in the lower-right panel indicates the slope inferred from the combined SLACS and SL2S samples by \citet{Sonnenfeld13b}; it clearly over-predicts $\gamma_{\rm tot}$ for the group- and cluster-scale lenses. This implies that $\gamma_{\rm tot}$ depends not only on properties of the stellar distribution, namely $\Sigma_*$, but also depends explicitly on properties of the DM halo.

To quantify this dependence and provide a reference for future observational studies and numerical simulations, we fit a linear regression
\begin{align}
\gamma_{\rm tot} = \gamma_{\rm tot, 0} &+ \frac{\partial \gamma_{\rm tot}}{\partial \log \Sigma_*} (\log \Sigma_* - 9.0)\nonumber\\
&+\frac{\partial \gamma_{\rm tot}}{\partial \log M_*} (\log M_* - 12.0)\nonumber\\
&+ \frac{\partial \gamma_{\rm tot}}{\partial \log M_{200}} (\log M_{200} - 14.0)\nonumber\\
& + \frac{\partial \gamma_{\rm tot}}{\partial z} (z - 0.3) + N(0, \sigma'_{\rm \gamma})\label{eqn:regression}
\end{align}
to the combined sample of galaxy, group, and cluster-scale lenses. Here $N(0, \sigma'_{\rm \gamma})$ represents Gaussian intrinsic scatter in $\gamma_{\rm tot}$ after accounting for these linear dependences. A Bayesian procedure was used to infer the linear dependences of $\gamma_{\rm tot}$ on each parameter. To account for small redshift differences among the samples, we take a Gaussian prior $N(-0.30, 0.25)$ on $\partial \gamma_{\rm tot} / \partial z$ from \citet{Sonnenfeld14}, but otherwise we use uniform priors. Individual halo masses are unavailable for the SLACS lenses. Therefore we introduce a parameter $\langle \log M_{200,\rm SLACS} \rangle$ with a Gaussian prior $N(13.2, 0.2)$, based on stacked weak lensing results \citep{Gavazzi07}; an uncertainty of 0.3~dex is assigned to the halo mass of each SLACS lens based on our estimate of the intrinsic scatter.\footnote{Likewise, for the two group-scale lenses lacking satellite kinematic data, we use $\log M_{200} = 14.0 \pm 0.2$ based on the mean and intrinsic scatter seen in the other 8 group-scale lenses.} We find the following linear dependences for $\gamma_{\rm tot}$:
\begin{align}
\gamma_{0, \rm tot} &= 1.79 \pm 0.06 \nonumber\\
\frac{\partial \gamma_{\rm tot}}{\partial \log \Sigma_*} &= 0.34 \pm 0.11 \nonumber\\
\frac{\partial \gamma_{\rm tot}}{\partial \log M_*} &= -0.06 \pm 0.12 \nonumber\\
\frac{\partial \gamma_{\rm tot}}{\partial \log M_{200}} &= -0.33 \pm 0.07 \nonumber\\
\sigma'_{\gamma} &= 0.12 \pm 0.10 \label{eqn:regressresults}
\end{align}
(The posterior distribution for $\partial \gamma_{\rm tot} / \partial z$ is consistent with its prior, as expected, and is omitted here.)

We recover the trend with $\Sigma_*$ seen in galaxy-scale lenses with a consistent value of $\partial \gamma_{\rm tot} / \partial \log \Sigma_*$ \citep{Sonnenfeld13b}, as well as the lack of any additional dependence on the stellar mass of the galaxy (i.e., $\partial \gamma_{\rm tot} / \partial \log M_*$ is consistent with zero). However, we also find a significant \emph{additional} dependence on halo mass: $\partial \gamma_{\rm tot} / \partial \log M_{200} = -0.33 \pm 0.07$ differs significantly from zero. The strength of this dependence on halo mass is roughly equal to the previously known one on $\Sigma_*$. Earlier work has shown systematic variations in the density profile of ETGs \cite[e.g.,][]{Sand08,Humphrey10,Newman13a,Sonnenfeld13b,Tortora14}, but to our knowledge this is the first measurement of the bivariate dependence on two parameters, one connected to the stellar distribution and the other to the DM halo. This is significant for the interpretation of the ``bulge--halo conspiracy'' as we will discuss in Section~\ref{sec:bhc}. These findings are a further demonstration of the non-homology of massive ETGs and show that their internal density structure cannot be predicted from the stellar light alone.
 
One caveat is that the masses and radii of massive galaxies are sensitive to the light profile and measurement technique, especially for BCGs. To test the possible effect on our results, we compared the $\Sigma_*$ of our BCGs to measurements by \citet{Kravtsov14}, who took particular care to fit the large-radius light profile of several BCGs including all intracluster light. Considering their 4 galaxies in halos with $M_{500} > 5 \times 10^{14} \msol$, we find possibly lower $\Sigma_*$ by $\sim0.3$~dex compared to our BCGs. Shifting the $\Sigma_*$ of our BCGs downward by this amount changes $\partial \gamma_{\rm tot} / \partial \log M_{200}$ by only $+0.08$. This is comparable to the random uncertainty and demonstrates that our basic result does not depend on the exact definition of the masses and radii of large galaxies.

\begin{figure*}
\centering
\includegraphics[width=0.8\linewidth]{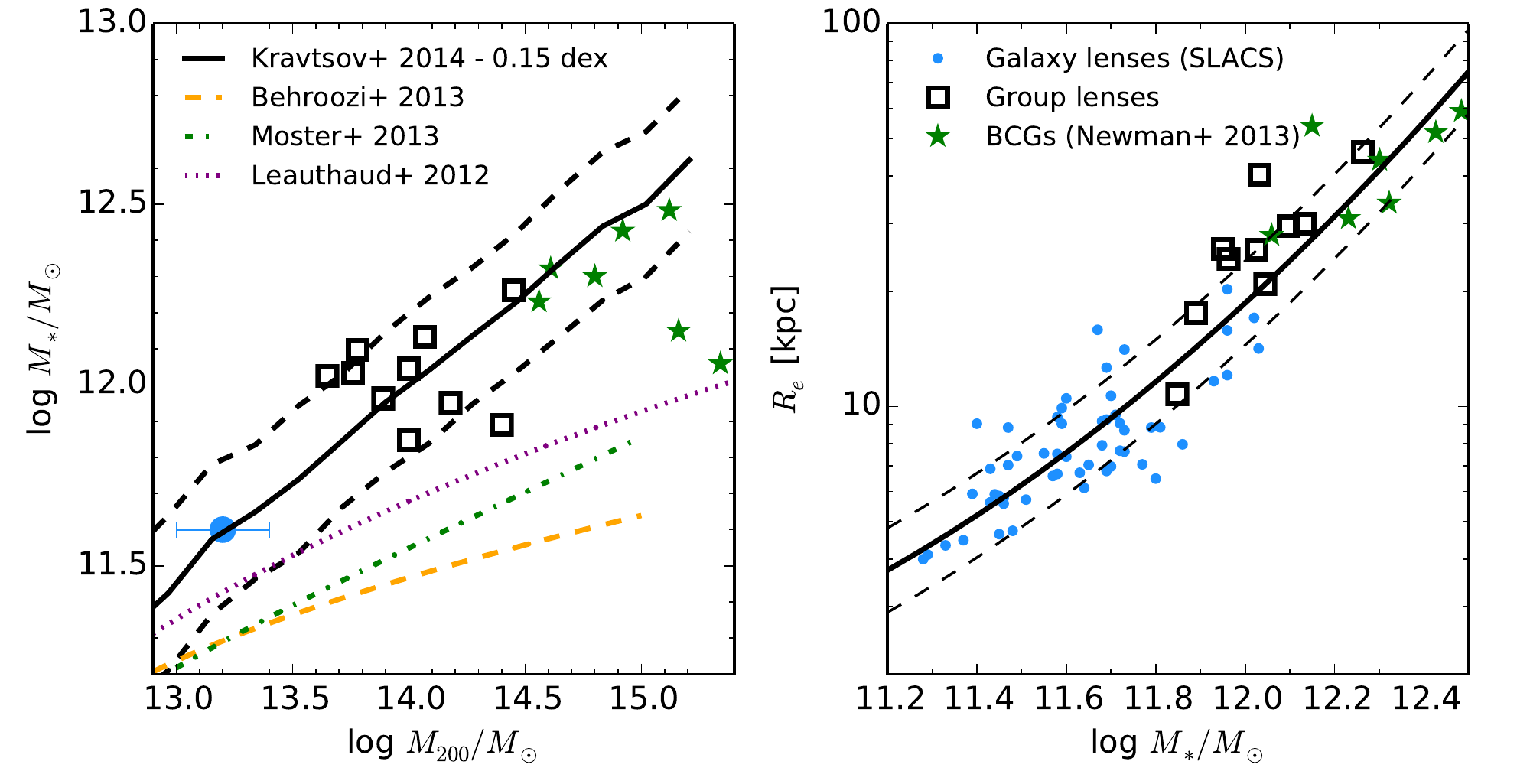}
\caption{Inputs to the halo occupation model described in Section~\ref{sec:cdmmodel}. \emph{Left:} The relation between halo mass and the stellar mass of the central galaxy for three lens samples: galaxy-scale SLACS lenses, based on the stacked weak lensing result of \citet{Gavazzi07}; the present group-scale lenses, based on satellite kinematics in individual lenses; and the N13 clusters, based on individual weak lensing measures. Several theoretical relations are overlaid. All stellar masses have been converted to a Salpeter IMF. The black line shows the abundance matching curve of \citet[][Appendix A]{Kravtsov14} shifted by $-0.15$~dex in $\log M_*$ to better match the lens samples; this is the relation used in our halo occupation models. Black dashed lines show the $1\sigma$ scatter at fixed $M_{200}$. \emph{Right:} The stellar mass--radius relation for lensing ETGs. A quadratic fit smoothly connects the galaxy, group, and cluster scale lenses.\label{fig:inputs}}
\end{figure*}

\section{Separating dark matter and stars within the effective radius}
\label{sec:separation}

Although the \emph{total} density profile within $R_e$ can be characterized well in individual ETG lenses (Section~\ref{sec:totslope}), the ultimate goal of separating the luminous and dark matter profiles is more challenging. This separation is well motivated because it entails (1) the distribution of DM on sub-galactic scales, which is sensitive to the poorly understood interplay between baryons and DM during galaxy formation (and possibly DM particle microphysics; see introduction), as well as (2) an absolute mass scale for the stellar population, which is sensitive to the IMF. Recent studies based on lensing and/or dynamics \citep{Auger10b,Treu10,Spiniello11,Cappellari13,Newman13b} or the analysis of weak surface gravity-sensitive absorption lines in the integrated galaxy light \citep{Conroy12,Ferreras13,LaBarbera13,Spiniello14} have both indicated that massive ETGs have a ``heavier'' IMF than the Milky Way, although the genuine degree of convergence between these techniques is debatable \citep{Smith14}.

We take two complementary approaches toward disentangling the stellar and DM profiles. In Section~\ref{sec:cdmmodel} we construct a simple, CDM-motivated halo occupation model that we use to forward model the distribution of $\gamma_{\rm tot}$ and $f_{*,\rm Salp}^{\rm 2D}$ within the ETG population. By varying the model ingredients and comparing with the observations presented in Section~\ref{sec:totslope}, we can constrain the small-scale DM distribution and the absolute stellar mass scale. This approach naturally encompasses the full range of galaxy- to cluster-scale ETGs and allows us to explore the origins of the trends discussed in Section~\ref{sec:totslope}. A disadvantage is that it does not incorporate all of the data, in particular the \emph{resolved} stellar kinematics for our group-scale ETGs. In Section~\ref{sec:dmimf} we therefore compare the results from the first approach with those obtained from direct mass modeling using the full data set collected for our group-scale lens sample.

\subsection{Interpreting Density Profile Trends with Halo Occupation Models}
\label{sec:cdmmodel}

Here we introduce a set of models in which CDM-motivated halos are populated with ETGs that have, by construction, the same stellar masses and sizes as the ETG lens samples. Since the  galaxy sizes and relative stellar masses are known empirically, they remained fixed throughout our set of models, whereas the inner DM profile and the absolute stellar masses (determined by the IMF) are varied. These ingredients are then constrained by studying the effect of their variation on the observables $\gamma_{\rm tot}$ and $f_{*, \rm Salp}^{\rm 2D}$. 

In short, we randomly sample halos from a halo mass function \citep{Angulo12} at $z=0.36$ and assume that they follow a theoretical MCR. Galaxies are then assigned to the halos based on an empirical stellar mass--halo mass relation. Their stellar distribution is assumed to follow the de Vaucouleurs' profile with $R_e$ sampled from the observed stellar mass--radius relation. \citet{DuttonTreu14} took a similar approach; the main difference with the present paper is that we consider trends over a wide range of halo masses, and we use scaling relations that are constructed from the lens samples themselves. Below we describe the main ingredients in more detail:

\begin{itemize}
\item \emph{Mass--concentration:} We use the \citet{Dutton14} MCR at $z = 0.36$, the mean redshift of the group lenses, with a scatter of 0.13~dex in $\log c_{200}$.

\item \emph{Stellar mass--halo mass:} The stellar masses and sizes of the model galaxies must be compatible with the lens samples. Since the luminosity and $R_e$ of massive ellipticals are sensitive to details of the measurement technique, we construct a stellar mass--halo mass relation using observations of the same lens galaxies under study. In the left panel of Figure~\ref{fig:inputs} we compare this to the relation presented in the Appendix of \citet{Kravtsov14}, which was constructed by abundance matching using the \citet{Bernardi13} stellar mass function. These authors paid particular attention to photometry of massive galaxies. A small shift of $-0.15$~dex in $\log M_*$ (after converting to a Salpeter IMF) brings this relation into reasonable agreement with the lens data. As emphasized by Kravtsov et al., many popular stellar mass--halo mass relations vastly underestimate $M_*$ at the high-mass end. We adopt a scatter of 0.2~dex in $M_*$ at fixed $M_{200}$.

\item \emph{Stellar mass--radius:} The right panel of Figure~\ref{fig:inputs} shows the relation defined by the lens samples. Since there is curvature in the relation at high masses, as noticed by earlier authors \citep[e.g.,][]{Hyde09}, we fit a quadratic relation: $\log R_e / {\rm kpc} = 0.45 + 0.56 \log M_{11} + 0.26 (\log M_{11})^2$, where $M_{11} = M_* / 10^{11} \msol$ and the scatter in $R_e$ is 0.11~dex at fixed $M_*$.\footnote{The group-scale lenses are slightly offset from this fit (Figure~\ref{fig:inputs}) because they are fit with free S\'{e}rsic profiles, whereas the earlier galaxy- and cluster-scale lenses were fit with $n = 4$ fixed. This affects the comparison between data and models mostly in $f_{*, \rm Salp}^{\rm 2D}$ (see slight offset for groups in Figure~\ref{fig:totslopes}, upper-left panel) but has a small effect on the main focus of our analysis, $\gamma_{\rm tot}$: using $n=6$ in the halo occupation models and increasing $R_e$ to better match the group-scale lenses shifts their $\langle \gamma_{\rm tot} \rangle$ by only 0.04.}
\end{itemize}

We vary the stellar IMF between those of \citet{Chabrier03} and \citet{Salpeter55}, as defined in the \citet{BC03} models, and consider three forms for the inner DM profile: unmodified NFW halos, adiabatically contracted halos following the \citet{Gnedin04} formulation, and cored NFW halos with a core radius $r_{\rm core} = 0.01 r_{200}$ (see Equation 3 of \citealt{Newman13a}).

For simplicity, we assume spherical symmetry and so do not model any orientation effects. For each mock galaxy we can then compute $\theta_{\rm Ein}$, $\gamma_{\rm tot}$, and $f_{*, \rm Salp}^{\rm 2D}$. When computing mean properties of model galaxies, we weight the galaxies by the product of the halo mass function (since the model galaxies are sampled uniformly in $\log M_{200}$) and $\theta_{\rm Ein}^2$. The second factor approximately accounts for the strong lensing cross-section, which is closely correlated with $\theta_{\rm Ein}$ \citep[e.g.,][although this factor has no effect when considering trends as a function of $\theta_{\rm Ein}$ as we typically do]{Meneghetti11}.

\begin{figure}
\includegraphics[width=\linewidth]{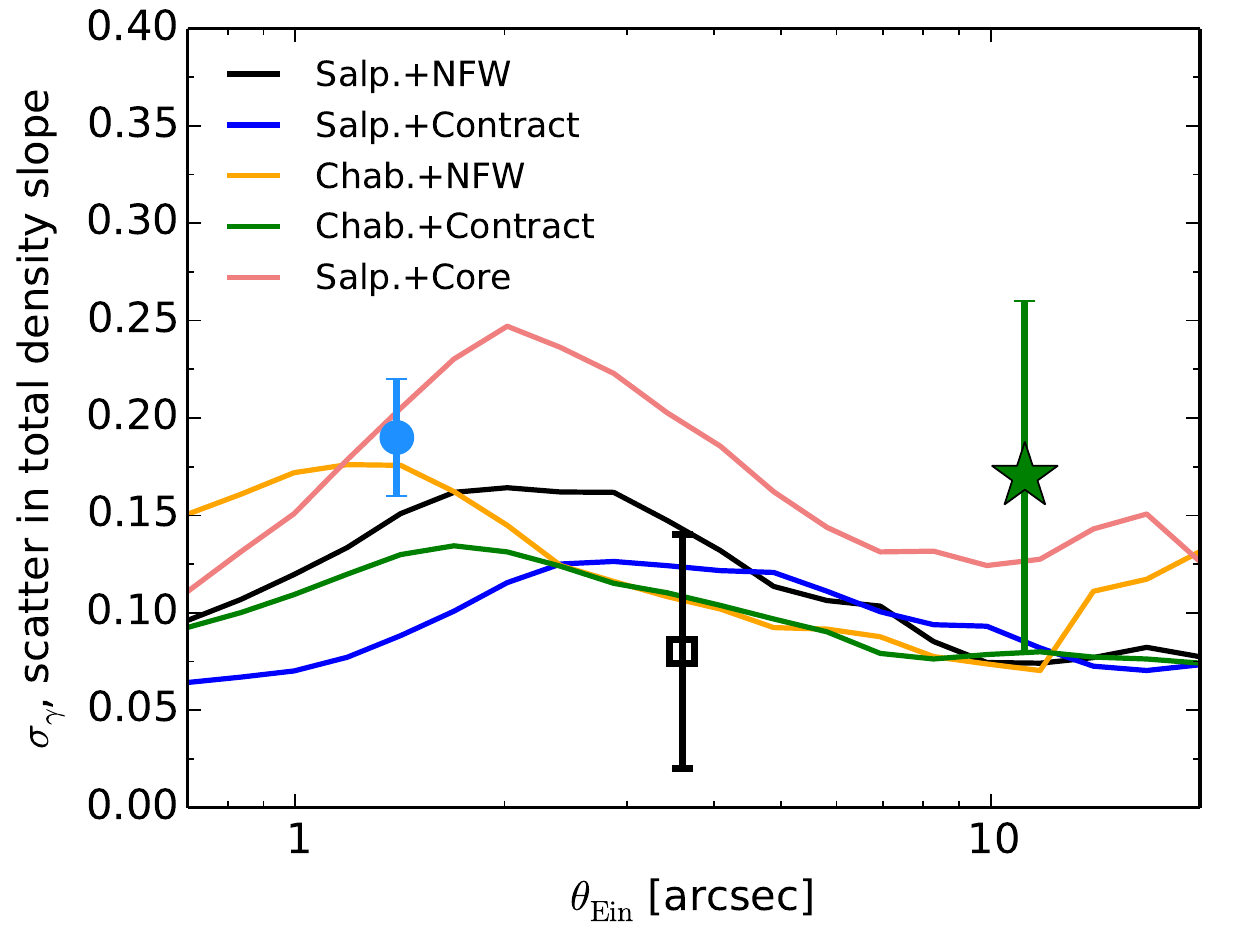}
\caption{Intrinsic scatter in $\gamma_{\rm tot}$ for the same data sets and halo occupation models plotted in Figure~\ref{fig:totslopes}, with matching colors and symbols. For the halo occupation models, we select galaxies having a $\theta_{\rm Ein}$ within $\pm 0.15$~dex of the indicated value, which approximates the width of the $\theta_{\rm Ein}$ distribution in the lens samples. (Note that $\sigma_{\gamma}$ is the total scatter and differs from $\sigma'_{\gamma}$ in Equation~\ref{eqn:regressresults}.)\label{fig:scatter}}
\end{figure}

Lines in the lower-left panel of Figure~\ref{fig:totslopes} show that all models predict shallower density profiles within increasing $\theta_{\rm Ein}$ or $M_{200}$. To first order this decline in $\gamma_{\rm tot}$ is driven by the sharp decrease in $f_{*,\rm Salp}^{\rm 2D}$ visible in the upper-left panel, which is common to all of the halo occupation models. The detailed behavior of $\gamma_{\rm tot}$, however, is sufficiently different to discriminate amongst these models. Considering the galaxy-scale lenses first, we see that the mean $\langle \gamma_{\rm tot} \rangle$ is best reproduced by the models with a Salpeter IMF. Both Chabrier-based models give too shallow slopes. This is consistent with earlier studies of the SLACS sample that assume either unmodified or contracted NFW halos \citep{Auger10b,Treu10}. The upper-left panel shows that the $f^{\rm 2D}_{*, \rm Salp}$ data disfavor contracted halos with a Salpeter IMF, on average, since the stellar fractions at small $\theta_{\rm Ein}$ are too low. Figure~\ref{fig:scatter} compares the \emph{scatter} in $\gamma_{\rm tot}$  in the halo occupation models to the intrinsic scatter measured in the three lens samples. The contracted Salpeter model predicts a scatter in $\gamma_{\rm tot}$ that is too small for the galaxy-scale lenses (although this scatter might increase if a distribution of parameters that describe the adiabatic contraction were included, e.g., \citealt{Gnedin11}). Of the 5 halo occupation models, therefore, those having a Salpeter IMF and unmodified or cored NFW halos best reproduce observations of the galaxy-scale lenses. This is consistent with recent work by \citet{DuttonTreu14}.

Moving on to the groups, the mean $\gamma_{\rm tot}$ is consistent with either unmodified NFW halos and a Salpeter IMF, or contracted halos and a Chabrier IMF. This degeneracy can be broken if we assume that the BGGs have a Salpeter-type IMF similar to the galaxy-scale lenses, which is reasonable if IMF variations are related to the central velocity dispersion \citep[e.g.,][]{Treu10,Cappellari13,Posacki15}.\footnote{If anything, the 0.06~dex higher average central dispersions of the BGGs relative to the SLACS lenses would suggest a \emph{heavier} IMF, although by $\Delta \log M_*/L \lesssim 0.1$, which would place more tension on the adiabatically contracted models.} In that case, we would conclude that nearly unmodified NFW halos are favored, on average, since the slope of the contracted model is too steep, while that of the cored model is too shallow. Formally $\langle \gamma_{\rm tot} \rangle$ is inconsistent with the contracted Salpeter model by $2.6\sigma$ and the cored model by $4.2\sigma$. We note that the $f_{*, \rm Salp}^{\rm 2D}$ data (upper-left panel) are not helpful for distinguishing models at the group and cluster scales.

At the cluster scale, models with NFW or contracted halos have steeper $\gamma_{\rm tot}$ than \emph{all} of the observations. This is consistent with N13, who inferred that the DM density profile is shallower than the NFW profile within $\simeq R_e$, whereas the \emph{total} density slope $\gamma_{\rm tot}$ is very close to that of a pure NFW halo (compare the green star and dashed line in Figure~\ref{fig:totslopes}). Our model with a core radius of $r_{\rm core} = 0.01 r_{200}$ fits the clusters, which is unsurprising since it was intended to approximate the N13 result. However, we see here that the same model does not match the group-scale lenses. This suggests that either cores are confined to the most massive halos, or at least are smaller as a fraction of $r_{200}$ in the groups.

The right panels in Figure~\ref{fig:totslopes} show that the halo mass dependence at fixed $\Sigma_*$ seen in the data is also present in the halo occupation models. Here we consider only the Salpeter+NFW model to clarify the trends, and we plot results (gray bands) in narrow intervals of halo mass that approximate those of the lens samples. At a given halo mass there is very little scatter in the relation between $\Sigma_*$ and $\gamma_{\rm tot}$ or $f^{\rm 2D}_{*, \rm Salp}$, which again indicates that the range of ETG properties seen at fixed $\Sigma_*$ is driven by differences in halo mass.

The halo occupation models allow us to estimate the effect that a strong lensing selection might have on the linear regression in Equation~\ref{eqn:regression}. Although strong lensing ETGs are representative of ETGs having a given $\theta_{\rm Ein}$ (or, nearly equivalently for the galaxy-scale lenses, a given $\sigma$), they are not strictly representative of ETGs of a given stellar or halo mass. We fit our ensemble of model galaxies using two different weights: (1) uniform in $\log M_{200}$, and (2) weighted to match the $\theta_{\rm Ein}$ distribution of the full observed lens sample. Compared to the uniformly weighted case, the lens analogs differ in $\gamma_{0, \rm tot}$ by $+0.03$, $\partial \gamma_{\rm tot} / \partial \log \Sigma_*$ by $+0.07$, and $\partial \gamma_{\rm tot} / \partial \log M_{200}$ by $+0.03$. These differences are less than the statistical errors, demonstrating that for the range of galaxies we consider and the quality of the available data, the observed trends are minimally affected by the strong lensing selection (see also \citealt{Sonnenfeld15}).

\subsection{Constraints on the Inner Dark Matter Profile from Resolved Stellar Kinematics}
\label{sec:dmimf}

\begin{figure*}
\centering
\includegraphics[width=0.7\linewidth]{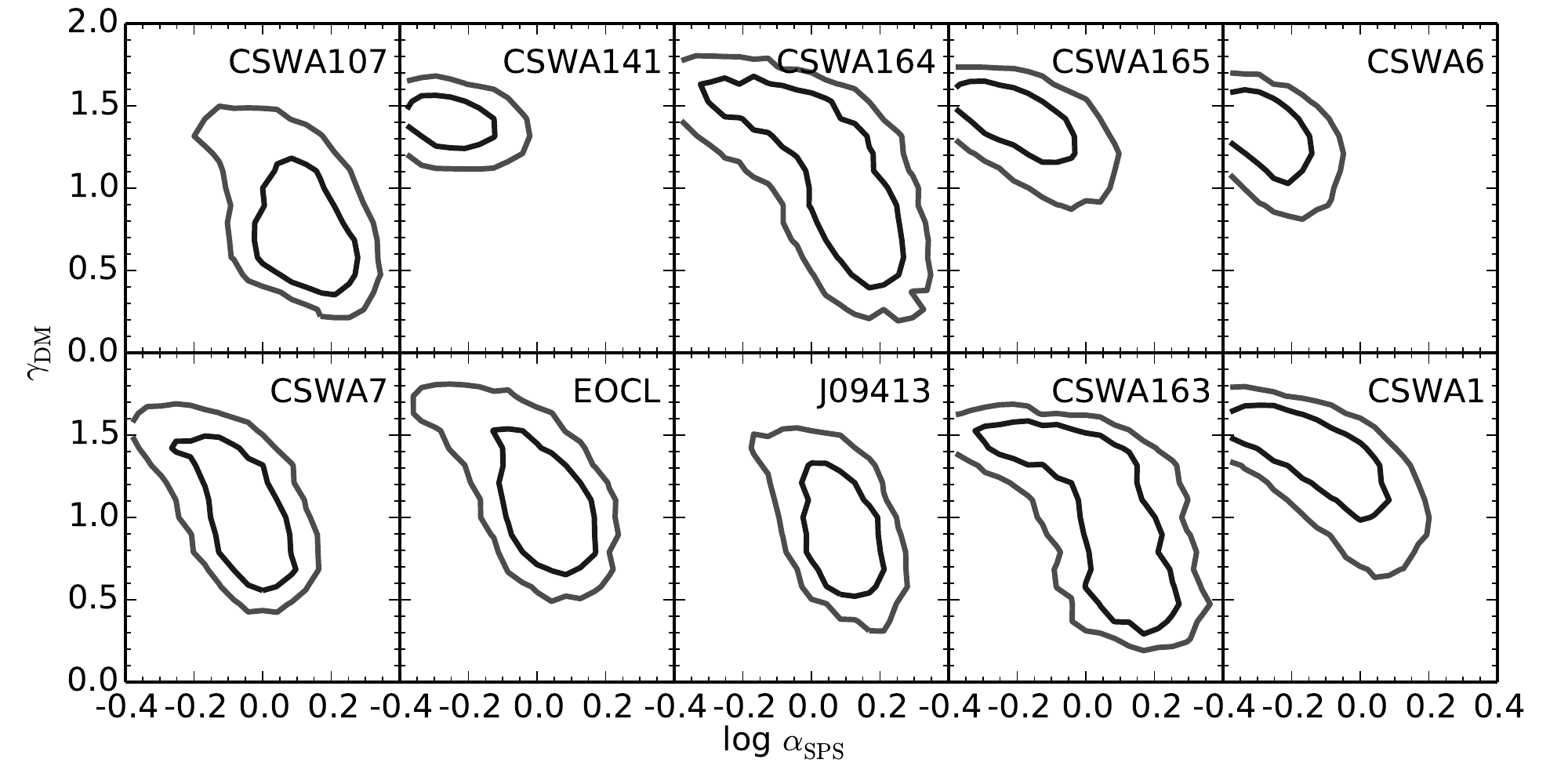}
\caption{Credible regions for the IMF mismatch parameter $\alpha_{\rm SPS}$ and $\gamma_{\rm DM}$, the average DM slope within $R_e$, for our 10 group-scale lenses. Contours enclose 68\% and 95\% of the posterior distribution.\label{fig:dm_imf_degen}}
\end{figure*}

The forward modeling approach presented in Section~\ref{sec:cdmmodel} relies on only two observables, $\gamma_{\rm tot}$ and $f_{*, \rm Salp}^{\rm 2D}$, and so does not make full use of the data collected for our group-scale lenses, especially the resolved $\sigma(R)$ profiles that are typically unavailable for galaxy-scale lenses. Here we consider the constraints on the mass distribution of the group-scale lenses that arise from direct mass modeling (Section~\ref{sec:modeling}) of the full data set presented in Sections~3--5.

To quantify the inner DM profile, we introduce a measure analogous to $\gamma_{\rm tot}$. Specifically, we define $\gamma_{\rm DM}$ as in Equation~\ref{eqn:gammatot}, but replace the total density and mass profiles $\rho(r)$ and $M(r)$ by those of the DM halo. This quantity represents the average DM density slope within $R_e$ and is better constrained by the data than the asymptotic slope $\beta$ that appears in the gNFW parametrization (Equation~\ref{eqn:gnfw}).

Figure~\ref{fig:dm_imf_degen} shows the constraints on $\gamma_{\rm DM}$ and $\alpha_{\rm SPS}$ for each lens. The lenses with the noisiest velocity dispersion profiles have broad contours indicating a nearly complete degeneracy between the DM profile and stellar mass normalization. As the quality of the data improve, they select preferred regions along this covariance. Broadly, the slope of the velocity dispersion profile is determined by the relative contribution of the stars and DM halo: the stellar mass profile contributes a roughly constant dispersion ($\rho_* \propto r^{-2}$, $\sigma(r) \propto r^0$ for isotropic orbits) whereas the DM halo has a shallower density profile and so contributes a rising velocity dispersion at small radii. (Radially variable velocity anisotropy can also affect the shape of the velocity dispersion profile, and we explore this issue further below.)

\begin{figure*}
\centering
\includegraphics[width=0.45\linewidth]{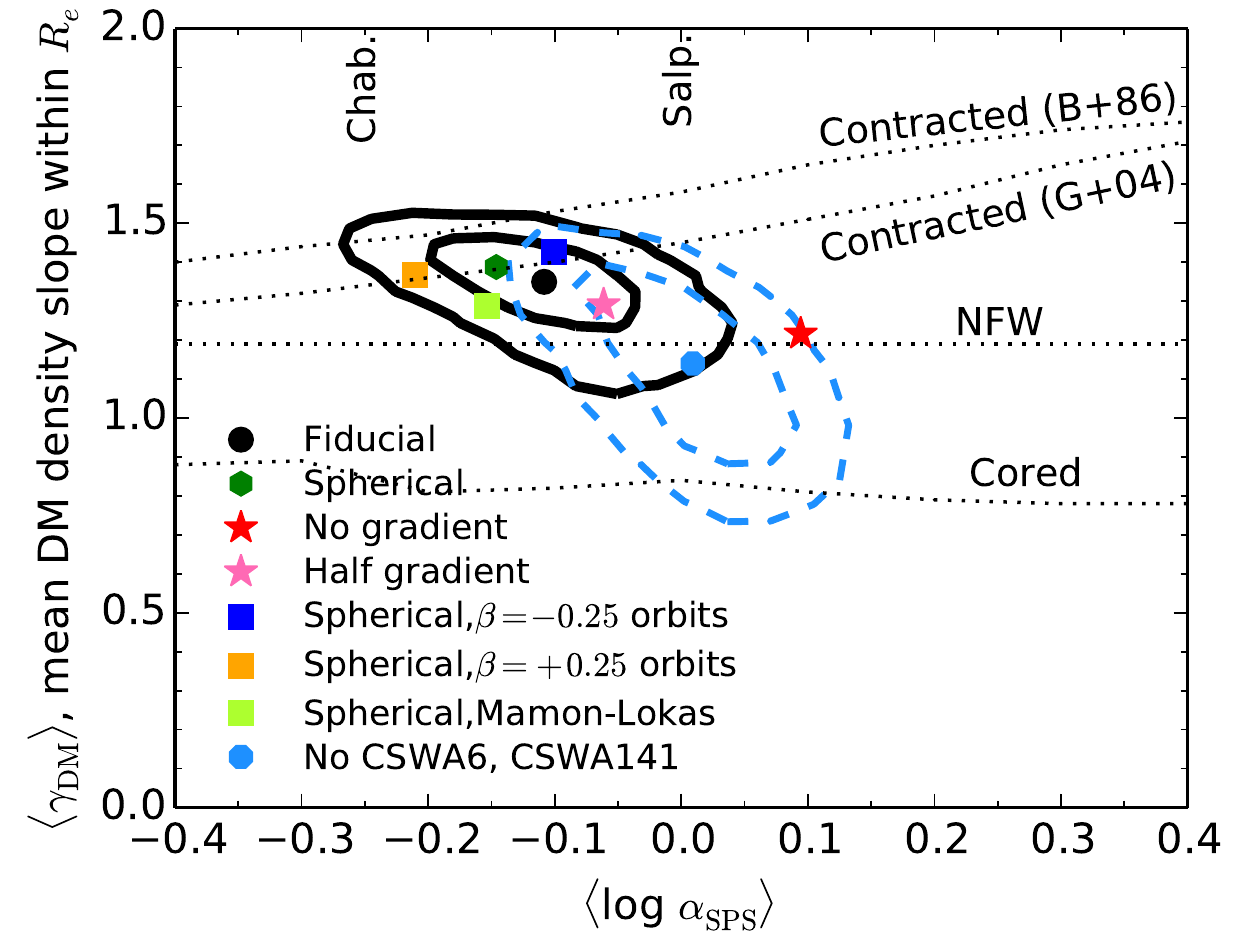}
\includegraphics[width=0.45\linewidth]{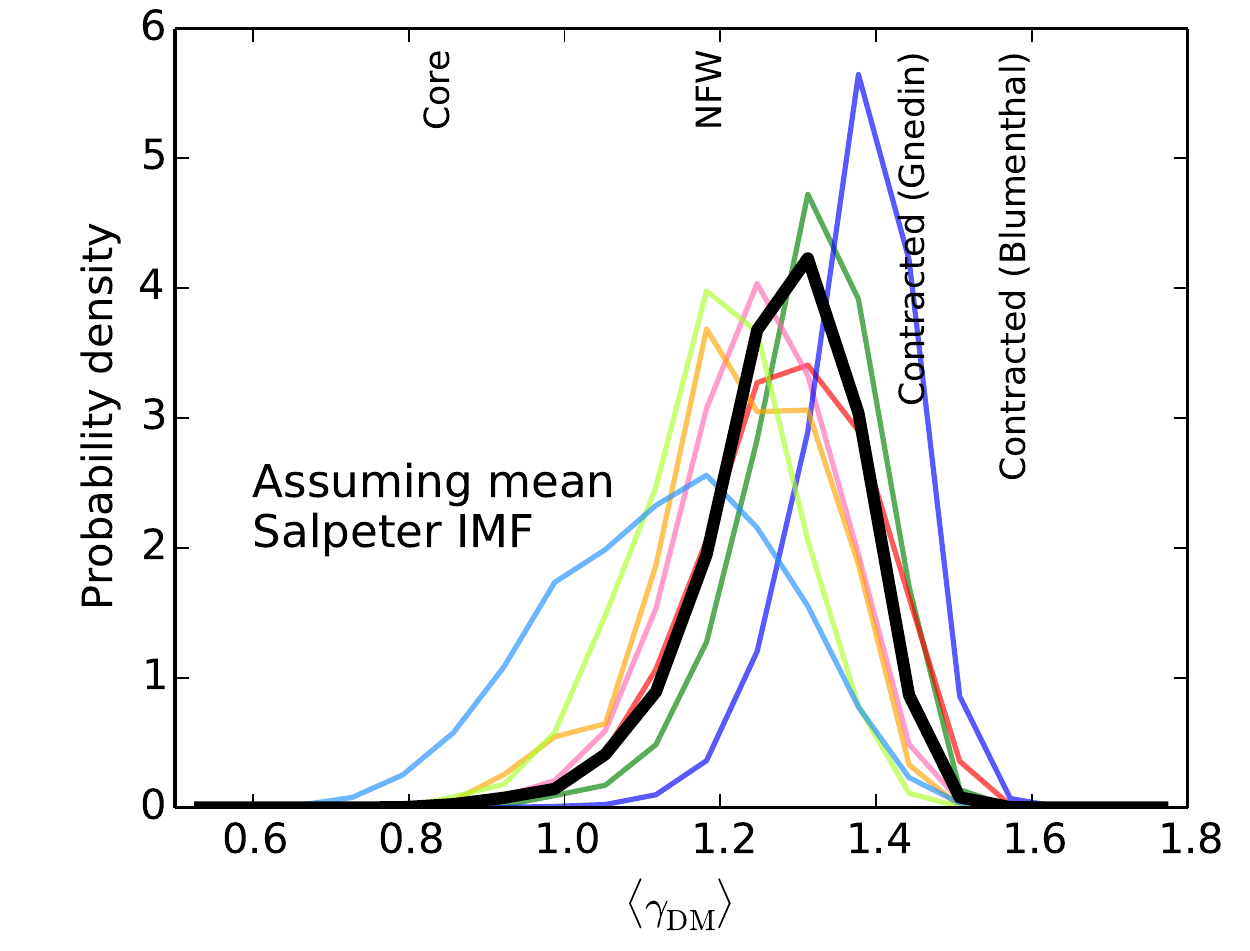}
\caption{\emph{Left:} Constraints on the mean $\langle \log \alpha_{\rm SPS} \rangle$ and $\langle \gamma_{\rm DM} \rangle$ for our ensemble of 10 group-scale lenses. The black circle and contours correspond to the fiducial mass model. Colored symbols show the systematic shifts that arise from various modifications to these model assumptions as indicated in the caption. Contours enclose the 68\% and 95\% credible regions. \emph{Right:} Marginalized constraints on the mean DM slope $\langle \gamma_{\rm DM} \rangle$ of the group lenses if a prior centered on a Salpeter IMF is adopted for $\langle \log \alpha_{\rm SPS} \rangle$. The thick black line represents our fiducial mass model, while colored lines have the same meaning as the corresponding symbols in the left panel. Labels on the top axis, as well as dotted lines in the left panel, show the $\gamma_{\rm DM}$ expected for various models of the inner DM distribution, derived by selecting lenses from our halo occupation models (Section~\ref{sec:cdmmodel}) matching the $\theta_{\rm Ein}$ distribution of the group lenses. We emphasize that $\gamma_{\rm DM}$ is averaged within $R_e$ and is not the same as the asymptotic inner slope $\beta$ appearing in the gNFW parametrization (Equation~\ref{eqn:gnfw}); in particular, $\gamma_{\rm DM} \neq 1$ for NFW halos.\label{fig:dm_imf_groups}}
\end{figure*}

Since the decomposition between DM and stars is noisy in individual objects, we seek to combine the sample to constrain the mean $\alpha_{\rm SPS}$ and $\gamma_{\rm DM}$ for the \emph{ensemble} of group-scale lenses. To do so, we use the framework developed in Appendix~B and find $\langle \log \alpha_{\rm SPS} \rangle = -0.11 \pm 0.06$ and $\langle \gamma_{\rm DM} \rangle = 1.35 \pm 0.09$ for our fiducial mass models.\footnote{Here $\alpha_{\rm SPS}$ is defined relative to a Salpeter IMF and so is lower than a Chabrier-based definition, as used by N13, by 0.25~dex.} These constraints are displayed in Figure~\ref{fig:dm_imf_groups}. In addition to the constraints from our fiducial mass models, we also plot the systematic shifts that arise from several modifications of our model assumptions described further below. These are indicated by the colored symbols. The dashed blue contours indicate the posterior distribution if CSWA6 and CSWA141 are excluded, based on the poorer fits to their $\sigma(R)$ profiles mentioned in Section~\ref{sec:fitqual}. 
To place these values in context, we select lenses with a matching $\theta_{\rm Ein}$ distribution in our halo occupation models and compute the mean $\gamma_{\rm DM}$ under the several models for the inner DM profile (dotted lines). We also include here a stronger adiabatic contraction prescription following \citet{Blumenthal86}.

Even after combining data from our 10 group lenses, there is a significant covariance between $\alpha_{\rm SPS}$ and $\gamma_{\rm DM}$. This particularly true when considering the systematic shifts that dominate the error budget. Including these, we find $\langle \gamma_{\rm DM} \rangle = 1.35 \pm 0.09~{\rm (stat.)}~{}^{+0.08}_{-0.21}~{\rm (sys.)}$. Figure~\ref{fig:dm_imf_groups} shows that $\alpha_{\rm SPS}$ is not robustly constrained by the group-scale lens data alone: it is particularly sensitive to the treatment of the radial $M_*/L$ gradient (compare the red and blue stars to the black circle) and to the inclusion of CSWA6 and CSWA141 (dashed contours).

More informative results on the inner DM profile are possible with a prior on the IMF motivated by external observations. Several studies, including our inference for group-scale lenses in Section~\ref{sec:cdmmodel} and for BCGs in \citet{Newman13b}, favor a Salpeter or even heavier IMF in massive ellipticals (see also references in Section~\ref{sec:separation}). Therefore, we also consider the result of imposing a \emph{mean} Salpeter IMF in our analysis: the right panel of Figure~\ref{fig:dm_imf_groups} shows the constraints on $\langle \gamma_{\rm DM} \rangle$ adopting a Gaussian prior $N(0.0, 0.05) $ on $\langle \log \alpha_{\rm SPS} \rangle$.

Under this assumption, the data favor an inner DM slope between that of an unmodified NFW halo and a mildly contracted one (i.e., contraction following \citealt{Gnedin04}). Stronger \citet{Blumenthal86} contraction is excluded, and the cored NFW profile preferred by the clusters ($r_{\rm core} / r_{200} = 0.01$) is disfavored. However, the significance of the latter conclusion is sensitive to the treatment of CSWA6 and CSWA141 (light blue line). When these systems are omitted, the discrepancy with the cored NFW profile decreases from $4.4\sigma$ to $1.9\sigma$. These results are consistent with the conclusions drawn from our halo occupation models in Section~\ref{sec:cdmmodel}, although the levels of significance differ. The present approach is more general, since it imposes no strong prior on halo concentrations and allows for galaxy-to-galaxy scatter in $\alpha_{\rm SPS}$. For this reason, we conclude that although the group-scale lenses disfavor the shallow DM profiles found in the centers of clusters, this difference is not yet definitive.

\subsection{Systematic Uncertainties}
\label{sec:sys}

Here we outline several tests designed to assess the sensitivity of $\langle \gamma_{\rm DM} \rangle$ and $\langle \log \alpha_{\rm SPS} \rangle$ to our model assumptions. Colored lines and symbols in Figure~\ref{fig:dm_imf_groups} show the inference obtained under each  modification to the fiducial model described below.

First, we switch from axisymmetric to spherical dynamical calculations, as used in most dynamical analyses of lenses, and adopt a prior of $N(0.0, 0.2)$ on the radial anisotropy parameter $\beta_r$ based on the results of \citet{Cappellari07}. Assuming that the axisymmetric models are more correct, the simplifying assumption of sphericity tends to make the IMF normalization slightly too light and the DM profile slightly too steep, but the effect is small (see Figure~\ref{fig:dm_imf_groups}). Second, we considered spherical models with fixed velocity anisotropy: both radially constant models with $\beta_r = 1 - \sigma_{\theta}^2/\sigma_r^2 = \pm 0.25$, which is intended to approximate the systematic differences in the mean anisotropy of massive ellipticals as derived by different authors \citep[e.g.,][]{Gerhard01,Cappellari07}, and a model with radially varying anisotropy following the \citet{Mamon05} parametrization with $r_a = 0.5 R_e$.\footnote{The anisotropy parameter increases from $\beta_r = 0$ in the center to $+0.25$ at $0.5 R_e$.} The radially variable model gives results very close to the spherical isotropic model. These tests are reassuring for N13 study of BCGs based on spherical dynamics and constant anisotropy models.

The biggest systematic uncertainties come not from the dynamical modeling, but from the magnitude of the $M_*/L$ gradient and the influence of CSWA6 and CSWA141. The red star in Figure~\ref{fig:dm_imf_groups} show the results using our fiducial dynamical model but a radially invariant $\Upsilon_V$, which we take as the mean $\langle \Upsilon_V^{\rm SPS} \rangle$ within $R_e$ listed in Table~\ref{tab:surfphot}. Neglecting the $M_*/L$ gradient entirely has a very large effect on $\langle \log \alpha_{\rm SPS} \rangle$, shifting it systematically by $+0.20$~dex. While this scenario is not plausible given the clearly observed color gradients, the test shows that the value of the gradient is important. We also considered a model in which the $\nabla \Upsilon_V$ prior is centered on $-0.08$, which corresponds the mean gradient generally found for large samples of ellipticals as discussed in Section~\ref{sec:sps}. This more realistic shift has a much milder effect on the resulting constraints (see pink star in Figure~\ref{fig:dm_imf_groups}). 

Finally, although ad hoc, it is useful to consider the influence of systems where the fit quality is poorer and the mass models may be inadequate. The exclusion of CSWA6 and CSWA141 has a significant effect on both $\langle \alpha_{\rm SPS} \rangle$ and $\langle \gamma_{\rm DM} \rangle$, as described in Section~9.2. 

Systematic uncertainties have a larger effect on $\alpha_{\rm SPS}$ and $\gamma_{\rm DM}$ than on the total density slope $\gamma_{\rm tot}$. The modifications to our mass models described above shift $\langle \gamma_{\rm tot} \rangle$ by only $\pm 0.07$, as quoted in Section~\ref{sec:totslope}. 

\section{Discussion}
\label{sec:discussion}

We have leveraged the discovery of large numbers of intermediate-separation strong lenses \citep[e.g.,][]{Limousin09,More12,Stark13} to study the mass distribution of ETGs central to $\simeq 10^{14} \msol$ group-scale halos. By combining strong lensing, satellite kinematics, and resolved stellar kinematics within the lens galaxy, we have constrained the mass distribution on multiple scales. This is the first study to apply these observational methods to a sample of group-scale lenses rather than to an individual system. It bridges a gap between earlier well-studied samples of lenses at the galaxy and cluster scales. Here we consider the implications of our findings for ETG formation scenarios and the dominant physical processes by which DM is redistributed during galaxy formation.

\subsection{Halo Concentrations at the Group Scale}

On large scales, we have measured the structure of the group-scale DM halos via the MCR (Section~\ref{sec:mcr}). We find a mean concentration that is consistent with expectations for unmodified CDM halos, with an uncertainty comparable to the variation among theoretical MCRs based on different cosmologies or halo selection criteria. This implies that baryons have minimally affected the structure of their halos on these scales, which is in good agreement with theoretical expectations but disagrees with some earlier observations at this mass scale. On the simulation front, \citet{Duffy10} found that the inclusion of baryons in group-scale halos increases their concentration by only $\Delta c = 0.1$ in their weak feedback scenario, while for stronger feedback prescriptions the concentrations were \emph{lower} than DM-only simulations. Likewise, \citet{Schaller15a} detect an increase of only $\Delta c_{200} = 0.4$ (i.e., 9\% or $\Delta \log c_{200} \simeq 0.04$~dex) when baryons are included in the EAGLE simulations.

In Section~\ref{sec:compareconc} we compared our findings with earlier observational constraints. Our results are consistent with the \citet{Buote07} X-ray--based study and with \citet{Auger13}, who studied a lens sample that overlaps ours but used different observables. Other lensing studies, however, have found higher concentrations at halo masses near $10^{14} \msol$. These discrepancies probably arise from different distributions of $\theta_{\rm Ein}$ in samples selected in different ways. Samples consisting only of lenses with $\theta_{\rm Ein} > 5''$ \citep[e.g.,][]{Oguri12} will necessarily show high concentrations for $\simeq 10^{14} \msol$ halos. Lenses with $\theta_{\rm Ein} \simeq 2\farcs5 - 5''$ are much more numerous. Our study focuses on these lenses, which may well be the dominant lens population in $\simeq 10^{14} \msol$ halos. Tremendous progress has been made by several groups (see Introduction) in building up large samples of these intermediate-separation lenses. Further progress in using them to constrain the MCR will likely require a more precise quantification of the selection functions of the surveys \citep[e.g.,][]{Gavazzi14}.

\subsection{Trends in the Total Density Profile Slope and Implications for the Bulge--Halo Conspiracy}
\label{sec:bhc}

A well-known and striking result from lensing and dynamical studies of massive ETGs is the similarity of the total density profile to the isothermal form, $\rho_{\rm tot} \propto r^{-2}$, which corresponds to a flat circular velocity curve \citep[e.g.,][]{Kronawitter00,Gerhard01,Treu04}. The proximity of the total density slope to the isothermal profile and its relatively small scatter have been interpreted as evidence for a ``bulge--halo conspiracy'' \citep[e.g.,][]{Koopmans06,Koopmans09,Gavazzi07}, in which some interaction between the stellar and DM components is hypothesized to drive the total mass density toward this form, e.g., perhaps as fundamental dynamical consequence of violent relaxation \citep{Treu02}.

These measurements pertain to ETGs in a fairly narrow range of halo masses ($\sigma_{\log M_{200}} \simeq 0.3$ for the SLACS galaxy-scale lenses; Section~\ref{sec:totslope}). By extending the same lensing and dynamical methods to the central galaxies of larger halos, we are able to better define the scope of the ``conspiracy.'' First, ETGs are not limited to nearly isothermal profiles, but display a wide range of density profiles within $R_e$ that spans $\gamma_{\rm tot} \simeq 1.0-2.4$ (Figure~\ref{fig:totslopes}, bottom left panel). Second, $\gamma_{\rm tot}$ depends jointly on two parameters, namely $\Sigma_*$ and $M_{200}$, that reflect properties of the stellar and DM distributions \emph{separately} (Figure~\ref{fig:totslopes}, bottom right panel, and Equation~\ref{eqn:regression}). Third, the small scatter in $\gamma_{\rm tot}$ at fixed $\theta_{\rm Ein}$ can be reproduced by a range of halo occupation models in which \emph{independent} scatter is assumed in the stellar mass--halo mass, stellar mass--radius, and halo mass--concentration relations (Figure~\ref{fig:scatter}). Our models without halo contraction explicitly have no ``cross-talk'' between the baryons and DM halo, other than what is contained in the stellar mass--halo mass relation, yet are able to reproduce the observed small scatter in $\gamma_{\rm tot}$. This third point was also found in recent similar models by \citet{DuttonTreu14}.

As pointed out by those authors, current cosmological, hydrodynamical simulations cannot match both the density slope and the star formation efficiency of ETGs \citep{Duffy10,Johansson12,Dubois13,Remus13}, which presumably reflects the inadequacy of current feedback implementations. Our measurement of the dependence of $\gamma_{\rm tot}$ on stellar structure and halo mass (Equation~\ref{eqn:regressresults}) can be used to test the performance of different feedback prescriptions in halos of widely varying mass. For instance, based on their suite of cosmological zoom-in simulations, \citet{Remus13} conclude the isothermal density profile is a ``natural attractor'' for the evolution of ETGs and that small deviations from it are compatible with the variation in assembly histories (e.g., number and gas content of mergers) seen in the simulations. Their density profiles, however, appear to be too homogeneous: nearly all of the simulated ETGs have $\gamma_{\rm tot} \gtrsim 1.9$, including systems as massive as BCGs, which is unlike the broad and systematic trends discussed in this paper. A more detailed comparison will require a careful matching of the definition of $\gamma_{\rm tot}$ used in simulations and observations.

\begin{figure}
\centering
\includegraphics[width=\linewidth]{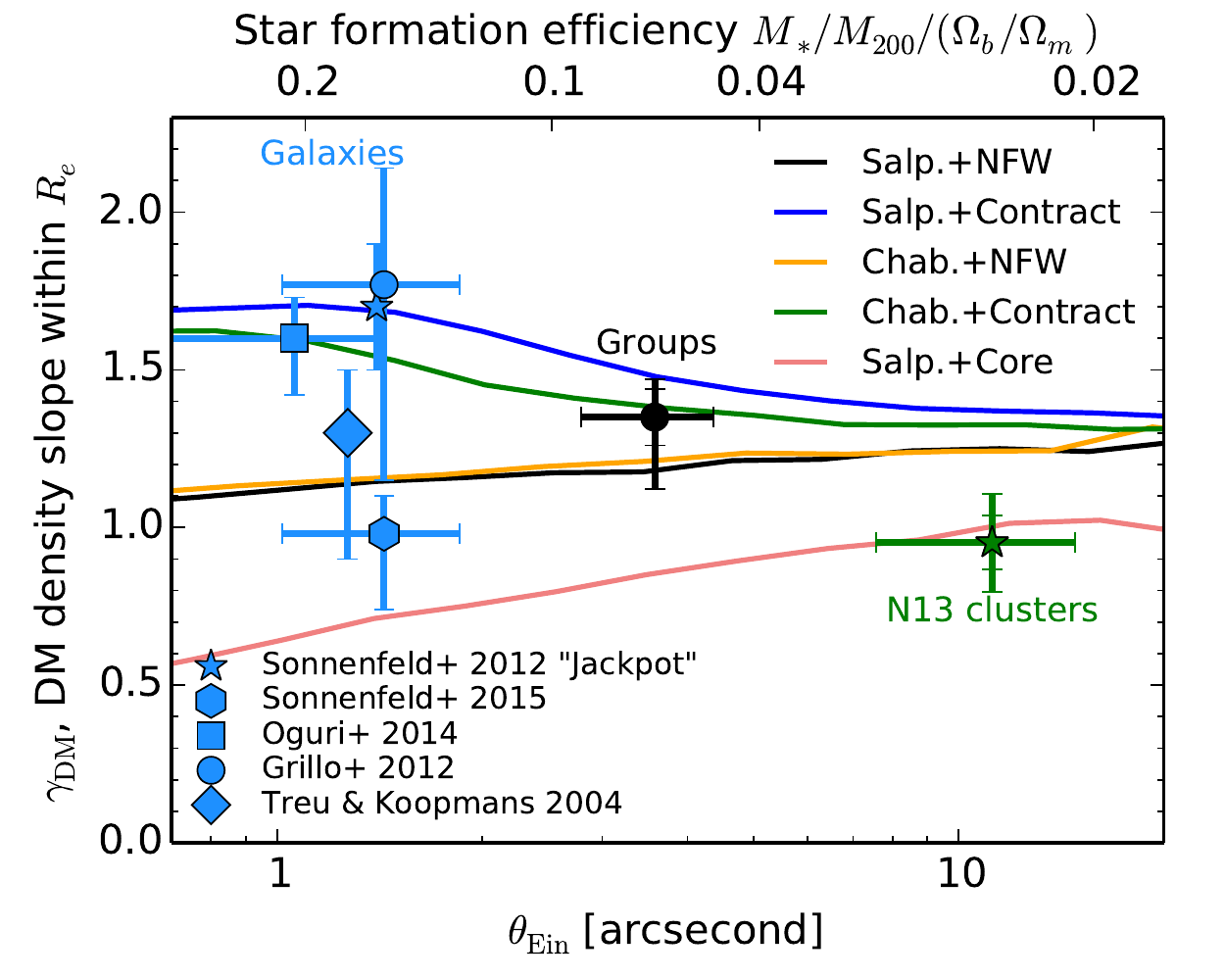}
\caption{Constraints on $\gamma_{\rm DM}$, the average DM density slope within $R_e$, for galaxy, group, and cluster-scale ETG lenses. For group and cluster lenses, $\gamma_{\rm DM}$ is computed following Equation~\ref{eqn:gammatot}, where $\rho$ and $M$ now refer to the DM profile; outer error bars include our systematic error estimate added in quadrature. We emphasize that $\gamma_{\rm tot}$ refers to the average slope within $R_e$ and cannot be directly compared to the asymptotic inner slope $\beta$ that appears in the gNFW profile (Equation~\ref{eqn:gnfw}) and is quoted in other works (including N13b). Results for galaxy-scale lenses are compiled as indicated in the caption. The \citet{Grillo12} constraint assumes a Salpeter IMF; the other results do not assume a particular IMF \emph{a priori}. The \citet{Sonnenfeld15} result is converted from the gNFW inner slope to our definition of $\gamma_{\rm DM}$. Lines follow the halo occupation models introduced in Section~\ref{sec:cdmmodel}.
\label{fig:dmslopes}}
\end{figure}

\subsection{The Dark Matter Density Profile within the Effective Radius: Galaxies, Groups, and Clusters}

Although the slope of the \emph{total} density profile clearly varies systematically with halo mass, assessing the universality of the \emph{dark matter} profile on small scales is more difficult. Figure~\ref{fig:dmslopes} summarizes the current constraints on the inner DM density profile of massive ETGs derived from lensing and dynamical observations.

The situation is most uncertain in the galaxy-scale lenses, where the high degree of baryon dominance makes it especially difficult to extract detailed information about the DM distribution on small scales (the ``bulge--halo degeneracy''). Our analysis in Section~\ref{sec:cdmmodel} disfavored contracted models. Similar analyses by \citet{DuttonTreu14} and \citet{Auger10b} likewise support nearly unmodified NFW halos, with mildly contracted ones permitted in the latter study. However, other studies instead favor contracted halos at this mass scale. \citet{Oguri14} incorporate microlensing constraints to infer a steep DM slope in galaxy-scale lenses, although the precision of the result has been questioned \citep{Schechter14}. \citet{Grillo12} construct a composite density profile based on an ensemble of galaxy-scale lenses and, under the assumption of a Salpeter IMF, derive a steep slope $\gamma_{\rm DM} = 1.77^{+0.37}_{-0.62}$. The uncertainty, however, is large, and \citet{DuttonTreu14} revised this analysis to a shallower slope of $1.40^{+0.15}_{-0.26}$. \citet{Sonnenfeld12} analyzed a rare double Einstein ring, allowing them to break the usual bulge--halo degeneracy and measure a slope $\gamma_{\rm DM} = 1.7 \pm 0.2$ that is indicative of a contracted halo, but it is unclear how representative this single lens is of the overall population of massive ETGs. Building on the methods of \citet{Treu04}, \citet{Sonnenfeld14} studied an ensemble of galaxy-scale lenses in the SLACS and SL2S surveys; although the DM slope cannot be constrained for individual objects with unresolved stellar kinematics, variations in the source redshift lead to variations in $R_{\rm Ein} / R_e$, which allows loose constraints to be placed on the DM slope. Their results favor NFW or even shallower DM profiles, although the authors stress the large uncertainties.

Greater progress has been made in constraining the DM profile in higher-mass halos owing to the reduced dominance of baryons in these systems and the richer observational constraints. N13a,b showed that the central galaxies in massive clusters have a DM profile that is shallower than the canonical NFW slope within $\simeq R_e$, a conclusion that is further supported by the halo occupation models presented here. Several observations guided the N13 interpretation of this result: the similarity of radial scale at which the DM profile deviates to the effective radius of the BCG, the remarkable similarity of the \emph{total} density profile to the NFW slope on these scales, and a tentative correlation between the inner DM profile and $R_e$ when individual objects are examined. These lines of evidence suggest that the flattening the DM profile within BCGs is connected with the assembly of stars in the galaxy. Such a connection might be achieved dynamically through the transfer of angular momentum from infalling satellites to the DM halo via dynamical friction \citep{ElZant01,ElZant04,Nipoti04,Tonini06,RomanoDiaz08,Jardell09,Johansson09,DelPopolo12}. \citet{Laporte15} recently used $N$-body simulations containing both DM and stars to study how these species mix during the assembly of a BCG. They verified that the DM profile can be flattened over time, provided that the assembly of the BCG is dominated by dissipationless stellar accretion following $z \simeq 2$ and that the progenitor galaxies at $z \simeq 2$ are realistic.

Our new observations of group-scale lenses, located in halos that are $\simeq 10$ times less massive than the N13 clusters, instead favor unmodified NFW halos. Mildly contracted halos following the \citet{Gnedin04} model are also allowed at the $1.4-2.6\sigma$ level, but stronger contraction prescriptions are ruled out if the stellar mass scale is indeed close to that of a Salpeter IMF. Cored DM halos with $r_{\rm core} / r_{200}$ as large as those seen in clusters are also disfavored by our data. They are not yet definitively ruled out: Figure~\ref{fig:dm_imf_groups} shows that the groups deviate from the cored halo model by $<2\sigma$ \emph{if} CSWA6 and CSWA141 (whose velocity dispersion profiles are less well fit) are excluded from the present sample.

\subsection{Interpreting Possible Trends in the Inner Dark Matter Density Profile}

These observations may support a scenario in which the DM profile is determined by the relative importance of ``two phases'' of massive galaxy formation \citep[e.g.,][]{Oser10}: early compression of the halo is associated with dissipative, in situ star formation fed by gas infall, while later reduction of the central density through dynamical friction occurs as dissipationless merging dominates the mass assembly. The balance of these processes, given by the fraction of stars formed in situ, is a strong function of halo mass: \citet{Behroozi13} find roughly 46\%, 24\%, and 7\% at $z=0.3$ for halos that correspond to our galaxy, group, and cluster lens samples. This may well lead to systematic trends in the inner DM profile as a function of halo mass, with the underlying physical driver being the variation in star formation efficiency $M_*/M_{200}$ and assembly history, or in other words, the balance of dissipational and dissipationless formation \citep{Lackner10}.

In this picture halo contraction becomes increasingly significant for ETGs in lower-mass halos, where more gas has cooled and been transformed into stars, whereas flattened DM profiles appear at the highest masses due to the rising importance of dissipationless mergers. Such a trend could explain why no single model in Figure~\ref{fig:dmslopes} is able to fit observations from galaxy to cluster scales. It is also possible that AGN feedback plays an increasing role in flattening the DM cusp in more massive systems \citep{Martizzi13}. This paper is an early step toward quantifying these trends observationally.

Although it might be supposed that absence of shallow DM profiles in at least some halos is a blow to interpretations that invoke the particle microphysics, such as self-interactions \citep[e.g.,][]{Peter13,Rocha13}, this is not necessarily the case. Appealing to a velocity-dependent scattering cross-section $\sigma_{\rm DM}$ does not help if $\sigma_{\rm DM}$ and $v$ vary inversely, as is usually presumed: we find stronger evidence for cores in \emph{higher} mass halos, which would have lower $\sigma_{\rm DM}$ and smaller cores. Instead, the main uncertainty is the behavior of self-interacting dark matter in a realistic setting with baryons. Further simulations are needed to test this.

\citet{Schaller15b} recently compared the N13 cluster observations with their EAGLE simulations. Unlike the simulations by \citet{Martizzi13} and \citet{Laporte15}, they found no evidence for shallow DM slopes and suggested instead that neglecting radial trends in velocity anisotropy could bias the observations. The volume of their cosmological simulation was too small to contain very massive systems analogous to the N13 clusters, so they selected 6 halos with masses intermediate between the N13 sample and the present group-scale lenses. Since we do find evidence for cores in $10^{14} \msol$ halos, we suggest that a mismatch of halo masses could be partially responsible. In Section~\ref{sec:sys} we considered a model with radially varying anisotropy and found the results to be very comparable to the range of constant anisotropy models that were considered by N13. This is reassuring, although a wider range for the anisotropy profile could be considered. Further progress can be made with future observations using the new integral field  spectrographs appearing on large telescopes (e.g., KCWI and MUSE), which will allow higher-order moments of the velocity distribution to be measured at large radii in these lenses.

\subsection{Future Improvements}

Larger samples of lenses with data of similar quality are needed to confirm the possible difference we find between the inner DM profiles of groups and clusters. Further progress will also come from improved understanding of the stellar populations. Some of our results for the group-scale lenses are sensitive to the radial gradient in $M_*/L$. This probably plays a lesser role in galaxy-scale lenses, which typically do not have resolved stellar kinematic data, and also in clusters, where we have verified that the BCGs in the N13 sample show weak color gradients (consistent with a mean of zero) over the relevant radial range. However, it has a surprisingly strong effect in the BGGs. Age and metallicity gradients could be better tested with near-infrared photometry, but it is also possible that the IMF itself varies with radius. Testing this requires high-quality spectroscopy at large radii, and currently we have little data to constrain the magnitude of such a gradient \citep[e.g.,][]{MartinNavarro15}. Fortunately, improved constraints are within reach of future observational programs.

\section{Summary}

\begin{enumerate}
\item We assembled a sample of 10 group-scale lenses ($\theta_{\rm Ein} = 2.5''-5.1''$) with high-quality strong lensing and kinematic data, including new Keck observations of 8 systems. This is the largest uniformly analyzed sample constructed with the aim of measuring the mass distribution over a wide range of scales, from within the central galaxy to the virial radius.

\item Measurements of the velocity dispersion of satellite galaxies indicate a mean halo mass of $\langle \log M_{200} \rangle = 14.0 \pm 0.1$, intermediate between earlier samples of galaxy- and cluster-scale lenses.

\item The mean concentration of the DM halos is $\langle c_{200} \rangle = 5.0 \pm 0.8$. Correcting for the expected bias arising from our selection in Einstein radius, relative to the underlying ETG population, yields $\langle c_{200} \rangle = 3.9 \pm 0.6$ (Section~\ref{sec:mcr}).

\item These concentrations are consistent with the current range of theoretical MCRs, suggesting that baryons have a minor effect on the structure of their halos at large radii near the scale radius. This is consistent with hydrodynamical simulations \citep[e.g.,][]{Duffy10,Schaller15a}; some earlier observational work, e.g., \citet{Oguri12}, was likely affected by incomplete treatment of how strong lenses were selected.

\item The average slope of the total density profile within $R_e$ is $\langle \gamma_{\rm tot} \rangle = 1.64 \pm 0.05~{\rm (stat.)} \pm 0.07~{\rm (sys.)}$ for our group-scale lenses (Section~\ref{sec:totslope}). By combining with earlier lens samples to examine trends over a factor of $\simeq 60$ in halo mass, we show that $\gamma_{\rm tot}$ depends on both the stellar surface density $\Sigma_*$, as shown in earlier lensing studies, but also on the halo mass. This reflects the strong non-homology of massive ETGs and shows that their internal density structure cannot be derived from the stellar light alone.

\item We built a set of simple halo occupation models based on the $M_*-M_{200}$ and $M_*-R_e$ relations observed in our lens compilation (Section~\ref{sec:cdmmodel}). These reproduce earlier findings that galaxy-scale lenses favor a Salpeter-type normalization of the IMF \citep[e.g.,][]{Auger10b,Treu10} and that the DM profile at the centers of massive clusters is shallower than the NFW profile \citep{Newman13a,Newman13b}.

\item Observations of the group lenses alone do not robustly resolve the covariance between $M_*/L$ (and therefore the IMF) and the inner DM profile. However, if the Salpeter-type IMF inferred for the galaxy- and cluster-scale lenses holds for the central galaxies of groups---as might be expected from the similarity of their central velocity dispersions---then the group lenses favor unmodified or weakly contracted halos. They disfavor an inner DM density profile as shallow as those found for clusters, although this difference is not yet definitive.

\item Combining our results with earlier studies, we suggest that the effect of baryons on the structure of their DM halos may vary with halo mass due to underlying trends with star formation efficiency and assembly history. Clusters show evidence for a reduction in the central DM density, which may arise from the dominant role of dissipationless merging in the assembly of BCGs. In group- and galaxy-scale halos, more of the gas has cooled and been transformed into stars, leading to a greater role for dissipation and adiabatic contraction. Larger lens samples with high-quality data are required to test this tentative variation in the small-scale distribution of DM.

\end{enumerate}

\acknowledgments

The authors recognize and acknowledge the very significant cultural role and reverence that the summit of Mauna Kea has always had within the indigenous Hawaiian community. We are most fortunate to have the opportunity to conduct observations from this mountain. T.T.~acknowledges support by the Packard Foundation in the form of a Packard Fellowship. T.T.~thanks the Observatory of Monteporzio Catone and the American Academy in Rome for their gracious hospitality during the writing of this manuscript.

\appendix
\section{Notes on Individual Lens Models\label{sec:modelnotes}}

CSWA107 has a naked-cusp configuration. Not much structure is visible in the current ground-based imaging, so we do not include external shear in this model. As for the other three-image lenses, we fix the center of mass to the light but allow its ellipticity and position angle to vary independently.

CSWA141 is the most ambiguous system. The bright blue clump is clearly singly imaged, while its fainter, extended tail may be composed of merging multiple images of a fainter part of the source. Furthermore, a second galaxy contributes significantly to the lensing potential. Given the increased uncertainties in this system, we fix the center, ellipticity, and position angle of the main deflector to follow the BGG light and omit external shear. We also increase the uncertainty in $\bar{\kappa}$ to 0.1.

CSWA164 is a nearly complete Einstein ring. Our measurement of $\theta_{\rm Ein}=3\farcs68$ and the low external shear $\Gamma = 0.022$ agree closely with measurements by \citet{KZ14}.

CSWA165 has a quad configuration that constrains the lens model well. Our $\theta_{\rm Ein} = 4\farcs33$ is slightly larger than the $3\farcs77^{+0.11}_{-0.16}$ measured by \citet{KZ14}, but their analysis was based on SDSS images with poorer depth and resolution than our DEIMOS imaging.

CSWA6 (The Clone) was discovered by \citet{Lin09}, who modeled the deflector using a single SIE with $\theta_{\rm Ein} = 3\farcs82$, smaller than our $4\farcs36$. \citet{Jones10}, however, analyzed \emph{HST} imaging and followed a procedure more analogous to our own, separating the contribution of the satellite galaxies. They measure Einstein radii of $3\farcs0 \pm 0\farcs3$, $0\farcs5 \pm 0\farcs3$, and $0\farcs3 \pm 0\farcs2$ for their G1, G2, and G3 deflectors, where G1 represents the BGG and G2 and G3 are satellites. These agree well with our measurements of $3\farcs2$, $0\farcs4$, and $0\farcs5$, respectively. Satellites in our model contribute 21\% of the mass within $\theta_{\rm Ein}$, larger than for the other lenses. Owing to the increased uncertainty in decomposing the mass components, we increase the error in $\bar{\kappa}$ to 0.1.

CSWA7 forms multiple images of two background galaxies, each presenting 3 images visible in the \emph{HST} data. Given the presence of the multiple sources and the compactness of their images, we decided to use {\tt glafic} \citep{Oguri10} to model this lens using the positions of the six images as constraints. This accounts for the different appearance of CSWA7 in Figure~\ref{fig:images}. As in the other naked cusp lenses, we fix the center of mass to that of the BGG. We find a fairly high external shear of $\Gamma = 0.149$ consistent with the prominent background structure seen in our redshift survey. The axis ratios of the mass and light agree ($q_{\rm mass}=0.67$, $q_{\rm light}=0.65$), suggesting that the ellipticity and shear have been successfully separated. \citet{Kubo09} noted that the arc spectrum shows superposed spectra at $z=1.411$ and $z=1.38$, and we see the same in our DEIMOS spectrum. By fixing the redshift of the outer brighter system to 1.411 and allowing that of the fainter inner system to vary in our lens model, we find a redshift of 1.38. This confirms that the fainter multiply imaged system is the galaxy seen in absorption in the spectrum of the brighter background galaxy. Our $\theta_{\rm Ein} = 2\farcs73$, which refers to $z_S = 1.411$, is smaller than the $3\farcs7$ estimated in the discovery paper by Kubo et al., but this can likely be attributed to their neglect of the (high) ellipticity.

EOCL (Eight O'Clock Arc) was discovered by \citet{Allam07}, who reported a preliminary SIE-based $\theta_{\rm Ein} = 2\farcs91$. Updated analyses with \emph{HST} observations revised this to $3\farcs32$ (\citealt{Dessauges11}, see also \citealt{Shirazi14}), very close to our $3\farcs29$. 

J09413--1100 has the most regular light distribution of the 13 group-scale lenses in the SL2S sample presented by \citet{Limousin09}, suggestive of a relaxed system. We find that external shear is essential to accurately fit the features seen in the \emph{HST} imaging. Reassuringly, when external shear is incorporated into the model, the axis ratio of the mass distribution ($q_{\rm mass} = 0.55$) closely matches that of the BGG light at the arc radius. (The overall axis ratio of the BGG in Table~\ref{tab:lensing} is higher due to the presence of an ellipticity gradient.) This arc is the only one in our sample that lacks a spectroscopic redshift. We detect its continuum throughout the DEIMOS wavelength range, but despite its blue color see no emission lines. Based on the non-detection of [\ion{O}{2}], we place a lower limit $z > 1.6$. Using the $uriz$ colors measured in public CFHT imaging, we use {\tt EAZY} \citep{Brammer08} to infer a 99\% upper limit on the photometric redshift of $z < 2.45$. This reflects the lack of a break in the $u$ photometry that would occur at higher redshifts due to absorption by the intergalactic medium. We use a fiducial $z_S = 2.0$ in our analysis and note that the lensing distance ratio $D_{\rm ds}/D_{\rm s}$ changes by only $\sim \pm 5\%$ over the allowed range $z_S = 1.6 - 2.45$. We consider this as an additional random error in $\bar{\kappa}$ that is added in quadrature.

CSWA163 is constrained only by SDSS imaging, which has relatively poor quality. We therefore omit external shear from this model. Our $\theta_{\rm Ein}=3\farcs49$ agrees closely with \citet{Deason13}.

CSWA1 (The Cosmic Horseshoe) has the largest Einstein radius in our sample and was studied in detail by \citet{Dye08}. The $\theta_{\rm Ein} = 4\farcs97$ obtained for their SIE model is within 2\% of our $5\farcs08$. In agreement with Dye et al., we find a very low contribution from external shear ($\Gamma = 0.022$).

\section{Hierarchical Bayesian Inference Method}

Here we describe the mathematical framework that we used to analyze the MCR. In this approach, we assume that the group lenses are drawn from a parent sample of lenses which is characterized by independent Gaussian distributions of $\log M_{200}$ and $\log c_{200}' = \log c_{200} + 0.07(\log M_{200} - 14)$. Here we account for the slope of MCR using the theoretically expected slope \citep{Duffy08}, since the range of masses in our sample is much too narrow to constrain it. With the slope of the MCR removed, we may consider $\log M_{200}$ and $\log c_{200}'$ to be independently distributed.

Our aim is to constrain the mean and intrinsic dispersion of this parent distribution---$\langle \log M_{200} \rangle$, $\langle \log c_{200}' \rangle$, $\sigma_{\log M}$, and $\sigma_{\log c'}$, which are known as the hyperparameters---using the posterior distributions of each of the 8 lenses (those with $M_{200}$ measured from satellite dynamics) in the $(M_{200}, c_{200})$ plane, after marginalizing over the other parameters. For brevity, we denote the hyperparameters as $\langle M \rangle$, $\langle c \rangle$, $\sigma_M$, and $\sigma_c$, respectively, and the set of these as $\omega$. Their posterior distribution is then
\begin{align}
P(\omega | D) &\propto Pr(\omega) P(D | \omega) \nonumber\\
&= Pr(\omega) \prod_i \iint dM dc~P_i(D_i | M, c) N(M, c; \omega) \nonumber\\
&\propto Pr(\omega) \prod_i \iint dM dc \frac{P_i(M, c | D_i)}{Pr_i(M, c)} N(M, c; \omega). \label{eqn:post}
\end{align}
Here $D_i$ represents the data for lens $i$ and $D$ the combined data set. $N(\omega)$ is the trial parent distribution, i.e., a bivariate Gaussian with mean $(\langle M \rangle, \langle c \rangle)$ and dispersion $(\sigma_M, \sigma_c)$. The posterior distribution $P_i(M, c | D_i)$ for each lens is constructed from its MCMC chains. Since both the prior $Pr_i(M, c)$ used in the first-level inference for each lens and the prior $Pr(\omega)$ on the hyperparameters $\omega$ are broad and uniform, Equation~\ref{eqn:post} amounts to a product of $i$ integrals of the trial parent distribution multiplied by the posterior for lens $i$. We sample $P(\omega | D)$ using the MCMC code {\tt emcee} \citep{emcee}.

Similar techniques have been used by, e.g., \citet{Bolton12}, N13a, and \citet{Auger13}. In Section~\ref{sec:dmimf} we apply the same framework to study the distributions of $\gamma_{\rm DM}$ and $\alpha_{\rm SPS}$.

\bibliographystyle{apj}
\bibliography{grouplens}

\begin{thebibliography}{165}
\expandafter\ifx\csname natexlab\endcsname\relax\def\natexlab#1{#1}\fi

\bibitem[{{Ahn} {et~al.}(2014){Ahn}, {Alexandroff}, {Allende Prieto}, {Anders},
  {Anderson}, {Anderton}, {Andrews}, {Aubourg}, {Bailey}, {Bastien}, \&
  et~al.}]{Ahn14}
{Ahn}, C.~P., {Alexandroff}, R., {Allende Prieto}, C., {et~al.} 2014, \apjs,
  211, 17

\bibitem[{{Allam} {et~al.}(2007){Allam}, {Tucker}, {Lin}, {Diehl}, {Annis},
  {Buckley-Geer}, \& {Frieman}}]{Allam07}
{Allam}, S.~S., {Tucker}, D.~L., {Lin}, H., {et~al.} 2007, \apjl, 662, L51

\bibitem[{{Angulo} {et~al.}(2012){Angulo}, {Springel}, {White}, {Jenkins},
  {Baugh}, \& {Frenk}}]{Angulo12}
{Angulo}, R.~E., {Springel}, V., {White}, S.~D.~M., {et~al.} 2012, \mnras, 426,
  2046

\bibitem[{{Auger} {et~al.}(2013){Auger}, {Budzynski}, {Belokurov}, {Koposov},
  \& {McCarthy}}]{Auger13}
{Auger}, M.~W., {Budzynski}, J.~M., {Belokurov}, V., {Koposov}, S.~E., \&
  {McCarthy}, I.~G. 2013, \mnras, 436, 503

\bibitem[{{Auger} {et~al.}(2007){Auger}, {Fassnacht}, {Abrahamse}, {Lubin}, \&
  {Squires}}]{Auger07}
{Auger}, M.~W., {Fassnacht}, C.~D., {Abrahamse}, A.~L., {Lubin}, L.~M., \&
  {Squires}, G.~K. 2007, \aj, 134, 668

\bibitem[{{Auger} {et~al.}(2009){Auger}, {Treu}, {Bolton}, {Gavazzi},
  {Koopmans}, {Marshall}, {Bundy}, \& {Moustakas}}]{Auger09}
{Auger}, M.~W., {Treu}, T., {Bolton}, A.~S., {et~al.} 2009, \apj, 705, 1099

\bibitem[{{Auger} {et~al.}(2010{\natexlab{a}}){Auger}, {Treu}, {Bolton},
  {Gavazzi}, {Koopmans}, {Marshall}, {Moustakas}, \& {Burles}}]{Auger10a}
---. 2010{\natexlab{a}}, \apj, 724, 511

\bibitem[{{Auger} {et~al.}(2010{\natexlab{b}}){Auger}, {Treu}, {Gavazzi},
  {Bolton}, {Koopmans}, \& {Marshall}}]{Auger10b}
{Auger}, M.~W., {Treu}, T., {Gavazzi}, R., {et~al.} 2010{\natexlab{b}}, \apjl,
  721, L163

\bibitem[{{Barnab{\`e}} {et~al.}(2011){Barnab{\`e}}, {Czoske}, {Koopmans},
  {Treu}, \& {Bolton}}]{Barnabe11}
{Barnab{\`e}}, M., {Czoske}, O., {Koopmans}, L.~V.~E., {Treu}, T., \& {Bolton},
  A.~S. 2011, \mnras, 415, 2215

\bibitem[{{Barnab{\`e}} {et~al.}(2013){Barnab{\`e}}, {Spiniello}, {Koopmans},
  {Trager}, {Czoske}, \& {Treu}}]{Barnabe13}
{Barnab{\`e}}, M., {Spiniello}, C., {Koopmans}, L.~V.~E., {et~al.} 2013,
  \mnras, 436, 253

\bibitem[{{Behroozi} {et~al.}(2013){Behroozi}, {Wechsler}, \&
  {Conroy}}]{Behroozi13}
{Behroozi}, P.~S., {Wechsler}, R.~H., \& {Conroy}, C. 2013, \apj, 770, 57

\bibitem[{{Bell} {et~al.}(2003){Bell}, {McIntosh}, {Katz}, \&
  {Weinberg}}]{Bell03}
{Bell}, E.~F., {McIntosh}, D.~H., {Katz}, N., \& {Weinberg}, M.~D. 2003, \apjs,
  149, 289

\bibitem[{{Belokurov} {et~al.}(2009){Belokurov}, {Evans}, {Hewett}, {Moiseev},
  {McMahon}, {Sanchez}, \& {King}}]{Belokurov09}
{Belokurov}, V., {Evans}, N.~W., {Hewett}, P.~C., {et~al.} 2009, \mnras, 392,
  104

\bibitem[{{Bernardi} {et~al.}(2013){Bernardi}, {Meert}, {Sheth}, {Vikram},
  {Huertas-Company}, {Mei}, \& {Shankar}}]{Bernardi13}
{Bernardi}, M., {Meert}, A., {Sheth}, R.~K., {et~al.} 2013, \mnras, 436, 697

\bibitem[{{Bernardi} {et~al.}(2003){Bernardi}, {Sheth}, {Annis}, {Burles},
  {Eisenstein}, {Finkbeiner}, {Hogg}, {Lupton}, {Schlegel}, {SubbaRao},
  {Bahcall}, {Blakeslee}, {Brinkmann}, {Castander}, {Connolly}, {Csabai},
  {Doi}, {Fukugita}, {Frieman}, {Heckman}, {Hennessy}, {Ivezi{\'c}}, {Knapp},
  {Lamb}, {McKay}, {Munn}, {Nichol}, {Okamura}, {Schneider}, {Thakar}, \&
  {York}}]{Bernardi03}
{Bernardi}, M., {Sheth}, R.~K., {Annis}, J., {et~al.} 2003, \aj, 125, 1849

\bibitem[{{Biviano} {et~al.}(2006){Biviano}, {Murante}, {Borgani}, {Diaferio},
  {Dolag}, \& {Girardi}}]{Biviano06}
{Biviano}, A., {Murante}, G., {Borgani}, S., {et~al.} 2006, \aap, 456, 23

\bibitem[{{Blanton} \& {Roweis}(2007)}]{Blanton07}
{Blanton}, M.~R., \& {Roweis}, S. 2007, \aj, 133, 734

\bibitem[{{Blumenthal} {et~al.}(1986){Blumenthal}, {Faber}, {Flores}, \&
  {Primack}}]{Blumenthal86}
{Blumenthal}, G.~R., {Faber}, S.~M., {Flores}, R., \& {Primack}, J.~R. 1986,
  \apj, 301, 27

\bibitem[{{Bolton} {et~al.}(2008){Bolton}, {Burles}, {Koopmans}, {Treu},
  {Gavazzi}, {Moustakas}, {Wayth}, \& {Schlegel}}]{Bolton08}
{Bolton}, A.~S., {Burles}, S., {Koopmans}, L.~V.~E., {et~al.} 2008, \apj, 682,
  964

\bibitem[{{Bolton} {et~al.}(2006){Bolton}, {Burles}, {Koopmans}, {Treu}, \&
  {Moustakas}}]{Bolton06}
{Bolton}, A.~S., {Burles}, S., {Koopmans}, L.~V.~E., {Treu}, T., \&
  {Moustakas}, L.~A. 2006, \apj, 638, 703

\bibitem[{{Bolton} {et~al.}(2012){Bolton}, {Brownstein}, {Kochanek}, {Shu},
  {Schlegel}, {Eisenstein}, {Wake}, {Connolly}, {Maraston}, {Arneson}, \&
  {Weaver}}]{Bolton12}
{Bolton}, A.~S., {Brownstein}, J.~R., {Kochanek}, C.~S., {et~al.} 2012, \apj,
  757, 82

\bibitem[{{Brammer} {et~al.}(2008){Brammer}, {van Dokkum}, \&
  {Coppi}}]{Brammer08}
{Brammer}, G.~B., {van Dokkum}, P.~G., \& {Coppi}, P. 2008, \apj, 686, 1503

\bibitem[{{Bruzual} \& {Charlot}(2003)}]{BC03}
{Bruzual}, G., \& {Charlot}, S. 2003, \mnras, 344, 1000

\bibitem[{{Buote} {et~al.}(2007){Buote}, {Gastaldello}, {Humphrey},
  {Zappacosta}, {Bullock}, {Brighenti}, \& {Mathews}}]{Buote07}
{Buote}, D.~A., {Gastaldello}, F., {Humphrey}, P.~J., {et~al.} 2007, \apj, 664,
  123

\bibitem[{{Cappellari}(2008)}]{Cappellari08}
{Cappellari}, M. 2008, \mnras, 390, 71

\bibitem[{{Cappellari} \& {Emsellem}(2004)}]{Cappellari04}
{Cappellari}, M., \& {Emsellem}, E. 2004, \pasp, 116, 138

\bibitem[{{Cappellari} {et~al.}(2007){Cappellari}, {Emsellem}, {Bacon},
  {Bureau}, {Davies}, {de Zeeuw}, {Falc{\'o}n-Barroso}, {Krajnovi{\'c}},
  {Kuntschner}, {McDermid}, {Peletier}, {Sarzi}, {van den Bosch}, \& {van de
  Ven}}]{Cappellari07}
{Cappellari}, M., {Emsellem}, E., {Bacon}, R., {et~al.} 2007, \mnras, 379, 418

\bibitem[{{Cappellari} {et~al.}(2013){Cappellari}, {McDermid}, {Alatalo},
  {Blitz}, {Bois}, {Bournaud}, {Bureau}, {Crocker}, {Davies}, {Davis}, {de
  Zeeuw}, {Duc}, {Emsellem}, {Khochfar}, {Krajnovi{\'c}}, {Kuntschner},
  {Morganti}, {Naab}, {Oosterloo}, {Sarzi}, {Scott}, {Serra}, {Weijmans}, \&
  {Young}}]{Cappellari13}
{Cappellari}, M., {McDermid}, R.~M., {Alatalo}, K., {et~al.} 2013, \mnras, 432,
  1862

\bibitem[{{Chabrier}(2003)}]{Chabrier03}
{Chabrier}, G. 2003, \pasp, 115, 763

\bibitem[{{Conroy} \& {van Dokkum}(2012)}]{Conroy12}
{Conroy}, C., \& {van Dokkum}, P.~G. 2012, \apj, 760, 71

\bibitem[{{Cooper} {et~al.}(2012){Cooper}, {Newman}, {Davis}, {Finkbeiner}, \&
  {Gerke}}]{Cooper12}
{Cooper}, M.~C., {Newman}, J.~A., {Davis}, M., {Finkbeiner}, D.~P., \& {Gerke},
  B.~F. 2012, {spec2d: DEEP2 DEIMOS Spectral Pipeline}, astrophysics Source
  Code Library

\bibitem[{{Deason} {et~al.}(2013){Deason}, {Auger}, {Belokurov}, \&
  {Evans}}]{Deason13}
{Deason}, A.~J., {Auger}, M.~W., {Belokurov}, V., \& {Evans}, N.~W. 2013, \apj,
  773, 7

\bibitem[{{Del Popolo}(2012)}]{DelPopolo12}
{Del Popolo}, A. 2012, \mnras, 424, 38

\bibitem[{{Dessauges-Zavadsky} {et~al.}(2011){Dessauges-Zavadsky},
  {Christensen}, {D'Odorico}, {Schaerer}, \& {Richard}}]{Dessauges11}
{Dessauges-Zavadsky}, M., {Christensen}, L., {D'Odorico}, S., {Schaerer}, D.,
  \& {Richard}, J. 2011, \aap, 533, A15

\bibitem[{{Diehl} {et~al.}(2009){Diehl}, {Allam}, {Annis}, {Buckley-Geer},
  {Frieman}, {Kubik}, {Kubo}, {Lin}, {Tucker}, \& {West}}]{Diehl09}
{Diehl}, H.~T., {Allam}, S.~S., {Annis}, J., {et~al.} 2009, \apj, 707, 686

\bibitem[{{Diemand} {et~al.}(2005){Diemand}, {Zemp}, {Moore}, {Stadel}, \&
  {Carollo}}]{Diemand05}
{Diemand}, J., {Zemp}, M., {Moore}, B., {Stadel}, J., \& {Carollo}, C.~M. 2005,
  \mnras, 364, 665

\bibitem[{{Dubois} {et~al.}(2013){Dubois}, {Gavazzi}, {Peirani}, \&
  {Silk}}]{Dubois13}
{Dubois}, Y., {Gavazzi}, R., {Peirani}, S., \& {Silk}, J. 2013, \mnras, 433,
  3297

\bibitem[{{Duffy} {et~al.}(2008){Duffy}, {Schaye}, {Kay}, \& {Dalla
  Vecchia}}]{Duffy08}
{Duffy}, A.~R., {Schaye}, J., {Kay}, S.~T., \& {Dalla Vecchia}, C. 2008,
  \mnras, 390, L64

\bibitem[{{Duffy} {et~al.}(2010){Duffy}, {Schaye}, {Kay}, {Dalla Vecchia},
  {Battye}, \& {Booth}}]{Duffy10}
{Duffy}, A.~R., {Schaye}, J., {Kay}, S.~T., {et~al.} 2010, \mnras, 405, 2161

\bibitem[{{Dutton} \& {Macci{\`o}}(2014)}]{Dutton14}
{Dutton}, A.~A., \& {Macci{\`o}}, A.~V. 2014, \mnras, 441, 3359

\bibitem[{{Dutton} \& {Treu}(2014)}]{DuttonTreu14}
{Dutton}, A.~A., \& {Treu}, T. 2014, \mnras, 438, 3594

\bibitem[{{Dye} {et~al.}(2008){Dye}, {Evans}, {Belokurov}, {Warren}, \&
  {Hewett}}]{Dye08}
{Dye}, S., {Evans}, N.~W., {Belokurov}, V., {Warren}, S.~J., \& {Hewett}, P.
  2008, \mnras, 388, 384

\bibitem[{{El-Zant} {et~al.}(2001){El-Zant}, {Shlosman}, \&
  {Hoffman}}]{ElZant01}
{El-Zant}, A., {Shlosman}, I., \& {Hoffman}, Y. 2001, \apj, 560, 636

\bibitem[{{El-Zant} {et~al.}(2004){El-Zant}, {Hoffman}, {Primack}, {Combes}, \&
  {Shlosman}}]{ElZant04}
{El-Zant}, A.~A., {Hoffman}, Y., {Primack}, J., {Combes}, F., \& {Shlosman}, I.
  2004, \apjl, 607, L75

\bibitem[{{Evans} {et~al.}(2003){Evans}, {Wilkinson}, {Perrett}, \&
  {Bridges}}]{Evans03}
{Evans}, N.~W., {Wilkinson}, M.~I., {Perrett}, K.~M., \& {Bridges}, T.~J. 2003,
  \apj, 583, 752

\bibitem[{{Faber} \& {Jackson}(1976)}]{Faber76}
{Faber}, S.~M., \& {Jackson}, R.~E. 1976, \apj, 204, 668

\bibitem[{{Faber} {et~al.}(2003){Faber}, {Phillips}, {Kibrick}, {Alcott},
  {Allen}, {Burrous}, {Cantrall}, {Clarke}, {Coil}, {Cowley}, {Davis}, {Deich},
  {Dietsch}, {Gilmore}, {Harper}, {Hilyard}, {Lewis}, {McVeigh}, {Newman},
  {Osborne}, {Schiavon}, {Stover}, {Tucker}, {Wallace}, {Wei}, {Wirth}, \&
  {Wright}}]{Faber03}
{Faber}, S.~M., {Phillips}, A.~C., {Kibrick}, R.~I., {et~al.} 2003, in Society
  of Photo-Optical Instrumentation Engineers (SPIE) Conference Series, Vol.
  4841, Instrument Design and Performance for Optical/Infrared Ground-based
  Telescopes, ed. M.~{Iye} \& A.~F.~M. {Moorwood}, 1657--1669

\bibitem[{{Fassnacht} {et~al.}(2008){Fassnacht}, {Kocevski}, {Auger}, {Lubin},
  {Neureuther}, {Jeltema}, {Mulchaey}, \& {McKean}}]{Fassnacht08}
{Fassnacht}, C.~D., {Kocevski}, D.~D., {Auger}, M.~W., {et~al.} 2008, \apj,
  681, 1017

\bibitem[{{Feroz} {et~al.}(2009){Feroz}, {Hobson}, \& {Bridges}}]{Feroz09}
{Feroz}, F., {Hobson}, M.~P., \& {Bridges}, M. 2009, \mnras, 398, 1601

\bibitem[{{Ferreras} {et~al.}(2013){Ferreras}, {La Barbera}, {de la Rosa},
  {Vazdekis}, {de Carvalho}, {Falc{\'o}n-Barroso}, \&
  {Ricciardelli}}]{Ferreras13}
{Ferreras}, I., {La Barbera}, F., {de la Rosa}, I.~G., {et~al.} 2013, \mnras,
  429, L15

\bibitem[{{Fo{\"e}x} {et~al.}(2014){Fo{\"e}x}, {Motta}, {Jullo}, {Limousin}, \&
  {Verdugo}}]{Foex14}
{Fo{\"e}x}, G., {Motta}, V., {Jullo}, E., {Limousin}, M., \& {Verdugo}, T.
  2014, \aap, 572, A19

\bibitem[{{Fo{\"e}x} {et~al.}(2013){Fo{\"e}x}, {Motta}, {Limousin}, {Verdugo},
  {More}, {Cabanac}, {Gavazzi}, \& {Mu{\~n}oz}}]{Foex13}
{Fo{\"e}x}, G., {Motta}, V., {Limousin}, M., {et~al.} 2013, \aap, 559, A105

\bibitem[{{Foreman-Mackey} {et~al.}(2013){Foreman-Mackey}, {Hogg}, {Lang}, \&
  {Goodman}}]{emcee}
{Foreman-Mackey}, D., {Hogg}, D.~W., {Lang}, D., \& {Goodman}, J. 2013, \pasp,
  125, 306

\bibitem[{{Gao} {et~al.}(2012){Gao}, {Navarro}, {Frenk}, {Jenkins}, {Springel},
  \& {White}}]{Gao12}
{Gao}, L., {Navarro}, J.~F., {Frenk}, C.~S., {et~al.} 2012, \mnras, 425, 2169

\bibitem[{{Gavazzi} {et~al.}(2014){Gavazzi}, {Marshall}, {Treu}, \&
  {Sonnenfeld}}]{Gavazzi14}
{Gavazzi}, R., {Marshall}, P.~J., {Treu}, T., \& {Sonnenfeld}, A. 2014, \apj,
  785, 144

\bibitem[{{Gavazzi} {et~al.}(2007){Gavazzi}, {Treu}, {Rhodes}, {Koopmans},
  {Bolton}, {Burles}, {Massey}, \& {Moustakas}}]{Gavazzi07}
{Gavazzi}, R., {Treu}, T., {Rhodes}, J.~D., {et~al.} 2007, \apj, 667, 176

\bibitem[{{Gerhard} {et~al.}(2001){Gerhard}, {Kronawitter}, {Saglia}, \&
  {Bender}}]{Gerhard01}
{Gerhard}, O., {Kronawitter}, A., {Saglia}, R.~P., \& {Bender}, R. 2001, \aj,
  121, 1936

\bibitem[{{Gnedin} {et~al.}(2011){Gnedin}, {Ceverino}, {Gnedin}, {Klypin},
  {Kravtsov}, {Levine}, {Nagai}, \& {Yepes}}]{Gnedin11}
{Gnedin}, O.~Y., {Ceverino}, D., {Gnedin}, N.~Y., {et~al.} 2011,
  arXiv:1108.5736

\bibitem[{{Gnedin} {et~al.}(2004){Gnedin}, {Kravtsov}, {Klypin}, \&
  {Nagai}}]{Gnedin04}
{Gnedin}, O.~Y., {Kravtsov}, A.~V., {Klypin}, A.~A., \& {Nagai}, D. 2004, \apj,
  616, 16

\bibitem[{{Greene} {et~al.}(2013){Greene}, {Murphy}, {Graves}, {Gunn},
  {Raskutti}, {Comerford}, \& {Gebhardt}}]{Greene13}
{Greene}, J.~E., {Murphy}, J.~D., {Graves}, G.~J., {et~al.} 2013, \apj, 776, 64

\bibitem[{{Grillo}(2012)}]{Grillo12}
{Grillo}, C. 2012, \apjl, 747, L15

\bibitem[{{Grillo} {et~al.}(2013){Grillo}, {Christensen}, {Gallazzi}, \&
  {Rasmussen}}]{Grillo13}
{Grillo}, C., {Christensen}, L., {Gallazzi}, A., \& {Rasmussen}, J. 2013,
  \mnras, 433, 2604

\bibitem[{{Humphrey} \& {Buote}(2010)}]{Humphrey10}
{Humphrey}, P.~J., \& {Buote}, D.~A. 2010, \mnras, 403, 2143

\bibitem[{{Hyde} \& {Bernardi}(2009)}]{Hyde09}
{Hyde}, J.~B., \& {Bernardi}, M. 2009, \mnras, 394, 1978

\bibitem[{{Jardel} \& {Sellwood}(2009)}]{Jardell09}
{Jardel}, J.~R., \& {Sellwood}, J.~A. 2009, \apj, 691, 1300

\bibitem[{{Jiang} \& {Kochanek}(2007)}]{Jiang07}
{Jiang}, G., \& {Kochanek}, C.~S. 2007, \apj, 671, 1568

\bibitem[{{Johansson} {et~al.}(2009){Johansson}, {Naab}, \&
  {Ostriker}}]{Johansson09}
{Johansson}, P.~H., {Naab}, T., \& {Ostriker}, J.~P. 2009, \apjl, 697, L38

\bibitem[{{Johansson} {et~al.}(2012){Johansson}, {Naab}, \&
  {Ostriker}}]{Johansson12}
---. 2012, \apj, 754, 115

\bibitem[{{Jones} {et~al.}(2010){Jones}, {Ellis}, {Jullo}, \&
  {Richard}}]{Jones10}
{Jones}, T., {Ellis}, R., {Jullo}, E., \& {Richard}, J. 2010, \apjl, 725, L176

\bibitem[{{Katgert} {et~al.}(2004){Katgert}, {Biviano}, \&
  {Mazure}}]{Katgert04}
{Katgert}, P., {Biviano}, A., \& {Mazure}, A. 2004, \apj, 600, 657

\bibitem[{{Keeton}(2001)}]{Keeton01}
{Keeton}, C.~R. 2001, astro-ph/0102341

\bibitem[{{Keeton} {et~al.}(2000){Keeton}, {Christlein}, \&
  {Zabludoff}}]{Keeton00}
{Keeton}, C.~R., {Christlein}, D., \& {Zabludoff}, A.~I. 2000, \apj, 545, 129

\bibitem[{{Koopmans} {et~al.}(2006){Koopmans}, {Treu}, {Bolton}, {Burles}, \&
  {Moustakas}}]{Koopmans06}
{Koopmans}, L.~V.~E., {Treu}, T., {Bolton}, A.~S., {Burles}, S., \&
  {Moustakas}, L.~A. 2006, \apj, 649, 599

\bibitem[{{Koopmans} {et~al.}(2009){Koopmans}, {Bolton}, {Treu}, {Czoske},
  {Auger}, {Barnab{\`e}}, {Vegetti}, {Gavazzi}, {Moustakas}, \&
  {Burles}}]{Koopmans09}
{Koopmans}, L.~V.~E., {Bolton}, A., {Treu}, T., {et~al.} 2009, \apjl, 703, L51

\bibitem[{{Kormann} {et~al.}(1994){Kormann}, {Schneider}, \&
  {Bartelmann}}]{Kormann94}
{Kormann}, R., {Schneider}, P., \& {Bartelmann}, M. 1994, \aap, 284, 285

\bibitem[{{Kostrzewa-Rutkowska} {et~al.}(2014){Kostrzewa-Rutkowska},
  {Wyrzykowski}, {Auger}, {Collett}, \& {Belokurov}}]{KZ14}
{Kostrzewa-Rutkowska}, Z., {Wyrzykowski}, {\L}., {Auger}, M.~W., {Collett},
  T.~E., \& {Belokurov}, V. 2014, \mnras, 441, 3238

\bibitem[{{Kravtsov} {et~al.}(2014){Kravtsov}, {Vikhlinin}, \&
  {Meshscheryakov}}]{Kravtsov14}
{Kravtsov}, A., {Vikhlinin}, A., \& {Meshscheryakov}, A. 2014, arXiv:1401.7329

\bibitem[{{Kronawitter} {et~al.}(2000){Kronawitter}, {Saglia}, {Gerhard}, \&
  {Bender}}]{Kronawitter00}
{Kronawitter}, A., {Saglia}, R.~P., {Gerhard}, O., \& {Bender}, R. 2000, \aaps,
  144, 53

\bibitem[{{Kubo} {et~al.}(2009){Kubo}, {Allam}, {Annis}, {Buckley-Geer},
  {Diehl}, {Kubik}, {Lin}, \& {Tucker}}]{Kubo09}
{Kubo}, J.~M., {Allam}, S.~S., {Annis}, J., {et~al.} 2009, \apjl, 696, L61

\bibitem[{{Kubo} {et~al.}(2010){Kubo}, {Allam}, {Drabek}, {Lin}, {Tucker},
  {Buckley-Geer}, {Diehl}, {Soares-Santos}, {Hao}, {Wiesner}, {West}, {Kubik},
  {Annis}, \& {Frieman}}]{Kubo10}
{Kubo}, J.~M., {Allam}, S.~S., {Drabek}, E., {et~al.} 2010, \apjl, 724, L137

\bibitem[{{Kuntschner} {et~al.}(2010){Kuntschner}, {Emsellem}, {Bacon},
  {Cappellari}, {Davies}, {de Zeeuw}, {Falc{\'o}n-Barroso}, {Krajnovi{\'c}},
  {McDermid}, {Peletier}, {Sarzi}, {Shapiro}, {van den Bosch}, \& {van de
  Ven}}]{Kuntschner10}
{Kuntschner}, H., {Emsellem}, E., {Bacon}, R., {et~al.} 2010, \mnras, 408, 97

\bibitem[{{Kurtz} \& {Mink}(1998)}]{rvsao}
{Kurtz}, M.~J., \& {Mink}, D.~J. 1998, \pasp, 110, 934

\bibitem[{{La Barbera} {et~al.}(2013){La Barbera}, {Ferreras}, {Vazdekis}, {de
  la Rosa}, {de Carvalho}, {Trevisan}, {Falc{\'o}n-Barroso}, \&
  {Ricciardelli}}]{LaBarbera13}
{La Barbera}, F., {Ferreras}, I., {Vazdekis}, A., {et~al.} 2013, \mnras, 433,
  3017

\bibitem[{{Lackner} \& {Ostriker}(2010)}]{Lackner10}
{Lackner}, C.~N., \& {Ostriker}, J.~P. 2010, \apj, 712, 88

\bibitem[{{Laporte} \& {White}(2015)}]{Laporte15}
{Laporte}, C.~F.~P., \& {White}, S.~D.~M. 2015, \mnras, 451, 1177

\bibitem[{{Limousin} {et~al.}(2009){Limousin}, {Cabanac}, {Gavazzi}, {Kneib},
  {Motta}, {Richard}, {Thanjavur}, {Foex}, {Pello}, {Crampton}, {Faure},
  {Fort}, {Jullo}, {Marshall}, {Mellier}, {More}, {Soucail}, {Suyu},
  {Swinbank}, {Sygnet}, {Tu}, {Valls-Gabaud}, {Verdugo}, \&
  {Willis}}]{Limousin09}
{Limousin}, M., {Cabanac}, R., {Gavazzi}, R., {et~al.} 2009, \aap, 502, 445

\bibitem[{{Lin} {et~al.}(2009){Lin}, {Buckley-Geer}, {Allam}, {Tucker},
  {Diehl}, {Kubik}, {Kubo}, {Annis}, {Frieman}, {Oguri}, \& {Inada}}]{Lin09}
{Lin}, H., {Buckley-Geer}, E., {Allam}, S.~S., {et~al.} 2009, \apj, 699, 1242

\bibitem[{{Mamon} {et~al.}(2013){Mamon}, {Biviano}, \& {Bou{\'e}}}]{Mamon13}
{Mamon}, G.~A., {Biviano}, A., \& {Bou{\'e}}, G. 2013, \mnras, 429, 3079

\bibitem[{{Mamon} \& {{\L}okas}(2005)}]{Mamon05}
{Mamon}, G.~A., \& {{\L}okas}, E.~L. 2005, \mnras, 363, 705

\bibitem[{{Mart{\'{\i}}n-Navarro} {et~al.}(2015){Mart{\'{\i}}n-Navarro},
  {Barbera}, {Vazdekis}, {Falc{\'o}n-Barroso}, \& {Ferreras}}]{MartinNavarro15}
{Mart{\'{\i}}n-Navarro}, I., {Barbera}, F.~L., {Vazdekis}, A.,
  {Falc{\'o}n-Barroso}, J., \& {Ferreras}, I. 2015, \mnras, 447, 1033

\bibitem[{{Martizzi} {et~al.}(2013){Martizzi}, {Teyssier}, \&
  {Moore}}]{Martizzi13}
{Martizzi}, D., {Teyssier}, R., \& {Moore}, B. 2013, \mnras, 432, 1947

\bibitem[{{McKean} {et~al.}(2010){McKean}, {Auger}, {Koopmans}, {Vegetti},
  {Czoske}, {Fassnacht}, {Treu}, {More}, \& {Kocevski}}]{McKean10}
{McKean}, J.~P., {Auger}, M.~W., {Koopmans}, L.~V.~E., {et~al.} 2010, \mnras,
  404, 749

\bibitem[{{Mehlert} {et~al.}(2003){Mehlert}, {Thomas}, {Saglia}, {Bender}, \&
  {Wegner}}]{Mehlert03}
{Mehlert}, D., {Thomas}, D., {Saglia}, R.~P., {Bender}, R., \& {Wegner}, G.
  2003, \aap, 407, 423

\bibitem[{{Meneghetti} {et~al.}(2011){Meneghetti}, {Fedeli}, {Zitrin},
  {Bartelmann}, {Broadhurst}, {Gottl{\"o}ber}, {Moscardini}, \&
  {Yepes}}]{Meneghetti11}
{Meneghetti}, M., {Fedeli}, C., {Zitrin}, A., {et~al.} 2011, \aap, 530, A17

\bibitem[{{More} {et~al.}(2012){More}, {Cabanac}, {More}, {Alard}, {Limousin},
  {Kneib}, {Gavazzi}, \& {Motta}}]{More12}
{More}, A., {Cabanac}, R., {More}, S., {et~al.} 2012, \apj, 749, 38

\bibitem[{{Mu{\~n}oz} {et~al.}(2013){Mu{\~n}oz}, {Motta}, {Verdugo}, {Garrido},
  {Limousin}, {Padilla}, {Fo{\"e}x}, {Cabanac}, {Gavazzi}, {Barrientos}, \&
  {Richard}}]{Munoz13}
{Mu{\~n}oz}, R.~P., {Motta}, V., {Verdugo}, T., {et~al.} 2013, \aap, 552, A80

\bibitem[{{Munari} {et~al.}(2013){Munari}, {Biviano}, {Borgani}, {Murante}, \&
  {Fabjan}}]{Munari13}
{Munari}, E., {Biviano}, A., {Borgani}, S., {Murante}, G., \& {Fabjan}, D.
  2013, \mnras, 430, 2638

\bibitem[{{Navarro} {et~al.}(1996{\natexlab{a}}){Navarro}, {Eke}, \&
  {Frenk}}]{Navarro96}
{Navarro}, J.~F., {Eke}, V.~R., \& {Frenk}, C.~S. 1996{\natexlab{a}}, \mnras,
  283, L72

\bibitem[{{Navarro} {et~al.}(1996{\natexlab{b}}){Navarro}, {Frenk}, \&
  {White}}]{NFW96}
{Navarro}, J.~F., {Frenk}, C.~S., \& {White}, S.~D.~M. 1996{\natexlab{b}},
  \apj, 462, 563

\bibitem[{{Newman} {et~al.}(2014){Newman}, {Ellis}, {Andreon}, {Treu},
  {Raichoor}, \& {Trinchieri}}]{Newman14}
{Newman}, A.~B., {Ellis}, R.~S., {Andreon}, S., {et~al.} 2014, \apj, 788, 51

\bibitem[{{Newman} {et~al.}(2011){Newman}, {Treu}, {Ellis}, \&
  {Sand}}]{Newman11}
{Newman}, A.~B., {Treu}, T., {Ellis}, R.~S., \& {Sand}, D.~J. 2011, \apjl, 728,
  L39

\bibitem[{{Newman} {et~al.}(2013b){Newman}, {Treu}, {Ellis}, \&
  {Sand}}]{Newman13b}
---. 2013b, \apj, 765, 25

\bibitem[{{Newman} {et~al.}(2013a){Newman}, {Treu}, {Ellis}, {Sand}, {Nipoti},
  {Richard}, \& {Jullo}}]{Newman13a}
{Newman}, A.~B., {Treu}, T., {Ellis}, R.~S., {et~al.} 2013a, \apj, 765, 24

\bibitem[{{Newman} {et~al.}(2009){Newman}, {Treu}, {Ellis}, {Sand}, {Richard},
  {Marshall}, {Capak}, \& {Miyazaki}}]{Newman09}
---. 2009, \apj, 706, 1078

\bibitem[{{Newman} {et~al.}(2013){Newman}, {Cooper}, {Davis}, {Faber}, {Coil},
  {Guhathakurta}, {Koo}, {Phillips}, {Conroy}, {Dutton}, {Finkbeiner}, {Gerke},
  {Rosario}, {Weiner}, {Willmer}, {Yan}, {Harker}, {Kassin}, {Konidaris},
  {Lai}, {Madgwick}, {Noeske}, {Wirth}, {Connolly}, {Kaiser}, {Kirby},
  {Lemaux}, {Lin}, {Lotz}, {Luppino}, {Marinoni}, {Matthews}, {Metevier}, \&
  {Schiavon}}]{JNewman13}
{Newman}, J.~A., {Cooper}, M.~C., {Davis}, M., {et~al.} 2013, \apjs, 208, 5

\bibitem[{{Nipoti} {et~al.}(2004){Nipoti}, {Treu}, {Ciotti}, \&
  {Stiavelli}}]{Nipoti04}
{Nipoti}, C., {Treu}, T., {Ciotti}, L., \& {Stiavelli}, M. 2004, \mnras, 355,
  1119

\bibitem[{{Oguri}(2010)}]{Oguri10}
{Oguri}, M. 2010, \pasj, 62, 1017

\bibitem[{{Oguri} {et~al.}(2012){Oguri}, {Bayliss}, {Dahle}, {Sharon},
  {Gladders}, {Natarajan}, {Hennawi}, \& {Koester}}]{Oguri12}
{Oguri}, M., {Bayliss}, M.~B., {Dahle}, H., {et~al.} 2012, \mnras, 420, 3213

\bibitem[{{Oguri} {et~al.}(2005){Oguri}, {Keeton}, \& {Dalal}}]{Oguri05}
{Oguri}, M., {Keeton}, C.~R., \& {Dalal}, N. 2005, \mnras, 364, 1451

\bibitem[{{Oguri} {et~al.}(2014){Oguri}, {Rusu}, \& {Falco}}]{Oguri14}
{Oguri}, M., {Rusu}, C.~E., \& {Falco}, E.~E. 2014, \mnras, 439, 2494

\bibitem[{{Oser} {et~al.}(2010){Oser}, {Ostriker}, {Naab}, {Johansson}, \&
  {Burkert}}]{Oser10}
{Oser}, L., {Ostriker}, J.~P., {Naab}, T., {Johansson}, P.~H., \& {Burkert}, A.
  2010, \apj, 725, 2312

\bibitem[{{Peter} {et~al.}(2013){Peter}, {Rocha}, {Bullock}, \&
  {Kaplinghat}}]{Peter13}
{Peter}, A.~H.~G., {Rocha}, M., {Bullock}, J.~S., \& {Kaplinghat}, M. 2013,
  \mnras, 430, 105

\bibitem[{{Planck Collaboration} {et~al.}(2014){Planck Collaboration}, {Ade},
  {Aghanim}, {Armitage-Caplan}, {Arnaud}, {Ashdown}, {Atrio-Barandela},
  {Aumont}, {Baccigalupi}, {Banday}, \& et~al.}]{Planck14}
{Planck Collaboration}, {Ade}, P.~A.~R., {Aghanim}, N., {et~al.} 2014, \aap,
  571, A16

\bibitem[{{Pontzen} \& {Governato}(2012)}]{Pontzen12}
{Pontzen}, A., \& {Governato}, F. 2012, \mnras, 421, 3464

\bibitem[{{Posacki} {et~al.}(2015){Posacki}, {Cappellari}, {Treu},
  {Pellegrini}, \& {Ciotti}}]{Posacki15}
{Posacki}, S., {Cappellari}, M., {Treu}, T., {Pellegrini}, S., \& {Ciotti}, L.
  2015, \mnras, 446, 493

\bibitem[{{Rawle} {et~al.}(2010){Rawle}, {Smith}, \& {Lucey}}]{Rawle10}
{Rawle}, T.~D., {Smith}, R.~J., \& {Lucey}, J.~R. 2010, \mnras, 401, 852

\bibitem[{{Remus} {et~al.}(2013){Remus}, {Burkert}, {Dolag}, {Johansson},
  {Naab}, {Oser}, \& {Thomas}}]{Remus13}
{Remus}, R.-S., {Burkert}, A., {Dolag}, K., {et~al.} 2013, \apj, 766, 71

\bibitem[{{Rocha} {et~al.}(2013){Rocha}, {Peter}, {Bullock}, {Kaplinghat},
  {Garrison-Kimmel}, {O{\~n}orbe}, \& {Moustakas}}]{Rocha13}
{Rocha}, M., {Peter}, A.~H.~G., {Bullock}, J.~S., {et~al.} 2013, \mnras, 430,
  81

\bibitem[{{Romano-D{\'{\i}}az} {et~al.}(2008){Romano-D{\'{\i}}az}, {Shlosman},
  {Hoffman}, \& {Heller}}]{RomanoDiaz08}
{Romano-D{\'{\i}}az}, E., {Shlosman}, I., {Hoffman}, Y., \& {Heller}, C. 2008,
  \apjl, 685, L105

\bibitem[{{Ruff} {et~al.}(2011){Ruff}, {Gavazzi}, {Marshall}, {Treu}, {Auger},
  \& {Brault}}]{Ruff11}
{Ruff}, A.~J., {Gavazzi}, R., {Marshall}, P.~J., {et~al.} 2011, \apj, 727, 96

\bibitem[{{Rusin} {et~al.}(2003){Rusin}, {Kochanek}, \& {Keeton}}]{Rusin03}
{Rusin}, D., {Kochanek}, C.~S., \& {Keeton}, C.~R. 2003, \apj, 595, 29

\bibitem[{{Salpeter}(1955)}]{Salpeter55}
{Salpeter}, E.~E. 1955, \apj, 121, 161

\bibitem[{{S{\'a}nchez-Bl{\'a}zquez} {et~al.}(2006){S{\'a}nchez-Bl{\'a}zquez},
  {Peletier}, {Jim{\'e}nez-Vicente}, {Cardiel}, {Cenarro},
  {Falc{\'o}n-Barroso}, {Gorgas}, {Selam}, \& {Vazdekis}}]{MILES}
{S{\'a}nchez-Bl{\'a}zquez}, P., {Peletier}, R.~F., {Jim{\'e}nez-Vicente}, J.,
  {et~al.} 2006, \mnras, 371, 703

\bibitem[{{Sand} {et~al.}(2002){Sand}, {Treu}, \& {Ellis}}]{Sand02}
{Sand}, D.~J., {Treu}, T., \& {Ellis}, R.~S. 2002, \apjl, 574, L129

\bibitem[{{Sand} {et~al.}(2008){Sand}, {Treu}, {Ellis}, {Smith}, \&
  {Kneib}}]{Sand08}
{Sand}, D.~J., {Treu}, T., {Ellis}, R.~S., {Smith}, G.~P., \& {Kneib}, J.-P.
  2008, \apj, 674, 711

\bibitem[{{Sand} {et~al.}(2004){Sand}, {Treu}, {Smith}, \& {Ellis}}]{Sand04}
{Sand}, D.~J., {Treu}, T., {Smith}, G.~P., \& {Ellis}, R.~S. 2004, \apj, 604,
  88

\bibitem[{{Schaller} {et~al.}(2015{\natexlab{a}}){Schaller}, {Frenk}, {Bower},
  {Theuns}, {Jenkins}, {Schaye}, {Crain}, {Furlong}, {Dalla Vecchia}, \&
  {McCarthy}}]{Schaller15a}
{Schaller}, M., {Frenk}, C.~S., {Bower}, R.~G., {et~al.} 2015{\natexlab{a}},
  \mnras, 451, 1247

\bibitem[{{Schaller} {et~al.}(2015{\natexlab{b}}){Schaller}, {Frenk}, {Bower},
  {Theuns}, {Trayford}, {Crain}, {Furlong}, {Schaye}, {Dalla Vecchia}, \&
  {McCarthy}}]{Schaller15b}
---. 2015{\natexlab{b}}, \mnras, 452, 343

\bibitem[{{Schechter} {et~al.}(2014){Schechter}, {Pooley}, {Blackburne}, \&
  {Wambsganss}}]{Schechter14}
{Schechter}, P.~L., {Pooley}, D., {Blackburne}, J.~A., \& {Wambsganss}, J.
  2014, \apj, 793, 96

\bibitem[{{Schlafly} \& {Finkbeiner}(2011)}]{Schlafly11}
{Schlafly}, E.~F., \& {Finkbeiner}, D.~P. 2011, \apj, 737, 103

\bibitem[{{Schramm}(1990)}]{Schramm90}
{Schramm}, T. 1990, \aap, 231, 19

\bibitem[{{Shirazi} {et~al.}(2014){Shirazi}, {Vegetti}, {Nesvadba}, {Allam},
  {Brinchmann}, \& {Tucker}}]{Shirazi14}
{Shirazi}, M., {Vegetti}, S., {Nesvadba}, N., {et~al.} 2014, \mnras, 440, 2201

\bibitem[{{Shu} {et~al.}(2015){Shu}, {Bolton}, {Brownstein}, {Montero-Dorta},
  {Koopmans}, {Treu}, {Gavazzi}, {Auger}, {Czoske}, {Marshall}, \&
  {Moustakas}}]{Shu15}
{Shu}, Y., {Bolton}, A.~S., {Brownstein}, J.~R., {et~al.} 2015, \apj, 803, 71

\bibitem[{{Smith}(2014)}]{Smith14}
{Smith}, R.~J. 2014, \mnras, 443, L69

\bibitem[{{Sonnenfeld} {et~al.}(2013{\natexlab{a}}){Sonnenfeld}, {Gavazzi},
  {Suyu}, {Treu}, \& {Marshall}}]{Sonnenfeld13a}
{Sonnenfeld}, A., {Gavazzi}, R., {Suyu}, S.~H., {Treu}, T., \& {Marshall},
  P.~J. 2013{\natexlab{a}}, \apj, 777, 97

\bibitem[{{Sonnenfeld} {et~al.}(2014){Sonnenfeld}, {Nipoti}, \&
  {Treu}}]{Sonnenfeld14}
{Sonnenfeld}, A., {Nipoti}, C., \& {Treu}, T. 2014, \apj, 786, 89

\bibitem[{{Sonnenfeld} {et~al.}(2012){Sonnenfeld}, {Treu}, {Gavazzi},
  {Marshall}, {Auger}, {Suyu}, {Koopmans}, \& {Bolton}}]{Sonnenfeld12}
{Sonnenfeld}, A., {Treu}, T., {Gavazzi}, R., {et~al.} 2012, \apj, 752, 163

\bibitem[{{Sonnenfeld} {et~al.}(2013{\natexlab{b}}){Sonnenfeld}, {Treu},
  {Gavazzi}, {Suyu}, {Marshall}, {Auger}, \& {Nipoti}}]{Sonnenfeld13b}
---. 2013{\natexlab{b}}, \apj, 777, 98

\bibitem[{{Sonnenfeld} {et~al.}(2015){Sonnenfeld}, {Treu}, {Marshall}, {Suyu},
  {Gavazzi}, {Auger}, \& {Nipoti}}]{Sonnenfeld15}
{Sonnenfeld}, A., {Treu}, T., {Marshall}, P.~J., {et~al.} 2015, \apj, 800, 94

\bibitem[{{Spergel} \& {Steinhardt}(2000)}]{Spergel00}
{Spergel}, D.~N., \& {Steinhardt}, P.~J. 2000, Physical Review Letters, 84,
  3760

\bibitem[{{Spiniello} {et~al.}(2011){Spiniello}, {Koopmans}, {Trager},
  {Czoske}, \& {Treu}}]{Spiniello11}
{Spiniello}, C., {Koopmans}, L.~V.~E., {Trager}, S.~C., {Czoske}, O., \&
  {Treu}, T. 2011, \mnras, 417, 3000

\bibitem[{{Spiniello} {et~al.}(2014){Spiniello}, {Trager}, {Koopmans}, \&
  {Conroy}}]{Spiniello14}
{Spiniello}, C., {Trager}, S., {Koopmans}, L.~V.~E., \& {Conroy}, C. 2014,
  \mnras, 438, 1483

\bibitem[{{Stark} {et~al.}(2013){Stark}, {Auger}, {Belokurov}, {Jones},
  {Robertson}, {Ellis}, {Sand}, {Moiseev}, {Eagle}, \& {Myers}}]{Stark13}
{Stark}, D.~P., {Auger}, M., {Belokurov}, V., {et~al.} 2013, \mnras, 436, 1040

\bibitem[{{Suyu} {et~al.}(2010){Suyu}, {Marshall}, {Auger}, {Hilbert},
  {Blandford}, {Koopmans}, {Fassnacht}, \& {Treu}}]{Suyu10}
{Suyu}, S.~H., {Marshall}, P.~J., {Auger}, M.~W., {et~al.} 2010, \apj, 711, 201

\bibitem[{{Takahashi} {et~al.}(2011){Takahashi}, {Oguri}, {Sato}, \&
  {Hamana}}]{Takahashi11}
{Takahashi}, R., {Oguri}, M., {Sato}, M., \& {Hamana}, T. 2011, \apj, 742, 15

\bibitem[{{Tamura} {et~al.}(2000){Tamura}, {Kobayashi}, {Arimoto}, {Kodama}, \&
  {Ohta}}]{Tamura00}
{Tamura}, N., {Kobayashi}, C., {Arimoto}, N., {Kodama}, T., \& {Ohta}, K. 2000,
  \aj, 119, 2134

\bibitem[{{Thanjavur} {et~al.}(2010){Thanjavur}, {Crampton}, \&
  {Willis}}]{Thanjavur10}
{Thanjavur}, K., {Crampton}, D., \& {Willis}, J. 2010, \apj, 714, 1355

\bibitem[{{Tonini} {et~al.}(2006){Tonini}, {Lapi}, \& {Salucci}}]{Tonini06}
{Tonini}, C., {Lapi}, A., \& {Salucci}, P. 2006, \apj, 649, 591

\bibitem[{{Tortora} {et~al.}(2014){Tortora}, {La Barbera}, {Napolitano},
  {Romanowsky}, {Ferreras}, \& {de Carvalho}}]{Tortora14}
{Tortora}, C., {La Barbera}, F., {Napolitano}, N.~R., {et~al.} 2014, \mnras,
  445, 115

\bibitem[{{Tortora} {et~al.}(2011){Tortora}, {Napolitano}, {Romanowsky},
  {Jetzer}, {Cardone}, \& {Capaccioli}}]{Tortora11}
{Tortora}, C., {Napolitano}, N.~R., {Romanowsky}, A.~J., {et~al.} 2011, \mnras,
  418, 1557

\bibitem[{{Treu}(2010)}]{Treu10ARAA}
{Treu}, T. 2010, \araa, 48, 87

\bibitem[{{Treu} {et~al.}(2010){Treu}, {Auger}, {Koopmans}, {Gavazzi},
  {Marshall}, \& {Bolton}}]{Treu10}
{Treu}, T., {Auger}, M.~W., {Koopmans}, L.~V.~E., {et~al.} 2010, \apj, 709,
  1195

\bibitem[{{Treu} \& {Ellis}(2014)}]{Treu14}
{Treu}, T., \& {Ellis}, R.~S. 2014, arXiv:1412.6916

\bibitem[{{Treu} {et~al.}(2009){Treu}, {Gavazzi}, {Gorecki}, {Marshall},
  {Koopmans}, {Bolton}, {Moustakas}, \& {Burles}}]{Treu09}
{Treu}, T., {Gavazzi}, R., {Gorecki}, A., {et~al.} 2009, \apj, 690, 670

\bibitem[{{Treu} {et~al.}(2006){Treu}, {Koopmans}, {Bolton}, {Burles}, \&
  {Moustakas}}]{Treu06}
{Treu}, T., {Koopmans}, L.~V., {Bolton}, A.~S., {Burles}, S., \& {Moustakas},
  L.~A. 2006, \apj, 640, 662

\bibitem[{{Treu} \& {Koopmans}(2002)}]{Treu02}
{Treu}, T., \& {Koopmans}, L.~V.~E. 2002, \apj, 575, 87

\bibitem[{{Treu} \& {Koopmans}(2004)}]{Treu04}
---. 2004, \apj, 611, 739

\bibitem[{{van de Ven} {et~al.}(2010){van de Ven}, {Falc{\'o}n-Barroso},
  {McDermid}, {Cappellari}, {Miller}, \& {de Zeeuw}}]{vandeVen10}
{van de Ven}, G., {Falc{\'o}n-Barroso}, J., {McDermid}, R.~M., {et~al.} 2010,
  \apj, 719, 1481

\bibitem[{{Verdugo} {et~al.}(2011){Verdugo}, {Motta}, {Mu{\~n}oz}, {Limousin},
  {Cabanac}, \& {Richard}}]{Verdugo11}
{Verdugo}, T., {Motta}, V., {Mu{\~n}oz}, R.~P., {et~al.} 2011, \aap, 527, A124

\bibitem[{{Verdugo} {et~al.}(2014){Verdugo}, {Motta}, {Fo{\"e}x},
  {Forero-Romero}, {Mu{\~n}oz}, {Pello}, {Limousin}, {More}, {Cabanac},
  {Soucail}, {Blakeslee}, {Mej{\'{\i}}a-Narv{\'a}ez}, {Magris}, \&
  {Fern{\'a}ndez-Trincado}}]{Verdugo14}
{Verdugo}, T., {Motta}, V., {Fo{\"e}x}, G., {et~al.} 2014, \aap, 571, A65

\bibitem[{{Weijmans} {et~al.}(2014){Weijmans}, {de Zeeuw}, {Emsellem},
  {Krajnovi{\'c}}, {Lablanche}, {Alatalo}, {Blitz}, {Bois}, {Bournaud},
  {Bureau}, {Cappellari}, {Crocker}, {Davies}, {Davis}, {Duc}, {Khochfar},
  {Kuntschner}, {McDermid}, {Morganti}, {Naab}, {Oosterloo}, {Sarzi}, {Scott},
  {Serra}, {Verdoes Kleijn}, \& {Young}}]{Weijmans14}
{Weijmans}, A.-M., {de Zeeuw}, P.~T., {Emsellem}, E., {et~al.} 2014, \mnras,
  444, 3340

\bibitem[{{Wiesner} {et~al.}(2012){Wiesner}, {Lin}, {Allam}, {Annis},
  {Buckley-Geer}, {Diehl}, {Kubik}, {Kubo}, \& {Tucker}}]{Wiesner12}
{Wiesner}, M.~P., {Lin}, H., {Allam}, S.~S., {et~al.} 2012, \apj, 761, 1

\bibitem[{{Williams} {et~al.}(2006){Williams}, {Momcheva}, {Keeton},
  {Zabludoff}, \& {Leh{\'a}r}}]{Williams06}
{Williams}, K.~A., {Momcheva}, I., {Keeton}, C.~R., {Zabludoff}, A.~I., \&
  {Leh{\'a}r}, J. 2006, \apj, 646, 85

\bibitem[{{Wu} {et~al.}(2005){Wu}, {Shao}, {Mo}, {Xia}, \& {Deng}}]{Wu05}
{Wu}, H., {Shao}, Z., {Mo}, H.~J., {Xia}, X., \& {Deng}, Z. 2005, \apj, 622,
  244

\bibitem[{{Zhang} {et~al.}(2011){Zhang}, {Andernach}, {Caretta}, {Reiprich},
  {B{\"o}hringer}, {Puchwein}, {Sijacki}, \& {Girardi}}]{Zhang11}
{Zhang}, Y.-Y., {Andernach}, H., {Caretta}, C.~A., {et~al.} 2011, \aap, 526,
  A105

\end{thebibliography}

\end{document}